\documentclass[prd,amsmath,aps,floats,amssymb, floatfix,
  superscriptaddress,nofootinbib,twocolumn,preprintnumbers]{revtex4-1}
\usepackage{tabularx}
\usepackage{amsmath,amssymb,natbib,latexsym,times}
\DeclareMathOperator\arctanh{arctanh}
\usepackage[]{lineno}
\usepackage[T1]{fontenc}
\usepackage{aecompl}

\usepackage{multirow}
\usepackage{rotating}

\newcommand{\aap}{A\&A}

\newcommand{\mnras}{MNRAS}

\newcommand{\jcap}{JCAP}
\newcommand{\apjl}{ApJ}
\newcommand{\pasp}{PASP}
\usepackage{verbatim}
\usepackage[usenames,dvipsnames]{xcolor}
\usepackage{tabulary}
\usepackage{tabularx}
\usepackage{todonotes}
\usepackage{hyperref}
\usepackage{ulem}

\newcommand{\maglim}{MagLim}
\newcommand{\redmagic}{redMaGiC}

\newcommand{\photoz}{photo-$z$}

\newcommand\lcdm{$\Lambda$CDM}
\newcommand\wcdm{$w$CDM}

\newcommand\mcal{{\textsc{metacalibration}}}

\newcommand{\LCDM}{$ \Lambda $CDM~}

\newcommand\be{\begin{equation}}
\newcommand\ee{\end{equation}}
\def\bea{\begin{eqnarray}}
\def\eea{\end{eqnarray}}
\allowdisplaybreaks
\begin{document}
\title{Dark Energy Survey Year 3 Results:\\ Cosmological Constraints from Galaxy Clustering and Weak Lensing}

\author{T.~M.~C.~Abbott}
\affiliation{Cerro Tololo Inter-American Observatory, NSF's National Optical-Infrared Astronomy Research Laboratory, Casilla 603, La Serena, Chile}
\author{M.~Aguena}
\affiliation{Laborat\'orio Interinstitucional de e-Astronomia - LIneA, Rua Gal. Jos\'e Cristino 77, Rio de Janeiro, RJ - 20921-400, Brazil}
\author{A.~Alarcon}
\affiliation{Argonne National Laboratory, 9700 South Cass Avenue, Lemont, IL 60439, USA}
\author{S.~Allam}
\affiliation{Fermi National Accelerator Laboratory, P. O. Box 500, Batavia, IL 60510, USA}
\author{O.~Alves}
\affiliation{Department of Physics, University of Michigan, Ann Arbor, MI 48109, USA}
\affiliation{Instituto de F\'{i}sica Te\'orica, Universidade Estadual Paulista, S\~ao Paulo, Brazil}
\affiliation{Laborat\'orio Interinstitucional de e-Astronomia - LIneA, Rua Gal. Jos\'e Cristino 77, Rio de Janeiro, RJ - 20921-400, Brazil}
\author{A.~Amon}
\affiliation{Kavli Institute for Particle Astrophysics \& Cosmology, P. O. Box 2450, Stanford University, Stanford, CA 94305, USA}
\author{F.~Andrade-Oliveira}
\affiliation{Instituto de F\'{i}sica Te\'orica, Universidade Estadual Paulista, S\~ao Paulo, Brazil}
\affiliation{Laborat\'orio Interinstitucional de e-Astronomia - LIneA, Rua Gal. Jos\'e Cristino 77, Rio de Janeiro, RJ - 20921-400, Brazil}
\author{J.~Annis}
\affiliation{Fermi National Accelerator Laboratory, P. O. Box 500, Batavia, IL 60510, USA}
\author{S.~Avila}
\affiliation{Instituto de Fisica Teorica UAM/CSIC, Universidad Autonoma de Madrid, 28049 Madrid, Spain}
\author{D.~Bacon}
\affiliation{Institute of Cosmology and Gravitation, University of Portsmouth, Portsmouth, PO1 3FX, UK}
% ...
\collaboration{DES Collaboration}

\date{\today}
\label{firstpage}
\begin{abstract}
\newpage
We present the first cosmology results from large-scale structure using the full 5000 deg$^2$ of imaging data from the Dark Energy Survey (DES) Data Release 1. 
We perform an analysis of large-scale structure combining three two-point correlation functions (3$\times$2pt): (i) cosmic shear using 100 million source galaxies, (ii) galaxy clustering, and (iii) the cross-correlation of source galaxy shear with lens galaxy positions, galaxy--galaxy lensing. 
To achieve the cosmological precision enabled by these measurements has required updates to nearly every part of the analysis from DES Year 1,  including the use of two independent galaxy clustering samples, modeling advances, and several novel  improvements in the calibration of gravitational shear and photometric redshift inference. The analysis was performed under strict conditions to mitigate confirmation or observer bias; we describe specific changes made to the lens galaxy sample following unblinding of the results and tests of the robustness of our results to this decision.
We model the data within the flat $\Lambda$CDM and $w$CDM cosmological models, marginalizing over 25 nuisance parameters. 
We find consistent cosmological results between the three two-point correlation functions;
their combination yields clustering amplitude $S_8=0.776^{+0.017}_{-0.017}$ and matter density $\Omega_{\mathrm{m}} = 0.339^{+0.032}_{-0.031}$ in $\Lambda$CDM, mean with 68\% confidence limits; $S_8=0.775^{+0.026}_{-0.024}$, $\Omega_{\mathrm{m}} = 0.352^{+0.035}_{-0.041}$, and dark energy equation-of-state parameter $w=-0.98^{+0.32}_{-0.20}$ in $w$CDM.
These constraints correspond to an improvement in signal-to-noise of the DES Year 3 3$\times$2pt data relative to DES Year 1 by a factor of 2.1, about 20\% more than expected from the increase in observing area alone.
This combination of DES data is consistent with the
prediction of the model favored by the \textit{Planck} 2018 cosmic microwave background (CMB) primary anisotropy data, which is quantified with a probability-to-exceed $p=0.13$ to $0.48$.
We find better agreement between DES 3$\times$2pt and \textit{Planck} than in DES Y1, despite the significantly improved precision of both. When combining DES 3$\times$2pt data with available baryon acoustic oscillation, redshift-space distortion, and type Ia supernovae data, we find $p=0.34$. 
Combining all of these data sets with \textit{Planck} CMB lensing yields joint parameter constraints of 
$S_8 = 0.812^{+0.008}_{-0.008}$, $\Omega_{\mathrm{m}} = 0.306^{+0.004}_{-0.005}$, $h=0.680^{+0.004}_{-0.003}$, and $\sum m_{\nu}<0.13 \;\mathrm{eV\; (95\% \;CL)}$ in $\Lambda$CDM; $S_8 = 0.812^{+0.008}_{-0.008}$, $\Omega_{\mathrm{m}} = 0.302^{+0.006}_{-0.006}$, $h=0.687^{+0.006}_{-0.007}$, and $w=-1.031^{+0.030}_{-0.027}$ in $w$CDM. 
\end{abstract}
\preprint{DES-2020-0617}
\preprint{FERMILAB-PUB-21-221-AE}
\maketitle

\section{Introduction}
\label{sec:intro}

The discovery of the accelerated expansion of the universe
\cite{riess98,perlmutter99} 
led to a new standard model of cosmology, which is dominated by a
spatially smooth component with negative pressure called dark energy. Over the
intervening two decades, the evidence for the presence of dark energy has
become much stronger thanks to data from an impressive variety of cosmological probes.
Modern cosmological measurements using type Ia supernovae
\cite{Astier:2005qq,WoodVasey:2007jb,Suzuki:2011hu,Guy:2010bc,Conley:2011ku,Betoule:2014frx,Rest:2013mwz,Scolnic:2017caz,
  Abbott:2018wog}, cosmic microwave background (CMB) fluctuations
\cite{Hinshaw:2012aka,Aghanim:2018eyx,Aiola:2020azj}, galaxy clustering
\cite{Cole:2005sx,Tegmark:2006az,Blake:2012pj,Aubourg:2014yra,Alam:2016hwk,Elvin-Poole:2017xsf,Alam:2020sor},
and weak gravitational lensing
\cite{2018ARA&A..56..393M,Schrabback:2009ba,Heymans:2013fya,Jee:2015jta,Troxel:2017xyo,Hikage:2018qbn,Heymans:2020gsg}
are in agreement with a spatially flat universe with about 30\% matter
(visible and dark) and 70\% dark energy.

However, the physical nature of the dark energy that causes 
accelerated expansion remains unknown. 
The simplest and best-known phenomenological model for dark energy is the
energy density of the vacuum, incorporated in the field equations of General
Relativity by the cosmological-constant term $\Lambda$
\cite{Einstein:1917ce}. The resulting $\Lambda$ Cold Dark Matter (\lcdm) model serves as a benchmark for
tests with current and future data. Beyond \lcdm, there exists a rich set of
other potential models to explain cosmic acceleration, including evolving scalar fields, modifications to
general relativity, and other physically-motivated possibilities. This has
spawned an active research area focused on describing and modeling dark energy
and its effects on the expansion rate and the growth of density fluctuations
\citep{Frieman:2008sn,Weinberg:2012es}.

The quest to understand dark energy has spawned a worldwide effort to better
measure the growth and evolution of cosmic structure in the universe.  The current
generation of observations is spearheaded by the so-called Stage-III
 dark energy experiments, which include the Dark Energy Survey
(DES)\footnote{\url{http://www.darkenergysurvey.org/}}
\cite{DECam,Abbott:2017wau,2016MNRAS.460.1270D}, the Hyper Suprime-Cam Subaru Strategic Program (HSC)\footnote{\url{https://www.naoj.org/Projects/HSC/}}
\cite{Aihara:2017paw,Hikage:2018qbn,Hamana:2019etx}, the Kilo-Degree Survey
(KiDS)\footnote{\url{http://kids.strw.leidenuniv.nl/}}
\cite{2015MNRAS.454.3500K,Heymans:2020gsg}, and 
the Extended Baryon Oscillation Spectroscopic Survey (eBOSS)\footnote{\url{https://www.sdss.org/surveys/eboss/}} \cite{2016AJ....151...44D}. These surveys have
demonstrated the feasibility of ambitious large-scale structure analyses, and
featured extensive tests of theory, development of state-of-the-art systematics 
calibration, and new rigor in protecting analyses against observer bias before the
results are revealed. These surveys have, thus far, provided constraints
consistent with the \lcdm\ model, and
contributed to tightening the constraints on several of the key cosmological parameters
related to dark matter and dark energy.

Large-scale structure (LSS) in the universe provides a powerful set of tools
to probe dark energy. The statistics and temporal growth of cosmic structure
complement the largely geometrical sensitivity to dark energy of type Ia
supernovae and the CMB.  For nearly half a century, measurements of the galaxy
two-point correlation function, a statistic describing the
spatial clustering of galaxies, have provided pioneering cosmological constraints and early evidence for the \LCDM model~\citep{1974ApJS...28...19P,Groth:1977gj,Blumenthal:1984bp,Maddox:1990yw,Baugh:1995kz,Maddox:1996vz,Eisenstein:1999jg,1992MNRAS.254..295C,Szapudi:1997jp,Huterer:2000uj,Saunders:2000af,Hamilton:2000du,Percival:2001hw,Cole:2005sx,Padmanabhan:2006cia,Tegmark:2006az},
as well as recent, high-precision constraints on the cosmological
parameters~\citep{Anderson:2013zyy,Ho:2012vy,Carrick:2015xza,Huterer:2016uyq,Nicola:2016eua,Adams:2017val,vanUitert:2017ieu,Kohlinger:2017sxk,Elvin-Poole:2017xsf,Loureiro:2018qva,Ivanov:2019hqk,Vagnozzi:2020zrh,Andrade-Oliveira:2021yio,Hadzhiyska:2021xbv}. Another
aspect of LSS that is sensitive to both dark matter and dark energy is cosmic
shear, slight distortions of the shapes of distant background galaxies due to weak
gravitational lensing of light passing through the structures between these
sources and us.  While the interpretation of galaxy clustering is complicated
by galaxy bias \citep{Kaiser:1984sw,Desjacques:2016bnm}, cosmic shear
measurements are more directly related to the distribution of mass. First
detections of cosmic shear
\cite{Bacon:2000sy,Kaiser:2000if,vanWaerbeke:2000rm,Wittman:2000tc} have been
followed by an impressive maturing of this probe, with increasingly more
competitive constraints on cosmological parameters
\cite{Jarvis:2005ck,Massey:2007gh,Schrabback:2009ba,Lin:2011bc,Heymans:2013fya,Huff:2011aa,Kilbinger:2012qz,Jee:2015jta,Hildebrandt:2016iqg,Joudaki:2017zdt}. Finally,
galaxy--galaxy lensing, the cross-correlation of lens galaxy positions and
source galaxy shapes, provides a link between galaxy clustering and cosmic
shear.  Galaxy--galaxy lensing measurements have also matured to the point
where their combination with galaxy clustering breaks degeneracies between the
cosmological parameters and bias, thereby helping to constrain dark energy
\cite{Brainerd:1995da,Fischer:1999zx,Sheldon:2003xj,Leauthaud:2011rj,Mandelbaum:2005nx,Johnston:2007uc,Cacciato:2008hm,Mandelbaum:2012ay,Choi:2012kf,Velander:2013jga,Clampitt:2016ljk,Leauthaud:2016jdb,Kwan:2016mcy,Prat:2017goa}.
The combination of galaxy clustering, cosmic shear, and galaxy--galaxy lensing
measurements powerfully constrains structure formation in the late universe,
while strongly self-calibrating many astrophysical or systematic parameters in the model.

The stakes have become higher with recent evidence for possible tensions
between parameters as measured by different cosmological probes. These
tensions may indicate new physics beyond $\Lambda$CDM --- or else could be due
to unaccounted-for systematics or an underestimation of uncertainty in some probes. 
Potentially most significant among these is the ``Hubble tension,''
indicated by a $\sim$4--6$\sigma$ discrepancy between measurements of the Hubble
 constant inferred from the primary CMB anisotropies \cite{Aghanim:2018eyx} and higher values measured from a local distance anchor, such as, most prominently the astronomical distance ladder (e.g., \cite{Riess:2020fzl,Huang:2020}) or masers \cite{Pesce:2020}, though some measurements also indicate a lower value \cite{Freedman:2020,Birrer:2020tdcosmoIV} in better agreement with the CMB.
 The Hubble tension may indicate new physics,
and it is crucial to improve measurements, revisit assumptions and systematics \citep[e.g.,][]{Efstathiou:2020, Birrer:2020tdcosmoIV} and invest in novel, independent methods and probes \citep{Chen:2017rfc, Feeney:2018mkj}.

Additionally, several
experiments that are sensitive to the
growth of structure have historically preferred, on average, lower
values of the parameter $S_8\equiv \sigma_8 (\Omega_{\mathrm{m}}/0.3)^{0.5}$ relative to 
that predicted by the CMB anisotropy, where the
amplitude of mass fluctuations $\sigma_8$ is scaled by the square root of matter
density $\Omega_{\mathrm{m}}$. This parameter is predicted to be higher with the CMB \cite{Aghanim:2018eyx} than is measured in lensing
 (e.g., \cite{Abbott:2017wau,Troxel:2018qll, Hikage:2018qbn,Asgari:2020wuj}). The difference has been claimed by 
other experiments to be as large as $2$--$3\sigma$. Other probes of the late
universe, in particular spectroscopic galaxy clustering
\citep{Troster:2020kai}, redshift-space distortions (RSD) \citep{Alam:2020sor}, and
the abundance of galaxy clusters \citep{Mantz:2014paa,Abbott:2020knk}, also
tend to favor a lower $S_8$ than that measured by the CMB on average (assuming the
\lcdm\ model).

Previously, the DES Collaboration analyzed data from its first year of
observations, which covered 1514 deg$^2$, and constrained 
cosmological parameters using galaxy clustering and gravitational lensing in
\LCDM and \wcdm\ \citep{Abbott:2017wau}, constrained beyond-\wcdm\ models
\citep{Abbott:2018xao}, and carried out numerous other tests of the
standard cosmological framework
\citep{Troxel:2018qll,Abbott:2018wzc,Omori:2018cid,Abbott:2018ydy,Abbott:2020knk,Huang:2020tpm,To:2020bhf,Muir:2020puy,Doux:2020duk,Chen:2020iwm,Andrade-Oliveira:2021yio}.
Along with the aforementioned KiDS and HSC observations and analyses, the DES Y1 analysis
emphasized redundancy using two shape measurement methods that
are independently calibrated, several photometric redshift estimation and
validation techniques, and two independent codes for predicting the measurements
and performing a likelihood analysis.

This paper presents key cosmological constraints from the first three years of
observations (henceforth Y3) of DES.  The DES Y3 data set analyzed here uses images covering
nearly 5000 sq.\ deg., or more than three times the area of Y1. It also
dramatically increases the number of source and lens galaxies, and introduces new
techniques for the analysis and treatment of statistical and systematic
errors. As in Y1, we rely on a key cosmological probe of photometric LSS surveys,
the so-called `3$\times$2pt' analysis, consisting of three two-point correlation
functions: (i) $w(\theta)$, the angular correlation function of the lens
galaxies; (ii) $\gamma_t(\theta)$, the correlation of the tangential shear of
sources with lens galaxy positions; and (iii) $\xi_\pm(\theta)$, the
correlation functions of different components of the ellipticities of the
source galaxies.  We use these measurements only on large angular scales, for
which we have verified that a relatively simple model describes the data,
although even with this restriction we must introduce 25 free parameters to
capture astrophysical and measurement-related systematic uncertainties.
The paper is built upon and uses tools and results from 29 accompanying papers \cite{y3-gold,y3-deepfields,y3-shapecatalog,y3-piff,y3-simvalidation,y3-imagesims,y3-balrog,y3-lenswz,y3-sompz,y3-sompzbuzzard,y3-sourcewz,y3-2x2ptaltlenssompz,y3-hyperrank,y3-shearratio,y3-generalmethods,y3-covariances,y3-blinding,y3-inttensions,y3-tensions,y3-samplers,y3-2x2maglimforecast,y3-galaxyclustering,y3-gglensing,y3-massmapping,y3-2x2ptmagnification,y3-2x2ptbiasmodelling,y3-2x2ptaltlensresults,y3-cosmicshear1,y3-cosmicshear2} that are summarized in App.~\ref{sec:papers}. We summarize in App.~\ref{changes} the major updates to the analysis that are different from the DES Y1 3$\times$2pt analysis.

The cosmological quantity that is best constrained by the 3$\times$2pt analysis is
the overall amplitude of matter clustering in the low redshift universe,
parameterized by $S_8$. The precise measurement of $S_8$ in this
paper allows a powerful test for consistency between the growth of
structure and the expansion history in the broad class of cosmic acceleration
models based on General Relativity (GR) and dark energy. Implementing this
test requires a CMB anchor for the matter clustering amplitude at high
redshift, and the test becomes sharper and more general when supernova and
baryon acoustic oscillation (BAO) data are used to constrain the
expansion history. DES probes matter clustering out to $z \approx 1$,
so it also constrains dark energy models on its own through the history
of structure growth over this redshift range. The degeneracy between
$\Omega_{\mathrm{m}}$ and $\sigma_8$ in $S_8$ is broken partly by this redshift
evolution and partly by the shape of the correlation functions,
and it can be broken more strongly using external data that are
sensitive to $\Omega_{\mathrm{m}}$. Lensing measurements depend on the expansion
history through the distance-redshift relation. This dependence affects
our analysis, but the geometric constraints from DES weak lensing
are not as strong as those from current supernova and BAO data.

In subsequent sections of the paper, we focus first on the DES data
sets in Sec.~\ref{sec:data} and measurements of the three two-point correlation functions in Sec.~\ref{sec:twopoint}.
We describe the modeling and analysis in Sec.~\ref{sec:method},
then turn to the primary results, tests, and parameter
constraints from combining these measurements with additional measurements
from DES, the CMB, and other external supernova, BAO, and RSD data in Sec.~\ref{sec:results}. We conclude in Sec.~\ref{sec:con}.

\section{Dark Energy Survey Data}\label{sec:data}

The Dark Energy Survey (DES) was a six-year observing program using the Dark Energy Camera (DECam \cite{DECam}) at the Blanco 4m telescope at the Cerro Tololo Inter-American Observatory (CTIO) in Chile. 
The survey covered 5000 deg$^2$ in $grizY$ bandpasses with approximately 10 overlapping dithered exposures in each filter (90 sec in $griz$, 45 sec in $Y$) covering the survey footprint.
In this paper, we utilize data taken during the first three years of DES operations (DES Y3), which made up DES Data Release 1 (DR1 \cite{2018ApJS..239...18A}). 
This analysis uses imaging data covering the full 5000 deg$^2$ survey footprint for the first time, at approximately half the full-survey integrated exposure time.  
Preparing the imaging data for cosmological analysis is an exacting,
multi-year process, and analysis of the final six-year data set is
now in its early stages.
The data is processed, calibrated, and coadded to produce a photometric data set of 399 million objects that is further refined to a `Gold' sample for cosmological use \cite{Burke18,2018PASP..130g4501M, y3-gold}. 
The Gold sample includes selection requirements (cuts) on minimal image depth and quality, additional calibration and deblending, and quality flags to identify problematic photometry and regions of the sky with substantial photometric degradation (e.g., around bright stars). 
The Gold galaxy sample extends to a signal-to-noise $>10$ (extended) limiting magnitude of 23 in $i$-band. The final Gold sample used in this work after all cuts contains 319 million objects. 

In addition to the wide-field Gold sample, we rely on data from the DES deep fields \cite{y3-deepfields} covering a subset of the 27 deg$^2$ DES transient search regions and the separate COSMOS field \cite{Laigle:2016jxn}. 
These images are taken in $ugrizY$ bandpasses with DECam, and also have overlapping VIDEO \cite{10.1093/mnras/sts118} or UltraVISTA \cite{2012A&A...544A.156M} imaging for near-IR photometry in $YJHK$ over most of the area.
Coadd images are constructed from the best images (i.e., with smallest point-spread function (PSF) full-width half-maximum (FWHM)) with a goal of attaining a depth approximately 10$\times$ the typical wide-field coadd image depth. 
From these coadd images, we produce a deep catalog of 2.8 million objects that has 10$\sigma$ limiting magnitude of 25 in $i$-band, and photometric variance 0.1$\times$ the typical wide-field variance. 
This catalog helps to validate and calibrate our wide-field data in several ways.
First, it is used to create an input model space for representative objects to draw onto wide-field-quality images in the Balrog \cite{y3-balrog} or weak lensing image simulations \cite{y3-imagesims}. 
Balrog is used to test the survey selection function, which describes the probability that an object type drawn from a complete galaxy population will be detected in our wide-field survey, and the weak lensing simulations are used to test our shear calibration. 
The deep catalog also serves as a stepping-stone in our redshift inference methodology \cite{y3-sompz}. 
It allows us to map available spectroscopic or many-band deep photometric observations into the $ugrizJHK$ bandpass space of our deep catalog, for which we have 1.68 million sources with matched near-infrared photometry covering an area of 5.88 deg$^2$. This is then mapped through Balrog onto wide-field galaxy information. 

\subsection{Source Galaxies}
\subsubsection{Shapes}

The DES Y3 shear catalog \cite{y3-shapecatalog} is derived using the \mcal\ pipeline \cite{Huff:2017qxu, Sheldon:2017szh}, which infers the ellipticity and similar photometric properties of objects using information from the \textit{r,i,z}-bands. The pipeline is similar to that used in the DES Y1 analysis \cite{2018MNRAS.481.1149Z}, but with a number of updates, including improved PSF solutions \cite{y3-piff}, improved astrometric solutions \cite{y3-gold}, and the inclusion of an inverse-variance weighting for the galaxies. \mcal\ is able to self-calibrate the initial estimate of the shear field from the measured galaxy shapes, including sample selection biases. The current \mcal\ implementation, however, does not correct for a shear-dependent detection bias \cite{2019arXiv191102505S} that is coupled with object blending effects, which we find to cause a multiplicative bias in the shear at the level of 2-3\%. This residual bias is calibrated using image simulations \cite{y3-imagesims}. Objects are included in the catalog if they pass a number of selection cuts designed to reduce potential systematic biases \citep{y3-shapecatalog}. After additional footprint masking to match the lens catalogs, the final DES Y3 shear catalog yields 100 million galaxies covering an area of 4143 deg$^2$, with a weighted effective number density $n_{\rm eff}=5.9$ per arcmin$^2$ and corresponding shape noise {$\sigma_e = 0.26$}.

The catalog has passed a variety of empirical tests \cite{y3-shapecatalog}, mostly aimed at identifying residual additive biases in the shear estimates. Systematic errors related to PSF modeling were shown to be negligible for the DES Y3 analysis, due to improved PSF modeling \cite{y3-piff}. The B-mode signal was also shown to be consistent with zero. Other tests included the dependence of the shear estimates on galaxy and survey properties.

While shear calibration is typically viewed as separable from redshift inference, which is described in the following section, we also account for the first time for how blending correlates the ensemble shear calibration in each redshift bin with corrections to the effective shape of the $n(z)$ of each of four redshift bins \cite{y3-imagesims}. These corrections stem from a blending--detection bias, which biases both the ensemble average shear (some blends will only be detected as separate objects depending on the shear) and redshift distribution in a potentially correlated way. One way to treat these effects coherently is to model the multiplicative shear calibration as a scaling of the total number density in each redshift bin, and fit these effects fully in redshift space. We would model then the shear calibration bias as $e^i_j =\int_0^{\infty} n_\gamma(z) \gamma^i_j(z)+c$, for additive shear bias $c$, observed ellipticity $e$, true ellipticity $\gamma$, shear component $j$, and redshift bin $i$.
In practice we continue to separate a scalar multiplicative ($m$) shear bias component to be compatible with existing codes, where $n_\gamma(z)\propto (1+m) n(z)$.

\begin{figure}
\centering
\resizebox{\columnwidth}{!}{\includegraphics{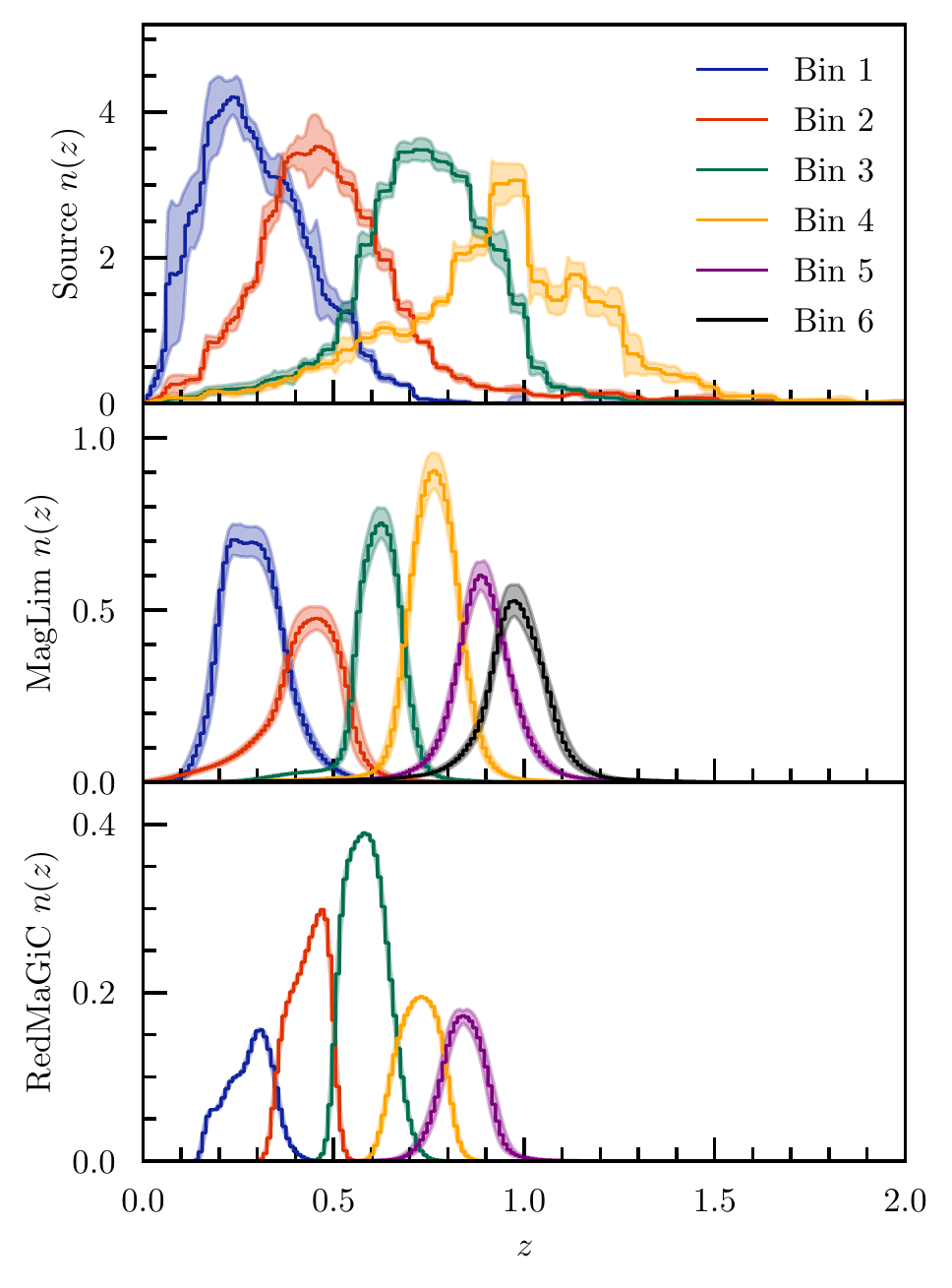}}
\caption{The source (top), \maglim\ lens (middle), and \redmagic\ lens (bottom) redshift
  distributions. The histograms are normalized to integrate to the total weighted galaxy density (arcmin$^{-2}$) in each
  tomographic bin. The equivalent 1$\sigma$ uncertainties on the redshift
  distributions are indicated by the shaded regions. The distributions have been
  corrected by non-zero mean and width offsets derived in the relevant
  photo-$z$ uncertainty models. We adopt \maglim\ as our fiducial
  lens sample in this work, and use only redshift bins 1--4.}
\label{fig:nofz}
\end{figure}

\subsubsection{Photometric redshifts}
 
The full redshift inference process for the source galaxies \cite{y3-sompz} relies on connecting information about deep-field galaxies to those in the wide field that are used for the cosmological analysis \cite{2019MNRAS.483.2801S,2019MNRAS.489..820B}. All galaxies in the deep fields with similar properties are clustered together into different `phenotypes' via a self-organizing map (SOM), while the same is also separately done for all galaxies in the wide field. The deep-field galaxies have much lower photometric noise and additional wavelength information (i.e., overlapping infrared photometry), so more phenotypes can be uniquely identified than for the wide-field galaxies. A redshift distribution is inferred for each of the deep-field galaxy phenotypes using overlapping spectroscopic \citep{Lilly09zcosmos,Masters2017,C3R2_DR2,vvds,vimos}, and photometric COSMOS \cite{Laigle:2016jxn} and PAUS \cite{2014MNRAS.442...92M,2020arXiv200711132A} redshift measurements. We then create a probabilistic mapping between the deep- and wide-field phenotypes using the Balrog simulation \cite{y3-balrog}. For example, if a given wide-field galaxy phenotype was mapped uniquely onto a single deep-field galaxy phenotype, its redshift distribution would be determined by the available redshift measurements of the deep-field galaxies that share that particular phenotype. In practice, the mapping is much more complicated: each wide-field phenotype has a non-zero probability of coming from many deep-field phenotypes, but the algorithm for generating an $n(z)$ for that galaxy phenotype is simply a weighted average. A given redshift bin is defined by a unique subset of many wide-field galaxy phenotypes, and its $n(z)$ then follows by averaging over these phenotypes. The four source redshift bins have edges $z\in[0.0,0.36,0.63,0.87,2.0]$.

The process we use to account for uncertainties accumulated in each step of this process is summarized in Ref.~\cite{y3-sompz}. These are due in part to shot-noise and cosmic variance in the redshift samples and deep fields \cite{2020MNRAS.498.2984S}, and photometric calibration uncertainty. At low redshift, this uncertainty is primarily due to uncertainties in the photometric calibration, while at high redshift it is due to a combination of cosmic variance and uncertainties in the redshift samples. Uncertainty in the $n(z)$ due to these effects are modeled or measured, and we generate many realizations of the redshift distribution, $n_i(z)$, that appropriately sample the joint space of this uncertainty without relying on a simple parameterization like mean and width. 

The emerging set of redshift realizations suffer from one further source of uncertainty that has not been explicitly included before: blending. Galaxies that are nearby one another when projected on the sky can actually be very far apart. Detection and measurement algorithms can misinterpret these blends and report not only incorrect shapes but also incorrect number densities or redshifts. To account for this effect, we created realistic simulations \cite{y3-imagesims} and  apply the same detection and measurement pipeline used for the DES data to obtain the ``observed'' number density, shape, and photometric redshift of a known simulated object population that matches our deep field data, from which the impact of blending can be inferred. The result is a likelihood model describing the impact of blending on the joint shear calibration and $n(z)$ shape that we add to the $n_i(z)$ \cite{y3-imagesims}.

We empirically constrain the likelihood of each $n_i(z)$ using information from galaxy clustering on small scales that is not used in the primary 3$\times$2pt observables~\cite{y3-sourcewz}. We know that galaxies are likely to be found near other galaxies due to gravitational clustering. Therefore, if there is a galaxy in a given direction whose redshift is known, it is likely that nearby galaxies on the sky are at a similar redshift. We make use of multiple galaxy samples with well-determined redshifts and cross-correlate them with the wide-field source sample, thereby obtaining a likelihood for each of the $n_i(z)$, which is jointly sampled with the models that produce the $n_i(z)$ before we account for blending effects. We produce several thousand $n_i(z)$ samples. We show the final redshift distributions and their uncertainties in the top panel of Fig.~\ref{fig:nofz}.

To sample over $n_i(z)$ in a likelihood analysis, we introduce a set of hyper parameters to our model that rank the $n_i(z)$ in multiple dimensions \cite{y3-hyperrank}. Ref.~\cite{y3-hyperrank} has demonstrated that our constraints on the variation in $n_i(z)$ (i.e., shown in Fig.~\ref{fig:nofz}) are sufficiently precise that uncertainty in higher-order modes in the $n_i(z)$, besides the mean redshift, were not expected to impact our cosmological constraints at a significant level in Y3. Thus in practice we simply sample the mean of the redshift distribution in the four redshift bins within a Gaussian prior based on the measured variance in the mean of each $n_i(z)$.

Finally, the measured two-point functions themselves further constrain the possible values of the redshift distributions via self-calibration in 3$\times$2pt. We further augment this by explicitly using a set of the ratios of the galaxy--galaxy lensing signal on small scales between source redshift bins sharing the same lens bin \cite{y3-shearratio}, which contains information not used in the standard 3$\times$2pt analysis. These scales are too difficult to model robustly in full, but the ratios are to first order independent of cosmological model and depend primarily on the redshift distribution and intrinsic alignment parameters, and to a lesser degree on any redshift dependent bias in shear calibration. This small-scale shear ratio likelihood is jointly sampled in the cosmological analyses.

\begin{figure*}
\centering
\resizebox{\textwidth}{!}{\includegraphics{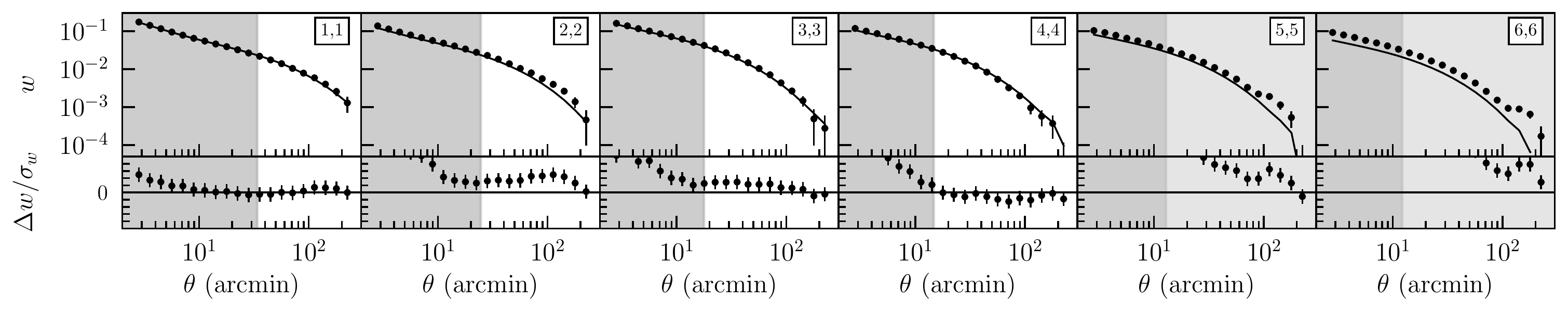}}
\caption{The measured $w(\theta)$ correlation functions for each
  tomographic bin $i$ of the \maglim\ lens galaxies (indicated by the $i,i$
  label in each panel). The best-fit $\Lambda$CDM model from the fiducial  3$\times$2pt 
  analysis is plotted as the solid line in the top part of each panel, while
  the bottom part of each  panel  shows the fractional difference between the
  measurements and the model prediction,
  $(w^{\textrm{obs.}}-w^{\textrm{th.}})/\sigma_w$ (with $y$-axis range $\pm
  5\sigma$). In both the top and bottom part of each panel, 1$\sigma$ error bars are
  shown.  Small angular scales where the linear galaxy bias assumption breaks
  down are not used in the cosmological analysis; these scales are indicated
  by grey shading. Bins 5 \& 6 are not used in the final analysis.}
\label{fig:wt}
\end{figure*}

\begin{figure*}
\centering
\resizebox{\textwidth}{!}{\includegraphics{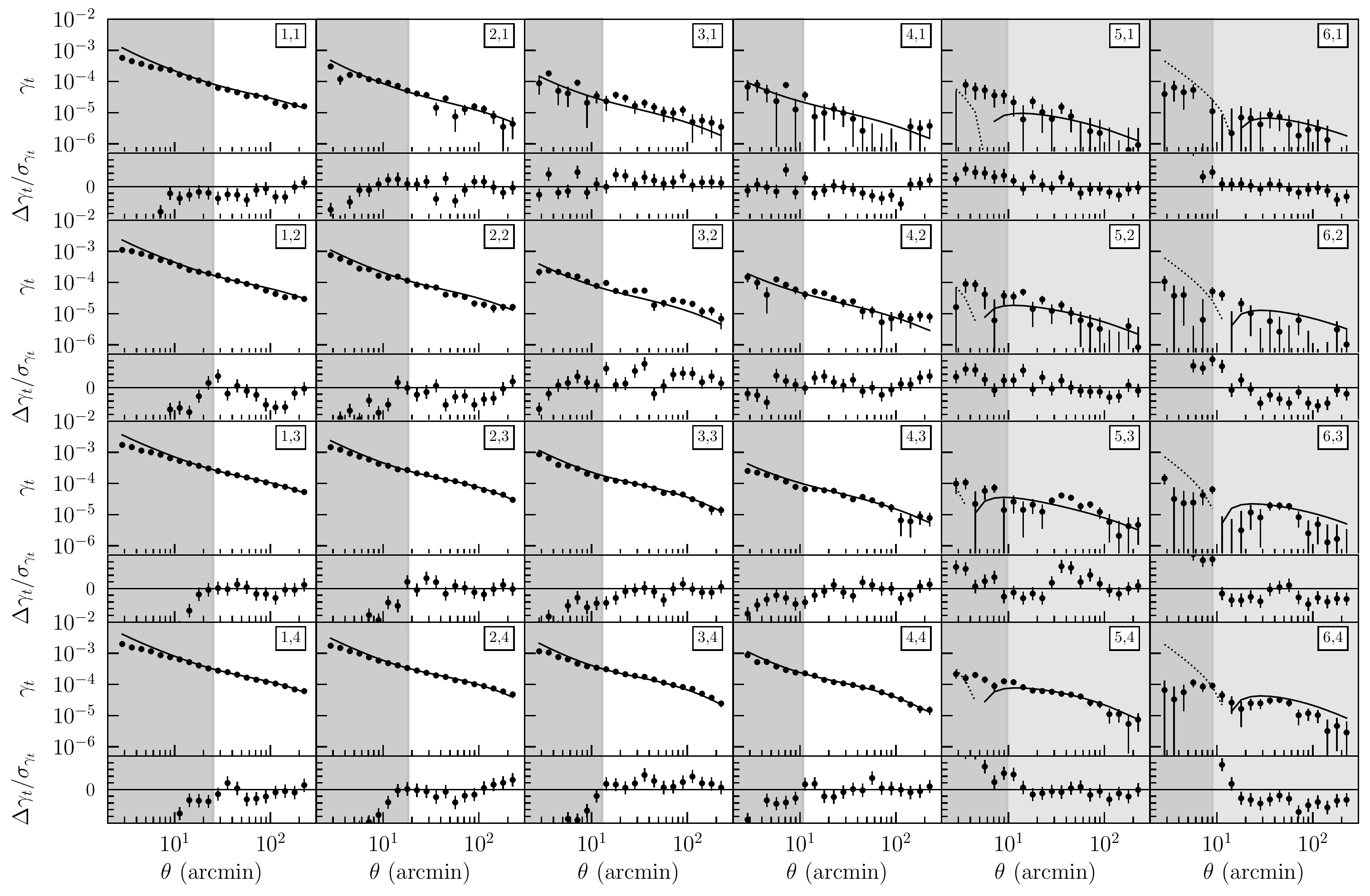}}
\caption{The measured $\gamma_{t}(\theta)$ correlation functions for each
  tomographic bin combination using the \maglim\ sample. In each panel, the
  label $i, j$ refers to \maglim\ lens tomographic bin $i$ and the source bin
  $j$ The best-fit $\Lambda$CDM model from the fiducial 3$\times$2pt analysis is plotted as
  the solid line in the top part of each panel, with dotted curves indicating
  a negative model fit. The bottom part of each panel shows the fractional
  difference between the measurements and the model prediction,
  $(\gamma_t^{\textrm{obs.}}-\gamma_t^{\textrm{th.}})/\sigma_{\gamma_t}$ (with
  $y$-axis range $\pm 5\sigma$). In both the top and bottom part of each panel, 1$\sigma$
  error bars are included. Small angular scales where the linear galaxy bias
  assumption breaks down are not used in the cosmological analysis; these
  scales are indicated by grey shading.  Bins 5 \& 6 are not used in the final
  analysis. }
\label{fig:gt}
\end{figure*}

\begin{figure}
\centering
\resizebox{\columnwidth}{!}{\includegraphics{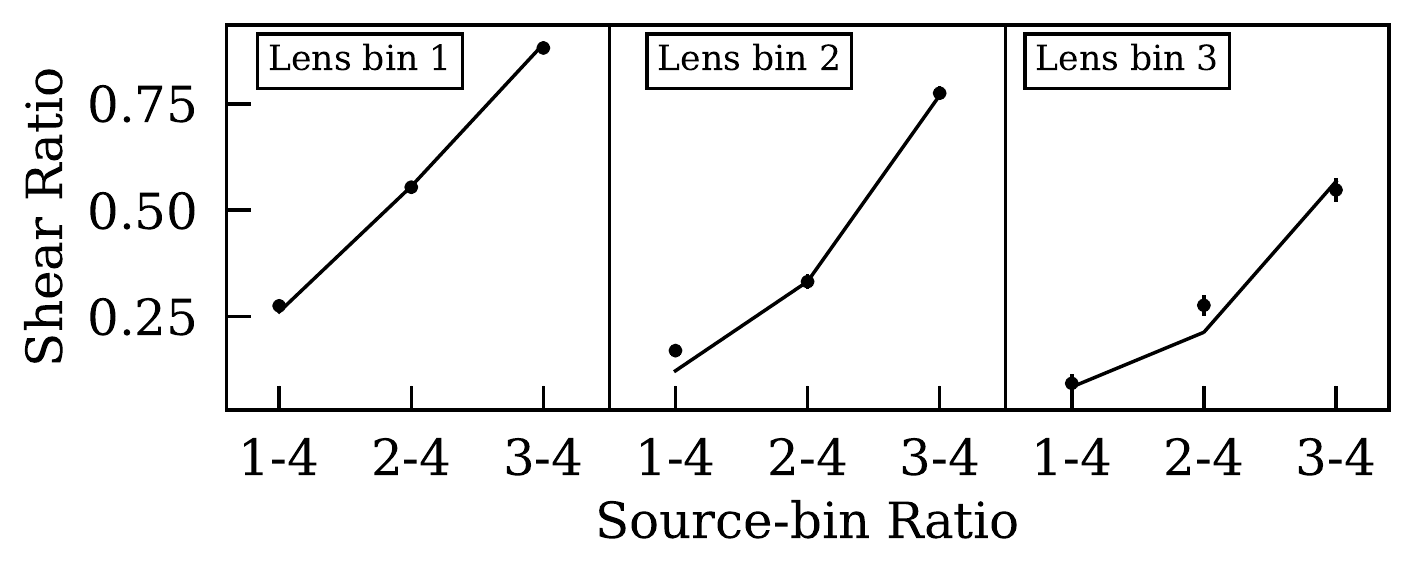}}
\caption{The measured small-scale shear ratio values for each tomographic bin combination using the \maglim\ sample, with 1$\sigma$ error bars indicated. The x-axis identifies the two source bins that make up the measured ratio. The best-fit cosmological model from the fiducial 3$\times$2pt analysis is over-plotted as the solid line for each set of lens-bin shear ratios.}
\label{fig:sr}
\end{figure}

\subsection{Lens Galaxies}

We have selected two galaxy populations (\maglim\ and \redmagic) that serve as
`lenses' in galaxy--galaxy lensing measurements and for galaxy clustering
measurements. The fiducial results presented in this work use the
\maglim\ sample. We now describe the two lens samples.

\subsubsection{\maglim\ sample}\label{sec:maglim}

We have selected a magnitude-limited lens sample
\cite{y3-2x2maglimforecast,y3-2x2ptaltlensresults}, which results in 10.7
million galaxies. This ``MagLim'' sample is defined with a magnitude cut in
the $i$-band that depends linearly on redshift, $i < 4z+ 18$, where $z$ is
the photometric redshift estimate from the Directional Neighborhood Fitting (DNF) algorithm
\cite{2016MNRAS.459.3078D,y3-gold}. This selection was optimized for $w$CDM
constraints \cite{y3-2x2maglimforecast}. 

The \maglim\ sample is divided into six tomographic bins from $z=0.2$ to $z=1.05$, with bin edges $z = [0.20, 0.40, 0.55, 0.70, 0.85, 0.95, 1.05]$. The redshift distributions from DNF are shown in the middle panel of Fig.~\ref{fig:nofz} and have been validated using galaxy clustering cross-correlations \cite{y3-lenswz}. Weights are derived to account for correlations in the number density with survey properties \cite{y3-galaxyclustering}. Further validation and characterization of the sample is described in Refs. \cite{y3-galaxyclustering,y3-2x2ptaltlensresults}. After unblinding, we discovered issues with the sample above $z=0.85$, which lead to disagreement between the galaxy clustering and galaxy--galaxy lensing signal, and contribute to a substantially poor model fit to any cosmological models considered in this work (i.e., the two right-most panels of Fig.~\ref{fig:wt}). This led us to remove these redshift bins in the fiducial analysis, which is discussed further in Secs.~\ref{lcdmsec} \& \ref{lenscomp}.

\subsubsection{\redmagic\ sample}\label{sec:redmagic}

This sample is selected with the
\redmagic\ algorithm \cite{Rozo:2015mmv}, which results in 2.6 million
galaxies. \redmagic\ selects Luminous Red Galaxies (LRGs) according to the
magnitude--color--redshift relation of red sequence galaxies, calibrated using
spectroscopic redshifts. The sample has a luminosity threshold $L_{\rm min}$
and approximately constant comoving density. The \redmagic\ sample has approximately 30\% narrower redshift distributions than \maglim, but approximately one-fourth the number of objects.

We split the \redmagic\ sample into five tomographic bins, selected on the \redmagic\ redshift point estimate quantity. The bin edges used are $z=[0.15, 0.35, 0.50, 0.65, 0.80, 0.90]$. The first three bins use a luminosity threshold of $L_{\min} > 0.5 L_{*}$ (the `high density' sample). The last two redshift bins use a luminosity threshold of $L_{\min} > 1.0 L_{*}$ (the `high luminosity' sample).
The redshift distributions are computed by stacking samples from a non-Gaussian redshift PDF of each individual \redmagic\ galaxy. Each distribution is built from several draws of the redshift PDF and are shown in the bottom panel of Fig.~\ref{fig:nofz}. The mean and RMS width of the redshift distributions are validated using galaxy clustering cross-correlations in Ref.~\cite{y3-lenswz}.

Weights are derived to account for correlations in the number density with survey properties \cite{y3-galaxyclustering}. Further validation and characterization of the sample is also described in Refs.~\cite{y3-galaxyclustering,y3-2x2ptbiasmodelling}.  We find a potential residual systematic in the \redmagic\ sample at all redshifts, which does not impact $\Lambda$CDM inference and is also discussed in Sec.~\ref{lenscomp}. 

\section{Two-point Measurements}
\label{sec:twopoint}

To extract cosmological information from the lens and source catalogs, we compute three sets of {\it two-point correlation functions}, which each measure information about how mass in the Universe is clustered. There are two fields representing the matter distribution that we can access with a galaxy survey: 1) the galaxy density field and 2) the weak lensing shear field. These two fields lead to these three sets of measured two-point functions.

\textit{Galaxy Clustering}: The two-point function between lens galaxy positions in redshift bins $i$ and $j$, $w^{ij}(\theta)$, describes the excess (over random) number of galaxy pairs separated by an angular distance $\theta$. The estimator for $w^{ij}(\theta)$ and its measurement and validation process are described in detail in Ref.~\cite{y3-galaxyclustering}. We only use the auto-correlations of the measured $w^{ii}(\theta)$ in our analysis; these are shown with their uncertainties in Fig.~\ref{fig:wt} for \maglim\ and in App. \ref{sec:redmagic} for \redmagic.

\textit{Galaxy--Galaxy Lensing}: The two-point function between lens galaxy positions and source galaxy tangential shear in redshift bins $i$ and $j$, $\gamma_t^{ij}(\theta)$, describes the over-density of mass around galaxy positions. The matter correlated with the lens galaxy alters the path of the light emitted by the source galaxy, thereby distorting its shape. The estimator for $\gamma_t^{ij}(\theta)$ and its measurement and validation process are described in detail in Ref.~\cite{y3-gglensing}. The measured $\gamma_t^{ij}(\theta)$ and their uncertainties are shown in Fig.~\ref{fig:gt} for \maglim\ and in App. \ref{sec:redmagic} for \redmagic. In addition, we include small-scale shear ratio information below the scale cuts used for $\gamma_t$. These ratios are constructed from $\gamma_t$ measurements using different source galaxy bins, while keeping the lens bin fixed. This effectively erases their dependence on the galaxy power spectrum, but keeps information about redshift calibration, shear calibration, and galaxy intrinsic alignment.  A detailed description of shear ratios and their validation can be found in Ref.~\cite{y3-shearratio}. Fig.~\ref{fig:sr} shows the shear-ratio measurement and uncertainties. This shear ratio data is included when analyzing all combinations of the three primary two-point functions in our analyses, unless otherwise noted.

\begin{figure*}
\centering
\resizebox{\textwidth}{!}{\includegraphics{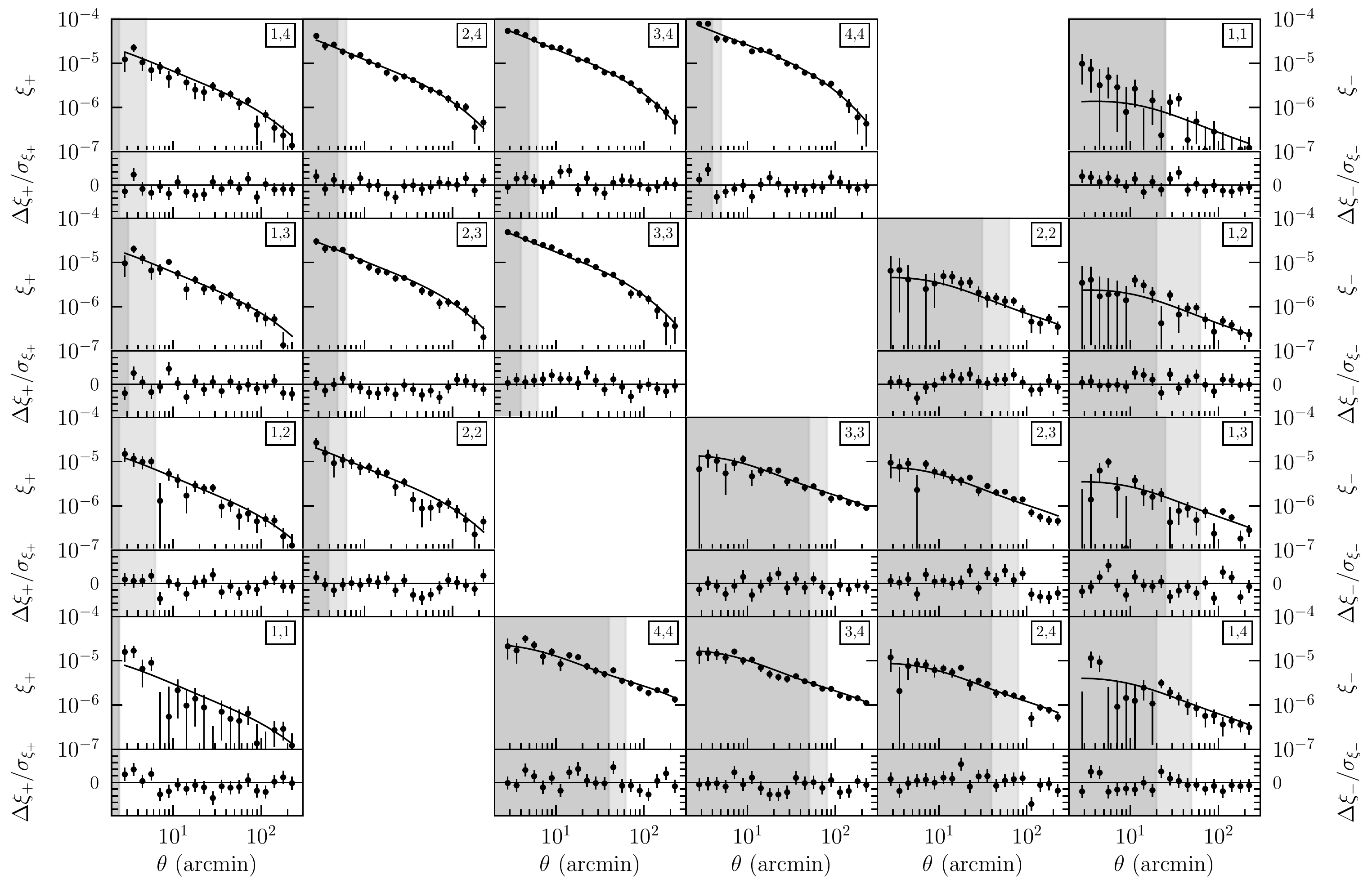}}
\caption{The measured $\xi_{\pm}(\theta)$ correlation functions for each
  tomographic bin combination, with labels as described in
  Fig.~\ref{fig:gt}. The best-fit $\Lambda$CDM model from the fiducial 3$\times$2pt 
  analysis is plotted as the solid line in the top part of each panel, while
  the bottom part of each panel shows the fractional difference between the
  measurements and the model prediction,
  $(\xi_{\pm}^{\textrm{obs.}}-\xi_{\pm}^{\textrm{th.}})/\sigma_{\xi_{\pm}}$
  (with $y$-axis range $\pm 5\sigma$). In both the top and bottom part of each
  panel, 1$\sigma$ error bars are included.  The shaded regions (both light and dark) indicate scales
  not used in the fiducial analysis, primarily due to uncertainties in the
  impact of baryonic effects. The lighter shaded regions indicate scales
  that are used in an $\Lambda$CDM-optimized analysis, which meets our
  criterion for scale cuts described in Sec.~\ref{sec:method} in $\Lambda$CDM
  only. }
\label{fig:xipm}
\end{figure*}

\textit{Cosmic Shear}: The correlation between source galaxy shears in redshift bins $i$ and $j$ is described by the two functions $\xi_{\pm}^{ij}(\theta)$, which are the sum and difference of the products of the tangential- and cross-components of the projected shear. The estimator for $\xi_{\pm}^{ij}(\theta)$ and its measurement and validation process are described in detail in Refs. \cite{y3-cosmicshear1,y3-cosmicshear2}. The measured $\xi_{\pm}^{ij}(\theta)$ and their uncertainties are shown in Fig.~\ref{fig:xipm}.

The total data vector includes measurements from five or six lens redshift bins and four source redshift bins, shown in Fig.~\ref{fig:nofz}, split into 20 logarithmic angular bins between 2.5 and 250 arcmin, for a total of 1300 elements (not including shear-ratio). After bin pair removal for $w(\theta)$, imposing a post-unblinding maximum lens redshift cut, and other scale cuts, 462 elements remain in the final 3$\times$2pt data vector. The scale cut choices and their validation are described in Refs.~\cite{y3-cosmicshear2,y3-2x2ptbiasmodelling,y3-generalmethods,y3-simvalidation}, but are generally set to control the impact of unmodeled non-linear effects (e.g., baryonic effects on the matter power spectrum or higher-order galaxy bias) to better than $0.3\sigma$ in the $\Omega_\mathrm{m}$--$S_8$ plane in $\Lambda$- and $w$CDM. A choice of scale cuts that require biases meet our fiducial requirements in $\Lambda$CDM-only ($\Lambda$CDM-optimized) leaves 508 data points. All measurements are made using TreeCorr \cite{Jarvis:2003wq}. We find a total signal-to-noise $S/N = 87$ for the 3$\times$2pt data vector after fiducial scale cuts, where $S/N\equiv \xi_{\mathrm{data}} C^{-1} \xi_{\mathrm{model}}/\sqrt{\xi_{\mathrm{model}} C^{-1} \xi_{\mathrm{model}}}$, with covariance matrix $C$ and best-fit model $\xi_{\mathrm{model}}$. This is a factor of 2.1 improvement over the DES Year 1 3$\times$2pt $S/N$.

All of these measurements are related to the underlying clustering of matter in the Universe, but in different ways. The relationship between the galaxy density and the underlying matter density is complex~\cite{y3-2x2ptbiasmodelling} and needs to be modeled with care. Alternately, the shape distortions depend more directly on the intervening matter, but the measurements themselves --- especially of shapes and the redshift distribution --- require greater care. Our work on this calibration is summarized in Appendix~\ref{sec:papers}. The advantage of using all of these measurements is that the systematic difficulties differ from one to another, but they all measure the same underlying matter field. Hence, by comparing the results from each set, we obtain a measure of consistency and additional ability to self-calibrate systematics, thereby giving confidence that we are correctly inferring information about the clustering of matter and the cosmological model. 
 
\section{Analysis}\label{sec:method}

To infer parameters $\mathbf p$ from the measured two-point functions, we
compare data organized in a ``data vector'' $\hat{\mathbf D}$,
\be
\hat{\mathbf D} \equiv\{\hat{w}^i(\theta),\hat{\gamma}_\mathrm{t}^{ij}(\theta),\hat{\xi}_\pm^{ij}(\theta)\} ,
\ee
 to a theoretical model prediction organized in a vector $\mathbf T_M$ of
   two-point correlation functions that are computed using the  parameters $\mathbf p$ of a given model ${M}$,
  \be
 \mathbf T_M(\mathbf{p}) \equiv \{w^i(\theta,\mathbf p),\gamma_\mathrm{t}^{ij}(\theta,\mathbf p),\xi_\pm^{ij}(\theta,\mathbf p)\},
 \ee
 assuming a Gaussian likelihood,
\be
\mathcal{L}(\hat{\mathbf D}|\mathbf p, M) \propto e^{-\frac{1}{2}\left[\left(\hat{\mathbf D}-\mathbf T_M(\mathbf p)\right)^{\mathrm{T}} \mathbf{C}^{-1}\left(\hat{\mathbf D}-\mathbf T_M(\mathbf p)\right)\right]}.
\ee
Here
$\mathbf{C}$ is the data covariance, which is obtained through analytic modeling as described and validated in Ref.~\cite{y3-covariances}. 

We construct a posterior probability distribution for the parameters  $\mathbf p$  of the theoretical model given the data $\hat{\mathbf D}$ as  
\be
P(\mathbf p |\hat{\mathbf D}, M) \propto \mathcal{L}(\hat{\mathbf D}|\mathbf{p}, {M})P(\mathbf{p}|{M}),
\ee
where $P(\mathbf{p}|{M})$ is a prior probability distribution on the parameters.
The proportionality constant is given by the inverse of the Bayesian Evidence
\be
P(\hat{\mathbf D}| M) = \int d\mathbf{p}\; \mathcal{L}(\hat{\mathbf D}|\mathbf{p}, {M})P(\mathbf{p}|{M}),
\ee
which corresponds to the marginalized probability of a dataset being produced under a given theoretical model.

This section summarizes the theoretical model and parameterization we use for $\mathbf
T_M(\mathbf p)$, which is described in more detail in
Ref.~\cite{y3-generalmethods} and validated in
Refs. \cite{y3-generalmethods,y3-2x2ptbiasmodelling,y3-2x2ptmagnification,y3-simvalidation}. For
clarity, we drop the parameter argument of the theoretical model predictions, such that,
e.g., the predicted clustering signal is simply denoted as $w^i(\theta)$. 

We report the mean in each parameter, along with the 68\% confidence limit (CL) of posterior volume around the mean.
For completeness, we also report the best-fit maximum posterior values.
We have used both a parameter-level and $\chi^2$ criterion for limiting the contribution
  of any systematic error to bias in the cosmological parameters. The threshold for this criterion
  is intended to limit the expected total bias in the 2D
  marginalized $\Omega_{\mathrm{m}}$--$S_8$ plane from several independent potential 
  sources of model bias to be contained within the
  68\% C.L.\ region \cite{y3-generalmethods} ($<0.3\sigma$ for any single contribution). The difference
between the mean and best-fit values can give an indication of the magnitude
of projection or non-Gaussian effects in the marginalized parameter
posteriors. The estimated impact of projection or volume effects in the DES Year 3 3$\times$2pt 
posteriors are tested and summarized in Ref.~\cite{y3-generalmethods}.
We also provide a 2D figure of merit (FoM) defined for
two parameters as FoM$_{p_1,p_2}=\left(\det
\mathrm{Cov}(p_1,p_2)\right)^{-1/2}$
\cite{PhysRevD.64.123527,PhysRevD.77.123525}. The FoM is
 proportional to the inverse area of the confidence region in the
space of the two parameters, and can be considered a summary
statistic that enables a straightforward comparison of constraining power of
experiments or analysis scenarios.

The analysis was designed and validated without access to the true cosmological results to protect against confirmation or observer bias. This process is described in detail in App.~\ref{sec:blind}.

\begin{table}
\caption{The model parameters and their priors used in the fiducial flat $\Lambda$CDM and $w$CDM analyses. The parameter $w$ is fixed to $-1$ in $\Lambda$CDM. The parameters are defined in Sec.~\ref{subsec:params}.}
\begin{center}
\begin{tabular*}{\columnwidth}{ l  @{\extracolsep{\fill}} c  c}
\hline
\hline
Parameter & \multicolumn{2}{c}{Prior}  \\  
\hline 
\multicolumn{2}{l}{{\bf Cosmology}} \\
$\Omega_{\mathrm{m}}$  &  Flat  & (0.1, 0.9)  \\ 
$10^{9}A_{\mathrm{s}}$ &  Flat  & ($0.5,5.0$)  \\ 
$n_{\mathrm{s}}$ &  Flat  & (0.87, 1.07)  \\
$\Omega_{\mathrm{b}}$ &  Flat  & (0.03, 0.07)  \\
$h$  &  Flat  & (0.55, 0.91)   \\
$10^{3}\Omega_\nu h^2$  & Flat  & ($0.60$, $6.44$) \\
$w$ &   Flat  & ($-2.0$, $-0.33$)   \\
\hline
\multicolumn{2}{l}{{\bf Lens Galaxy Bias} } \\
$b_{i} (i\in[1,4])$   & Flat  & (0.8, 3.0) \\
\hline
\multicolumn{2}{l}{{\bf Lens magnification} } \\
$C_{\rm l}^1 $ & Fixed &  $0.42$ \\
$C_{\rm l}^2 $ & Fixed &  $0.30$ \\
$C_{\rm l}^3 $ & Fixed &  $1.76$ \\
$C_{\rm l}^4 $ & Fixed &  $1.94$ \\
\hline
\multicolumn{2}{l}{{\bf Lens \photoz\ } } \\
$\Delta z^1_{\rm l} \times 10^{2}$  & Gaussian  & ($-0.9, 0.7$) \\
$\Delta z^2_{\rm l} \times 10^{2}$  & Gaussian  & ($-3.5, 1.1$) \\
$\Delta z^3_{\rm l} \times 10^{2}$  & Gaussian  & ($-0.5, 0.6$) \\
$\Delta z^4_{\rm l} \times 10^{2}$  & Gaussian  & ($-0.7, 0.6$) \\
$\sigma^1_{z,\rm l}$  & Gaussian  & ($0.98, 0.06$) \\
$\sigma^2_{z,\rm l}$  & Gaussian  & ($1.31, 0.09$) \\
$\sigma^3_{z,\rm l}$  & Gaussian  & ($0.87, 0.05$) \\
$\sigma^4_{z,\rm l}$  & Gaussian  & ($0.92, 0.05$) \\
\hline
\multicolumn{2}{l}{{\bf Intrinsic Alignment}} \\
$a_{i}$ ($i\in [1,2]$)   & Flat &  ($-5,5$) \\
$\eta_{i}$ ($i\in [1,2]$)  & Flat  & ($-5,5$) \\
$b_{\mathrm{TA}}$   & Flat  & ($0,2$) \\
$z_0$ & Fixed  & 0.62 \\
\hline
\multicolumn{2}{l}{{\bf Source \photoz}} \\
$\Delta z^1_{\rm s} \times 10^{2}$  & Gaussian  & ($0.0, 1.8$) \\
$\Delta z^2_{\rm s} \times 10^{2}$  & Gaussian  & ($0.0, 1.5$) \\
$\Delta z^3_{\rm s} \times 10^{2}$  & Gaussian  & ($0.0, 1.1$) \\
$\Delta z^4_{\rm s} \times 10^{2}$  & Gaussian  & ($0.0, 1.7$) \\
\hline
\multicolumn{2}{l}{{\bf Shear calibration}} \\
$m^1 \times 10^{2}$ & Gaussian  & ($-0.6, 0.9$)\\
$m^2 \times 10^{2}$ & Gaussian  & ($-2.0, 0.8$)\\
$m^3 \times 10^{2}$ & Gaussian  & ($-2.4, 0.8$)\\
$m^4 \times 10^{2}$ & Gaussian  & ($-3.7, 0.8$)\\
\hline
\hline
\end{tabular*}
\end{center}
\label{tab:params}
\vspace{-0.9cm}
\end{table}

\subsection{Model} \label{sec:model}
We model the observed projected (lens) galaxy density contrast $\delta_\mathrm{obs}^i(\hat{\mathbf n})$ as a combination of projected galaxy density contrast and modulation by magnification, $\delta_\mu$,
\be
\delta_\mathrm{obs}^i(\hat{\mathbf n}) = \delta_\mathrm{g}^i(\hat{\mathbf n}) + \delta_\mu^i(\hat{\mathbf n})\,
\label{eq:deltag}
\ee
for position vector $\hat{\mathbf n}$, where $i$ and $j$ represent the redshift bin. The observed shear signal $\gamma$ is modeled as the sum of gravitational shear, $\gamma_{\mathrm G}$, and intrinsic alignments, $\epsilon_\mathrm{I}$,
\be
 \gamma_{\alpha}^{j}(\hat{\mathbf n}) = \gamma_{\mathrm G,\alpha}^{j}(\hat{\mathbf n})+\epsilon_{\mathrm{I},\alpha}^{j}(\hat{\mathbf n})\,,
\ee
with $\alpha$ the shear components.
While B-modes produced by higher-order weak lensing effects are negligible for our analysis, it is important to account for B-modes generated by intrinsic alignments in the computation of cosmic shear two-point correlation functions. In Fourier space, this decomposition can be written as 
\begin{align}
\gamma^j_{\mathrm E}(\mathbf\ell)=\kappa^j(\mathbf\ell)+\epsilon_{\mathrm{I, E}}^{j}(\mathbf\ell)\,, && \gamma^j_{\mathrm B}(\mathbf\ell) = \epsilon_{\mathrm{I, B}}^{j}(\mathbf\ell)\,, 
\end{align}
with the convergence field
\be
\kappa^j(\hat{\mathbf n})=\int d\chi \,W_{\kappa}^j(\chi)\delta_\mathrm{m}\left(\hat{\mathbf n} \chi, \chi\right)\,,
 \ee
where $\delta_\mathrm{m}$ is the 3D matter density contrast. The galaxy density contrast $\delta_{\mathrm{g}}$ is related to $\delta_{\mathrm{m}}$ via a linear galaxy bias $b_i$. The tomographic lens efficiency is
 \be
 W_{\kappa}^j (\chi)= \frac{3\Omega_{\mathrm{m}} H_0^2}{2}\int_\chi^{\chi_H} d\chi' n_{\mathrm{s}}^j(\chi')
 \frac{\chi}{a(\chi)}\frac{\chi'-\chi}{\chi'}\,.
 \ee
 $\chi$ is the comoving distance, $\chi_H$ the comoving distance to the horizon, $n_s(\chi)$ the source galaxy number density distribution, and $a(\chi)$ the scale factor. 

\subsubsection{Two-point statistics}
The angular power spectra $C(\ell)$ of these observed fields can be written as 
\begin{align}
\nonumber C^{ij}_{\mathrm{EE}}(\ell) = & C^{ij}_{\kappa\kappa}(\ell) + C^{ij}_{\kappa \mathrm{I_E}}(\ell) + C^{ji}_{\kappa \mathrm{I_E}}(\ell)+C^{ij}_{\mathrm{I_E}\mathrm{I_E}}(\ell) \\
\nonumber C^{ij}_{\mathrm{BB}}(\ell) = & C^{ij}_{\mathrm{I_B}\mathrm{I_B}}\\
\nonumber C^{ij}_{\delta_{\mathrm{obs}}\mathrm{E}}(\ell) = &C^{ij}_{\delta_{\mathrm{g}}\kappa}(\ell) + C^{ij}_{\delta_{\mathrm{g}}\mathrm{I_E}}(\ell) +  C^{ij}_{\delta_{\mu}\kappa}(\ell) + C^{ij}_{\delta_{\mu}\mathrm{I_E}}(\ell)\\
\nonumber C^{ii}_{\delta_{\mathrm{obs}}\delta_{\mathrm{obs}}}(\ell) =&C^{ii}_{\delta_{\mathrm{g}}\delta_{\mathrm{g}}}(\ell)+C^{ii}_{\delta_{\mu}\delta_{\mu}}(\ell) +C^{ii}_{\delta_{\rm RSD}\delta_{\rm RSD}}(\ell) \\ &+ 2C^{ii}_{\delta_{\mathrm{g}}\delta_{\mu}}(\ell)+2C^{ii}_{\delta_{\rm g}\delta_{\rm RSD}}(\ell)+2C^{ii}_{\delta_{\rm RSD}\delta_{\mu}}(\ell)\,.
\end{align}
With the exception of the galaxy clustering power spectra $C_{\delta_{\mathrm{obs}}\delta_{\mathrm{obs}}}$, which are evaluated using the method described in Ref.~\cite{Fang2020}, we calculate the angular cross-power spectrum between two fields $A,B$ using the Limber approximation 
 \begin{align}
     C_{AB}^{ij}(\ell) = \int d\chi \frac{W_A^i(\chi)W_B^j(\chi)}{\chi^2}P_{AB}\left(k = \frac{\ell+\frac{1}{2}}{\chi},z(\chi)\right)\,,
 \end{align}
 with $P_{AB}$ the corresponding three-dimensional power spectrum, which is specified by the parameterization choices summarized in \ref{subsec:params}.
 The kernels $W^{ij}_{A,B}$ correspond to $W^j_\kappa$ for shear and the lens galaxy density $n^i_l$ for position.
The two-point correlation functions within an angular bin $[\theta_{\rm min},\theta_{\rm max}]$ are related to the projected power spectra as
\begin{align}
\nonumber w^i(\theta) = &\sum_{\ell} \mathcal G_0\left(\ell,\theta_{\rm min},\theta_{\rm max}\right) C^{ii}_{\delta_{\mathrm{obs}}\delta_{\mathrm{obs}}}(\ell)\\
\gamma_\mathrm{t}^{ij}(\theta) = &\sum_{\ell} \mathcal G_2\left(\ell,\theta_{\rm min},\theta_{\rm max}\right) C^{ij}_{\delta_{\mathrm{obs}}\mathrm{E}}(\ell) \\
\nonumber \xi_{\pm}^{ij}(\theta) = &\sum_{\ell} \mathcal G_{4,\pm}\left(\ell,\theta_{\rm min},\theta_{\rm max}\right) \left[  C^{ij}_{\mathrm{EE}}(\ell)\pm C^{ij}_{\mathrm{BB}}(\ell)\right]\,,
\end{align}
with $\mathcal G_n$ analytic functions detailed in Refs.~\cite{y3-covariances,y3-generalmethods}\,.
\subsection{Parameterization and Priors}
\label{subsec:params}

We sample the posterior of these measurements in two cosmological models: flat $\Lambda$CDM and $w$CDM, with the sum of the three neutrino masses as a free parameter, where the impact of neutrino mass on the power spectrum is modeled via a fitting function \cite{2012MNRAS.420.2551B}. $\Lambda$CDM contains three energy densities in units of the critical density: the total matter density $\Omega_{\mathrm{m}}$, the baryonic density $\Omega_{\mathrm{b}}$, and the massive neutrino density $\Omega_{\nu}$. We vary $\Omega_{\nu}h^2$, where $h$ is the Hubble parameter, as a free parameter, while noting that it is often fixed in other cosmological analyses to be zero or to the minimum mass allowed by oscillation experiments $m_{\nu}=0.06$ eV \cite{pdg}.

The other cosmological parameters we vary within $\Lambda$CDM are the Hubble parameter $h$, the amplitude of primordial scalar density perturbations $A_{\mathrm{s}}$, and the spectral index $n_{\mathrm{s}}$ of the power spectrum. We assume a flat model, with $\Omega_{\Lambda} = 1-\Omega_{\mathrm{m}}$. In $w$CDM, we allow for a free dark energy equation-of-state parameter $w$ that is constant in time (in $\Lambda$CDM, this is fixed to $w=-1$, corresponding to a cosmological constant). Thus $\Lambda$CDM includes six free cosmological parameters and $w$CDM contains seven. The prior ranges for cosmological parameters in Table \ref{tab:params} are either motivated by physical constraints (e.g., an accelerating universe requires $w<-1/3$), or for parameters that are not strongly constrained by the DES data, typically given a range that encompasses five times the 68\% C.L. from relevant external constraints. In analyses that sample external CMB likelihoods, we include the optical depth $\tau$ as a free parameter.

We will typically refer to the amplitude of density perturbations at $z=0$ in terms of the RMS amplitude of mass on scales of $8h^{-1}$ Mpc in linear theory, $\sigma_8$. The constraints on the amplitude and density of matter fluctuations are degenerate in our analysis, and we will also refer to the parameter $S_8$,
which describes the width of the posterior in the direction roughly orthogonal to the primary degeneracy direction for cosmic shear in the $\sigma_8$--$\Omega_{\mathrm{m}}$ plane \cite{jain97}, though this does not hold exactly for 3$\times$2pt and changes with effective redshift.

In addition to these cosmological parameters, our fiducial analysis includes an additional 25 free parameters, for a total of 31 (32) parameters in $\Lambda$CDM ($w$CDM). These additional parameters describe astrophysical and systematic contributions to the measured signal. The effective linear galaxy bias of lens galaxies in each redshift bin is parameterized by a scalar $b^i$. We also test and apply a nonlinear galaxy bias model (with one extra free parameter per redshift bin) \cite{y3-2x2ptbiasmodelling,y3-simvalidation, y3-2x2ptaltlensresults}, which is described in App.~\ref{nlbias}.  The intrinsic alignment of galaxies \cite{Troxel:2014dba,2015SSRv..193....1J} is modeled with the Tidal Alignment and Tidal Torquing (TATT) model \cite{2019PhRvD.100j3506B}, which is parameterized by an amplitude $a_i$ and redshift power-law $\eta_i$ parameter (with redshift pivot $z_0=0.62$), for each of the (1) tidal alignment- and (2) tidal torquing-sourced terms in the model, as well as an effective source galaxy bias parameter $b_{\mathrm{TA}}$, which is described in further detail in Refs.~\cite{y3-generalmethods,2019PhRvD.100j3506B}. The TATT model contains the commonly employed non-linear linear alignment (NLA) model in the $a_2 =b_{\mathrm{TA}}=0$ subspace. The amplitude of the lens magnification term in Eq.~\ref{eq:deltag} depends on the slope of the lens sample's luminosity and size distribution at the sample detection limit. The corresponding parameter 
$C^i_{\mathrm{l}}$
 is calibrated from the data, as described in Ref.~\cite{y3-2x2ptmagnification}, and held fixed to that value.  
 Nonlocal effects in $\gamma_t$ can significantly contaminate larger angular scales with nonlinear information due to integration of the projected mass within a given angular separation from the center of the halo. This is mitigated by analytically marginalizing over a free point-mass contribution to $\gamma_t$ in all analyses \cite{2020MNRAS.491.5498M}.

Photometric redshift systematics are parameterized by an additive shift to the mean redshift of each bin, $\Delta z_{\mathrm{l}}^i$ for lenses and $\Delta z_{\mathrm{s}}^i$ for sources, where the true redshift distribution is related to the photometric redshift distribution $n_{pz}$ such that 
 \be
n^i(z) = n^i_{pz}(z-\Delta z^i).
 \ee
In addition, differences in the width of the lens redshift distribution are important at DES Y3 precision, which we parameterize by a stretch $\sigma_{z}^i$, such that
 \be
n^i(z) = \sigma_z^i n^i_{pz}\left(\sigma_z^i[z-\langle z\rangle]+\langle z\rangle\right).
 \ee
Finally, uncertainty in the shear calibration bias is parameterized by $m^i$, where the measured ellipticity $e_j$ is related to the true shear $\gamma_j$ in each bin by
 \be
e^i_j = (1+m^i)\gamma^i_j. 
 \ee
 The full set of parameters (cosmological, astrophysical, and systematic) and their priors are summarized in Table \ref{tab:params}.
 
 Differences in the \redmagic\ analysis are described in App.~\ref{sec:redmagic}.

\subsection{Likelihood Analysis}

Our likelihood analysis uses two independently developed analysis and inference pipelines, \textsc{CosmoSIS} \citep{Zuntz:2014csq} and \textsc{CosmoLike} \citep{Krause:2016jvl}, which have been validated against one another to ensure they produce consistent predictions of the observables and final cosmological constraints. A comparison of the theory predictions from \textsc{CosmoSIS} and \textsc{CosmoLike} is presented in Ref.~\cite{y3-generalmethods}. The residual offset of $\chi^2 < 0.2$ between 3$\times$2pt model data vectors obtained from both codes in this analysis at a reference cosmology is found to have negligible impact on parameter constraints and we conclude that both pipelines can be used interchangeably.

\subsubsection{CosmoSIS}
This pipeline uses the \textsc{CAMB}
Boltzmann code \citep{Lewis:1999bs,2012JCAP...04..027H} to compute
underlying background quantities and the linear matter power spectrum, and the \textsc{Halofit} \cite{2003MNRAS.341.1311S} version
presented in Ref.~\cite{Takahashi:2012em} for the non-linear power spectrum.  It then generates theory predictions
following the model described in section
\ref{sec:model}, and using the Fast-PT method \citep{FastPT} 
for non-linear galaxy bias and the TATT model for intrinsic alignments.
Non-Limber integrals are computed following the method of \cite{Fang2020}. Accuracy parameters throughout the pipeline are chosen by requiring the log-likelihood to differ by less than 0.05 from a high precision calculation.
For chains including \textit{Planck} CMB measurements \citep{Aghanim:2018eyx}, we use the \textit{Planck} 2018 public
likelihood code \citep{Aghanim:2019ame}.\footnote{While running the \textit{Planck} \textsc{clik} lensing likelihood from plc-3.0, discrepancies between the constraints obtained using the likelihood code and the publicly released chains were found. The disagreement has been identified to originate from the treatment of the linear correction term to the theory $C_{L}^{\phi\phi}$ spectrum. This has been corrected in the upstream plc-3.01 release.}

The version of \textsc{CosmoSIS}\footnote{\url{https://bitbucket.org/joezuntz/CosmoSIS} \\ \url{https://bitbucket.org/joezuntz/CosmoSIS-standard-library}} used for the analysis may be found in the \texttt{des-y3} branch of the repositories.
The \textsc{CosmoSIS} runs presented here use the \texttt{PolyChord} sampling method \citep{Polychord1,Polychord2},
for both posterior samples and Bayesian evidence. The shear calibration values $m_i$ are used as fast parameters.
The \texttt{PolyChord} parameters we use for our fiducial runs are: fast\_fraction $ = 0.1$,
live\_points $=500$, num\_repeats $=60$, tolerance $=0.01$, and boost\_posteriors $=10.0$.

Analyses with the \textsc{CosmoSIS} pipeline also use the non-Gaussian covariance matrix from \textsc{CosmoLike} described below.

\subsubsection{CosmoLike}
This pipeline uses the \textsc{CLASS}
Boltzmann code \citep{CLASSII} to compute
underlying background quantities and linear and non-linear matter power spectra, using the \textsc{Halofit} version
presented in Ref.~\cite{Takahashi:2012em} for the latter. The theory predictions are calculated using the model described in section
\ref{sec:model}, relying on the \textsc{Fast-PT} method \citep{FastPT,2017JCAP...02..030F} to evaluate integrals over perturbation theory kernels for non-linear galaxy bias and the TATT intrinsic
alignment model. The computation of non-Limber integrals for galaxy clustering further employs the FFTLog implementation of Ref.~\cite{Fang2020}.

The evaluation time of angular two-point statistics in \textsc{CosmoLike} is optimized through a series of interpolation schemes, for which runtime-optimized accuracy settings are validated through comparison to high-accuracy evaluations with slow runtime. Most DES Y3 likelihood analyses with \textsc{CosmoLike} employ the \textsc{emcee} \citep{emcee} sampler, c.f.~\cite{Miranda2020} for a detailed comparison of sampler configurations for \textsc{CosmoLike}  likelihood analyses.

The \textsc{CosmoCov} module \citep{CosmoCov} of \textsc{CosmoLike} is used to generate covariances for DES Y3 analyses, which include Gaussian and non-Gaussian terms \citep{Krause:2016jvl} and account for the effect of the survey geometry on shape and shot-noise terms \citep{Troxel:2018qll}.

\subsection{Tests on Simulations}\label{sec:sim}

Our model and many other components of our analysis have been validated 
end-to-end on a suite of 18 cosmological simulations\footnote{Each simulation assumes a flat \LCDM cosmology with $\Omega_{\mathrm{m}}=0.286$, $\Omega_{\mathrm{b}}=0.046$, $h=0.7$, $n_{\mathrm{s}}=0.96$, and $\sigma_8=0.82$.} \cite{y3-simvalidation, Wechsler2021, DeRose2021, Becker2013}.   
Validation is performed on the mean of measurements from all 18 of these 
simulations without shape noise, including photo-$z$s and marginalizing over all cosmological and nuisance parameters. 
We have verified that we can recover the correct cosmology with our 
fiducial analysis to within approximately $0.3 \sigma$ in the 2D $\sigma_8$--$\Omega_{\mathrm{m}}$ plane ($\Lambda$CDM) and $w$--$\Omega_{\mathrm{m}}$ plane ($w$CDM). 
We have shown that for a more stringent test in the 
absence of photometric redshift and shear calibration uncertainties (using true redshifts) our model is able to 
reproduce the mean $\xi_{\pm}$ and $w+\gamma_t$ measurements from our 18
simulations with a $\chi^2$ of 1.4 for cosmic shear (207 data points), 4.5 
for $w(\theta)$ (53 data points), and 9.1 for $\gamma_t(\theta)$ (232 data 
points). These $\chi^2$ numbers are relative to the fiducial covariance for a single DES Y3 realizations, but with a measurement that is the average of 18 realizations without shape noise. Thus, they represent the potential systematic $\chi^2$ contribution due to model inaccuracies, and shouldn't be interpreted as a goodness-of-fit metric.

In addition to these model tests, we have also investigated the systematic uncertainty inherent to our redshift inference process using these simulations. We have shown that the three independent source redshift $n(z)$ estimates -- SOMPZ, source--lens clustering, and shear ratios -- produce consistent constraints on the redshift distribution. 
We have also performed our fiducial analysis using a \redmagic-like lens sample, assuming source redshift distributions that are calibrated using the same three methods and lens redshift distributions estimated from \redmagic.  We found that the constraints from this analysis are consistent with those that use the true redshift distributions from the simulation.  
The final constraining power is similar between \redmagic\ and \maglim, so we do not repeat the simulated analysis twice.

\subsection{Quantifying internal and external consistency}\label{consistencya}

To quantify consistency of internal and external data sets, we define a priori
a process to guide decisions and conclusions before seeing the cosmological
constraints. For internal consistency, we calculate the Posterior Predictive
Distribution (PPD) \cite{y3-inttensions} and derive a (calibrated)
probability-to-exceed $p$. In short, the idea is to draw realizations of a
particular subset of the data vector for model parameters drawn from the
posterior of the same subset (goodness-of-fit tests) or a disjoint subset
(consistency tests). These realizations are then compared to actual
observations and a distance metric is computed in data space, which is then used to compute the $p$-value. We test the goodness-of-fit for the two combinations of two-point functions, $\xi_{\pm}$ and $w+\gamma_t$, and the combination of all three after confirming they are mutually consistent. In all cases, we require as part of the unblinding criteria defined in App.~\ref{sec:blind} that $p>0.01$. The validation of the use of PPD for these tests is described in Ref.~\cite{y3-inttensions}.

To quantify consistency with external experiments, we have explored a variety of metrics in order to calibrate expectations, which are described in Ref.~\cite{y3-tensions}. We studied the particular case of quantifying consistency between DES 3$\times$2pt and \textit{Planck} CMB using both simulated DES Y3 data and real Y1 data. The consistency metrics can be divided into two categories: parameter-based, which measure relative deviations in the multi-dimensional parameter space, and Evidence-based, which also account for how well the individual and combined data sets fit the model. We discuss results in terms of at least one metric from each category: the parameter difference and Suspiciousness \cite{PhysRevD.100.043504}, along with the Evidence ratio to compare to DES Y1. These metrics can produce a probability-to-exceed, and we require the same criterion of $p<0.01$ to conclude there exists evidence for inconsistency between probes. 

Detailed results from these consistency studies are shown in App.~\ref{consistency}.

\section{DES Y3 Results: Parameter Constraints}\label{sec:results}

\subsection{\lcdm}\label{lcdmsec}

\begin{figure}
\centering
\resizebox{\columnwidth}{!}{\includegraphics{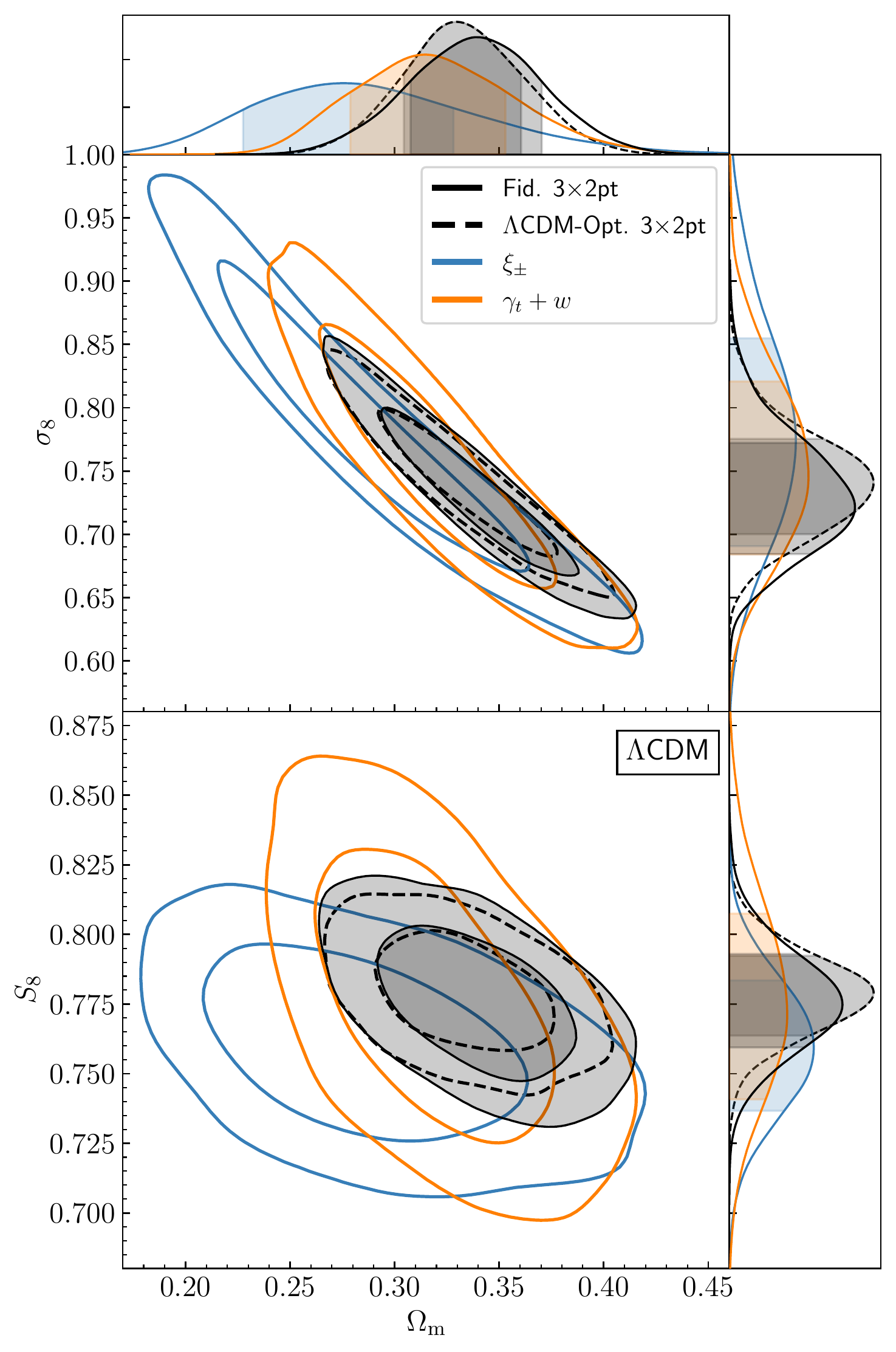}}
\caption{Marginalized constraints on the three parameters $\sigma_8$, $S_8=\sigma_8 \sqrt{\Omega_{\mathrm{m}}/0.3}$, and $\Omega_{\mathrm{m}}$ in the $\Lambda$CDM model from cosmic shear ($\xi_{\pm}$, blue), galaxy clustering and galaxy--galaxy lensing ($\gamma_t+w(\theta)$, orange) and their combination (3$\times$2pt, solid black). We also show a $\Lambda$CDM-optimized 3$\times$2pt analysis that is valid for $\Lambda$CDM using smaller angular scales in cosmic shear (dashed black). The marginalized contours in this and further figures below show the 68\% and 95\% confidence levels. The top and side panels show 1D marginalized constraints with the 68\% confidence region indicated. 
\label{lcdm}}
\vspace{-0.5cm}
\end{figure}

\begin{figure*}
\centering
\resizebox{.8\textwidth}{!}{\includegraphics{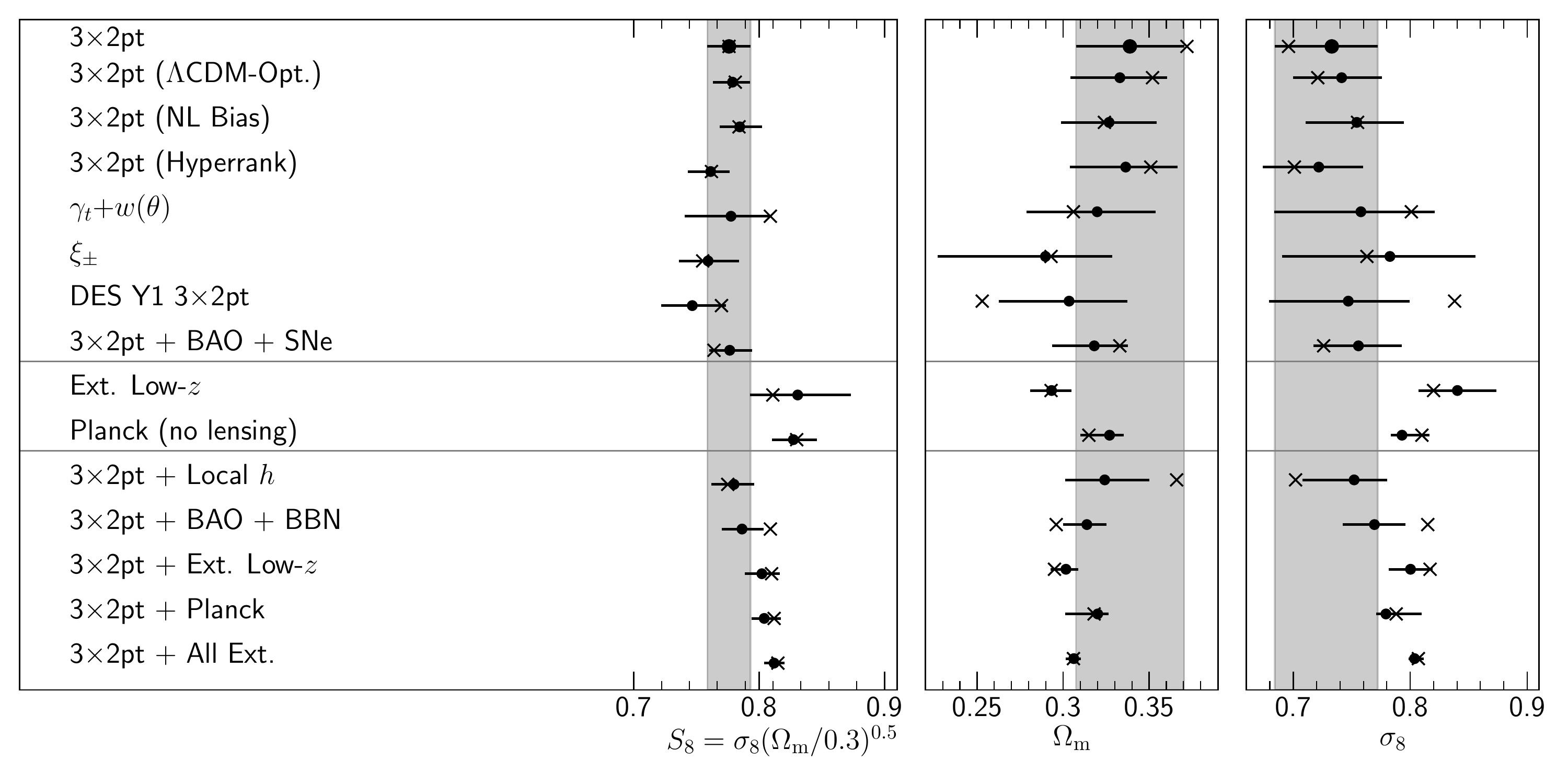}}
\caption{Summary of marginalized constraints (mean and 68\% CL) and maximum posterior values (crosses) on $S_8$, $\Omega_{\mathrm{m}}$, and $\sigma_8$ in $\Lambda$CDM. `Ext. Low-$z$' data consists of external SNe Ia, BAO, and RSD, while `All Ext.' data consists of external SNe Ia, BAO, RSD, and \textit{Planck} CMB with lensing. The top section shows constraints using only DES data, the middle section only external data, and the bottom section combinations of DES and external data. \label{lcdmtabfig}}
\vspace{0cm}
\end{figure*}

\begin{figure}
\centering
\resizebox{\columnwidth}{!}{\includegraphics{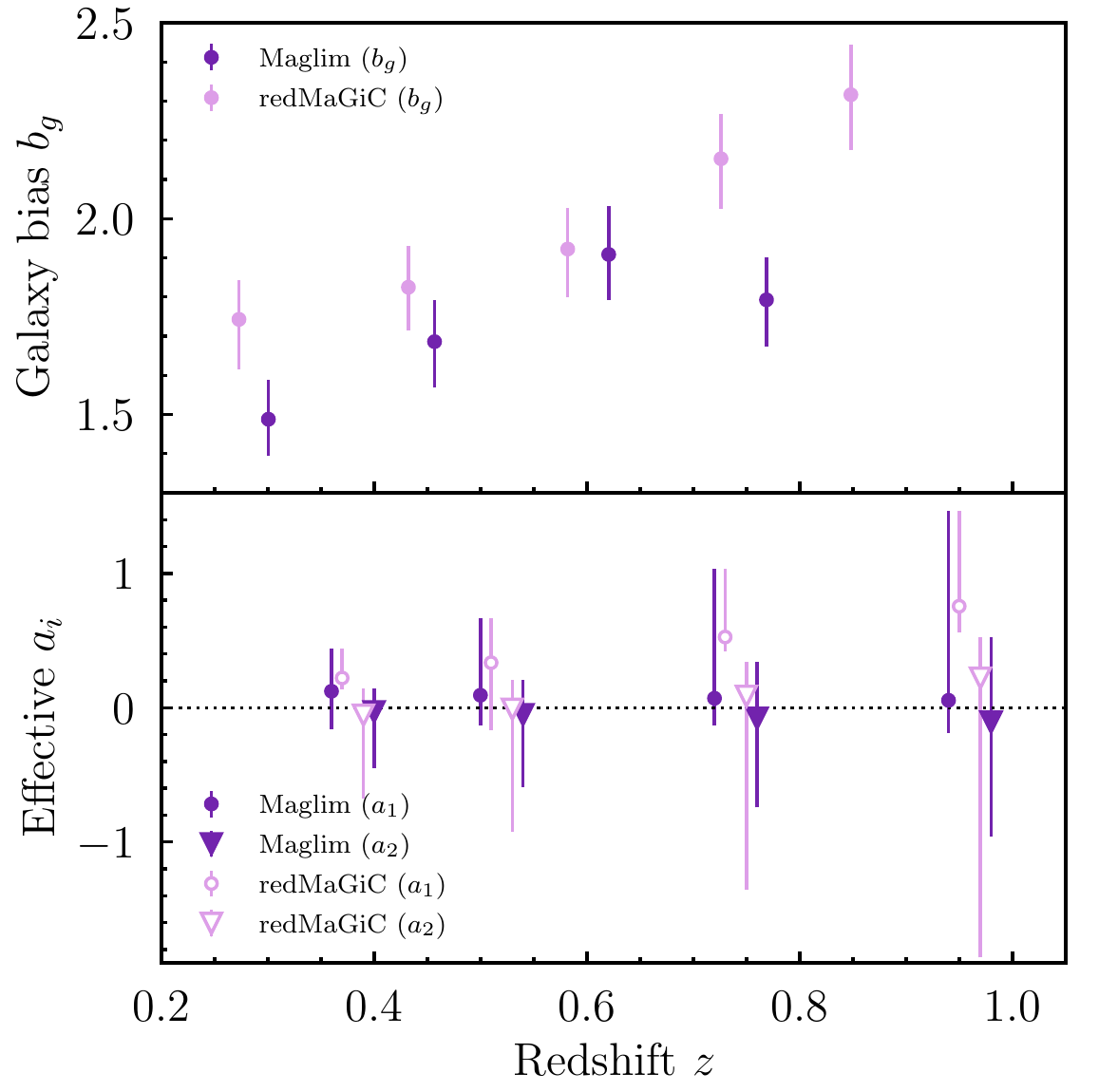}}
\caption{Constraints on the galaxy bias ($b_g$) and effective intrinsic alignment (IA) amplitude from tidal alignment ($a_1$) and tidal torquing ($a_2$) are shown per redshift bin. Constraints using both lens samples (\maglim\ and \redmagic) are shown. The galaxy bias is expected to be different for both lens samples, but the IA amplitude constraints, which are a property of the source galaxy sample, are consistent. We do not necessarily expect $a_1$ and $a_2$ to be consistent with one another. We sample over a power-law evolution of the IA amplitude, so the redshift evolution is forced to be smooth in $a_i$. \label{biasia}}
\vspace{0cm}
\end{figure}

The principal cosmological test of the DES Y3 data is to compare our data to
the currently favored $\Lambda$CDM model. The model has six
cosmological parameters, but also 25 nuisance parameters for a total of 31
free parameters (listed in Table \ref{tab:params}) in our fiducial analysis
with the \maglim\ lens sample. Recall also that, in the fiducial analysis, we
use all four source-galaxy bins, but only the first four (out of six total)
\maglim\ lens-galaxy bins.

We first concentrate on comparing two subsets of 3$\times$2pt measurements: those from
cosmic shear ($\xi_{\pm}$) and those from the combination of galaxy clustering
and galaxy--galaxy lensing ($w+\gamma_t$). It is logical to compare these two
subsets because each can constrain the $\Lambda$CDM parameters to a similar
precision, yet the constraints from the two subsets contain independent information, since they are sensitive to the underlying matter density
fluctuations in different ways. Moreover, comparing the two subsets of the
3$\times$2pt measurements provides an internal consistency test.

We must first check that the $\xi_{\pm}$ and $w+\gamma_t$ measurements
are a good fit to the data, mutually consistent, and that their combination
(3$\times$2) is a good fit to the model.  For each of the two measurements
individually, we find a PPD result for model goodness-of-fit
$p(\xi_{\pm})=0.21$ and $p(w+\gamma_t)=0.02$. The PPD
result for consistency between the two model constraints is
$p(\xi_{\pm}|w+\gamma_t)=0.30$, meaning that it is appropriate to combine
$\xi_\pm$ and $w+\gamma_t$.  The joint 3$\times$2pt goodness-of-fit is
$p(\xi_{\pm}+\gamma_t+w)=0.04$. Finally, the shear-ratio data has
goodness-of-fit $p=0.03$ in this joint best-fit model.  All of these $p$ values
meet our original criterion of $p>0.01$ defined in App.~\ref{sec:blind}.

The marginalized constraints from each probe and the 3$\times$2pt
combination in the parameter space spanned by $\sigma_8$, $S_8$, and
$\Omega_{\mathrm{m}}$ are shown in Fig.~\ref{lcdm}.  This is also summarized in Table \ref{tab:post} and
Fig.~\ref{lcdmtabfig}, which show the numerical constraints on these three
parameters. The DES Y3 3$\times$2pt constraints on the key parameters are
\begin{equation}
  \begin{aligned}
    S_8              &= 0.776^{+0.017}_{-0.017} \;\;(0.776)\\[0.2cm]
    \Omega_{\mathrm{m}}&= 0.339^{+0.032}_{-0.031}\;\; (0.372)\\[0.2cm]
    \sigma_8         &= 0.733^{+0.039}_{-0.049} \;\;(0.696).
  \end{aligned}
\end{equation}
The 3$\times$2pt contours in these parameters are not centered on the overlap of $\xi_{\pm}$ and $\gamma_t+w$ due to degeneracies in the higher dimensional parameter space.
  
We also perform two alternative 3$\times$2pt analyses that use smaller scales. First, we perform a $\Lambda$CDM-optimized
analysis that includes smaller-scale information in cosmic shear. This analysis meets our parameter bias requirements in
$\Lambda$CDM (i.e., Sec.~\ref{sec:method}), but not $w$CDM. The 3$\times$2pt results from the optimized
analysis are shown in the row labeled `$\Lambda$CDM-Opt.' in Table
\ref{tab:post} and Fig.~\ref{lcdmtabfig}.  The optimized results are consistent with the fiducial
analysis, but are about 30\% more constraining in the 2D marginalized
$\Omega_{\mathrm{m}}$--$\sigma_8$ plane.  The second alternative analysis
utilizes a more complicated nonlinear bias model in order to model smaller
scale information in $\gamma_t + w(\theta)$, and is described in
App. \ref{sec:robustness}. The 3$\times$2pt results from the nonlinear analysis,
shown in the row labeled `NL bias' in Table
\ref{tab:post} and Fig.~\ref{lcdmtabfig}, are consistent with the fiducial analysis and lead to an
 increase of 15\% in constraining power in the
$\Omega_{\mathrm{m}}$--$\sigma_8$ plane.

While we found no significant evidence of internal inconsistency with the $\Lambda$CDM model using the final \maglim\ lens selection, we have identified potential systematic modes in the data at high redshift for the \maglim\ sample and at all redshifts for the \redmagic\ sample. 
We had agreed before seeing any cosmological results that we would pursue potential systematics in the case where the results failed to sufficiently fit any of the models considered in this work ($\Lambda$CDM and $w$CDM) at $p<0.01$. Including \maglim\ lens bins 5 and 6 caused a very poor model fit to both models, with $p\approx 5 \times 10^{-4}$.
Based on this criterion, we applied a high-$z$ cut to limit the \maglim\ sample to approximately the same redshift range of \redmagic\ post-unblinding. 
This change is discussed further in App. \ref{sec:blind}.
The two lens samples are compared and further details of this are discussed in Sec.~\ref{lenscomp}, but all issues that have been uncovered appear to be mostly orthogonal to the 3$\times$2pt $\Lambda$CDM parameter dimensions --- that is, they do not significantly impact the inferred cosmological parameters, and the cosmological parameters inferred from the two lens samples are consistent. 
This resilience of the 3$\times$2pt combination of data and its ability to self-calibrate potential systematics in a subset of the two-point functions is one of the main motivations for pursuing this cosmological probe for large-scale structure.

We find
that the DES Y3 3$\times$2pt analysis is able to add information beyond
the prior for 15 parameter dimensions in the model, three of which are
cosmological. The cosmological modes that DES 3$\times$2pt most improves with respect to the prior 
are obtained with the Karhunen-Lo\`{e}ve decomposition of the posterior and prior covariance, and are: 
\begin{equation}
  \begin{aligned}
    p_1&=\sigma_8\Omega_{\mathrm{m}}^{0.77} = 0.317^{+0.015}_{-0.014},\\[0.2cm]
    p_2&=\Omega_{\mathrm{m}}\sigma_8^{-1.16} = 0.49^{+0.16}_{-0.15} ,\\[0.2cm]
    p_3&= h n_{\mathrm{s}}^{1.24}\Omega_{\mathrm{b}}^{-0.39} = 2.11^{+0.45}_{-0.42} . 
  \end{aligned}
\end{equation}
The combined 3$\times$2pt data is also able to simultaneously constrain a variety of `astrophysical' parameters that encode how galaxies are connected to the underlying dark matter perturbation field, namely the linear and nonlinear bias parameters and intrinsic alignment of galaxies. 
Constraints for these model parameters are shown in Fig.~\ref{biasia}. We find slightly higher galaxy bias constraints for \redmagic\ galaxies than in the DES Y1 analysis using a similar \redmagic\ sample. 
We find a preference for a slightly smaller intrinsic alignment amplitude than DES Y1. This value is consistent with the DES Y1 analysis, but is also consistent with zero intrinsic alignment.

\begin{figure}
\centering
\resizebox{\columnwidth}{!}{\includegraphics{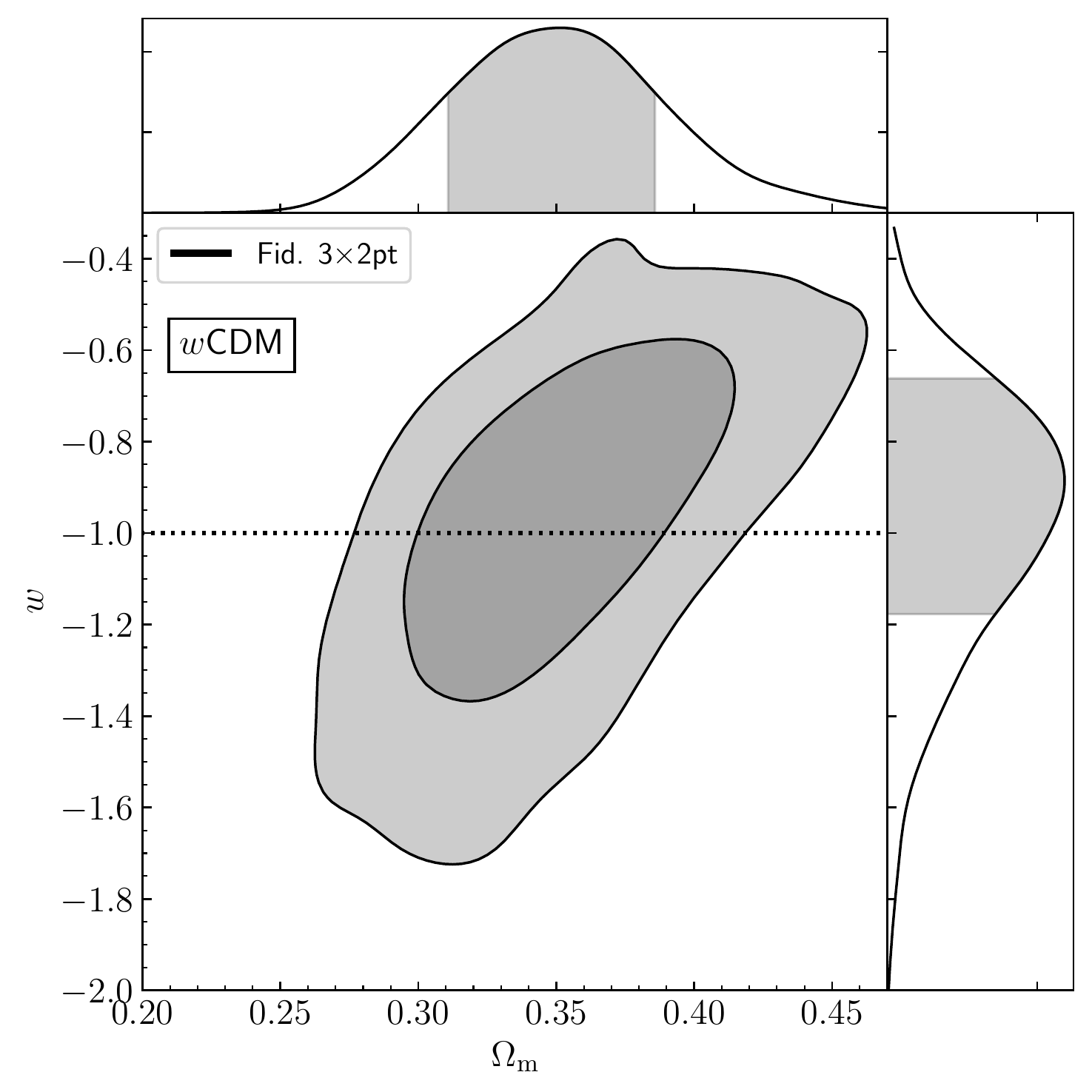}}
\caption{Marginalized constraints on the two parameters $\Omega_{\mathrm{m}}$ and $w$ in the $w$CDM model from DES Y3 3$\times$2pt. A dotted line indicates $w=-1$ as given by the cosmological constant. \label{wcdm}}
\vspace{0cm}
\end{figure}

\begin{figure*}
\centering
\resizebox{.8\textwidth}{!}{\includegraphics{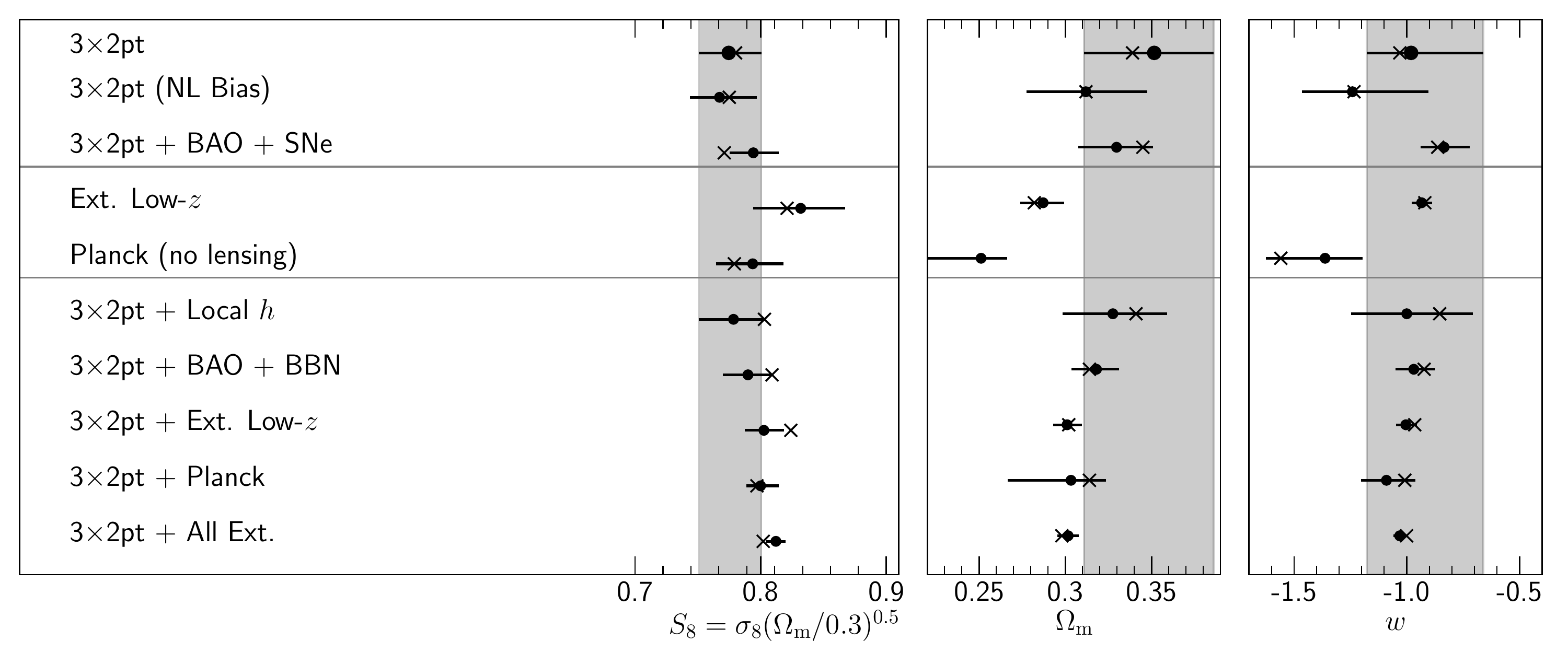}}
\caption{Summary of marginalized constraints (mean and 68\% CL) and maximum posterior values (crosses) on $S_8$, $\Omega_{\mathrm{m}}$, and $w$ in $w$CDM. `Ext. Low-$z$' data consists of external SNe Ia, BAO, and RSD, while `All Ext.' data consists of external SNe Ia, BAO, RSD, and \textit{Planck} CMB with lensing. The top section shows constraints using only DES data, the middle section only external data, and the bottom section combinations of DES and external data.\label{wcdmtabfig}}
\vspace{0cm}
\end{figure*}

\begin{table*}
\setlength{\extrarowheight}{7pt}
\caption{\label{tab:post} Summary of marginalized parameter constraints in \lcdm.  The mean and 68\% CL are provided for each cosmological parameter, followed by the maximum posterior value in parentheses, except for neutrino mass, for which the 95\% upper bound is given. Parameters that are not significantly constrained are indicated by a dash. All data have been re-analyzed with model and prior choices matching the DES Y3 3$\times$2pt analysis. }
\begin{tabular}{lcccccccc}
\hline
\hline
\textbf{$\Lambda$CDM} & $S_8$ & $\Omega_{\mathrm{m}}$ & $\sigma_8$ & $\Omega_{\mathrm{b}}$ & $n_{\mathrm{s}}$ & $h$ & $\sum m_{\nu}$ (eV)  & FoM$_{\sigma_8,\Omega_{\mathrm{m}}}$ \\ [0.2cm]
\hline
 \vspace{0.1cm}
{\textbf{DES Data}} &&&& \vspace{0.15cm}  \\
\hline
\multirow{2}{*}{3$\times$2pt} & $0.776^{+0.017}_{-0.017}$  & $0.339^{+0.032}_{-0.031}$  & $0.733^{+0.039}_{-0.049}$  & --  & --  & -- & --  & \multirow{2}{*}{2068}  \\ [-0.2cm] 
 & $(0.776)$  & $(0.372)$  & $(0.696)$  & --  &--  & -- & -- \\ 
 \hline\multirow{2}{*}{3$\times$2pt ($\Lambda$CDM-Opt.)} & $0.779^{+0.014}_{-0.015}$  & $0.333^{+0.028}_{-0.029}$  & $0.741^{+0.034}_{-0.042}$  & -- & --  & -- &--  & \multirow{2}{*}{2765}  \\ [-0.2cm] 
 & $(0.781)$  & $(0.352)$  & $(0.721)$  & --  &--  & --  & --  \\ 
 \hline\multirow{2}{*}{3$\times$2pt (NL Bias)} & $0.785^{+0.018}_{-0.016}$  & $0.327^{+0.028}_{-0.028}$  & $0.754^{+0.040}_{-0.044}$  &--  & --  & -- & --   & \multirow{2}{*}{2379}  \\ [-0.2cm] 
 & $(0.784)$  & $(0.324)$  & $(0.755)$  &--  & --  & --  \\ 
 \hline\multirow{2}{*}{$\gamma_t$+$w(\theta)$} & $0.778^{+0.031}_{-0.037}$  & $0.320^{+0.034}_{-0.041}$  & $0.758^{+0.063}_{-0.074}$  &--  & --  & --   & -- & \multirow{2}{*}{927}  \\ [-0.2cm] 
 & $(0.809)$  & $(0.306)$  & $(0.801)$ &--  & --  & --  \\ 
 \hline\multirow{2}{*}{$\xi_{\pm}$} & $0.759^{+0.025}_{-0.023}$  & $0.290^{+0.039}_{-0.063}$  & $0.783^{+0.073}_{-0.092}$  &--  & --  & -- & --  & \multirow{2}{*}{740}  \\ [-0.2cm] 
 & $(0.755)$  & $(0.293)$  & $(0.763)$  &--  & --  & --   \\ 
 \hline\multirow{2}{*}{DES Y1 3$\times$2pt} & $0.747^{+0.027}_{-0.025}$  & $0.303^{+0.034}_{-0.041}$  & $0.747^{+0.052}_{-0.068}$  &--  & --  & -- & --   & \multirow{2}{*}{1085}  \\ [-0.2cm] 
 & $(0.770)$  & $(0.253)$  & $(0.838)$  &--  & --  & --   \\ 
 \hline\multirow{2}{*}{3$\times$2pt + BAO + SNe} & $0.777^{+0.018}_{-0.017}$  & $0.318^{+0.020}_{-0.025}$  & $0.756^{+0.037}_{-0.039}$  & $0.041^{+0.004}_{-0.010}$  &--  & --  &--  & \multirow{2}{*}{2942}  \\ [-0.2cm] 
 & $(0.765)$  & $(0.333)$  & $(0.726)$  & $(0.031)$  & -- & --  & --  \\ 
 \hline \vspace{0.1cm}
{\textbf{External Data}} &&&& \vspace{0.15cm} \\
\hline
\multirow{2}{*}{Ext. BAO + BBN} & --  & $0.295^{+0.014}_{-0.017}$  &--  & $0.049^{+0.001}_{-0.001}$  & -- & $0.667^{+0.009}_{-0.010}$  & --  & \multirow{2}{*}{--}\\ [-0.2cm] 
 & --  & $(0.293)$  & --  & $(0.050)$  & -- & $(0.661)$  & --  &  \\ 
 \hline\multirow{2}{*}{Ext. Low-$z$} & $0.831^{+0.042}_{-0.038}$  & $0.293^{+0.012}_{-0.012}$  & $0.840^{+0.033}_{-0.033}$  & $0.053^{+0.014}_{-0.006}$  & --  & --  & --  & \multirow{2}{*}{2325}  \\ [-0.2cm] 
 & $(0.811)$  & $(0.293)$  & $(0.820)$  & $(0.032)$  & --  & --  & --  \\ 
 \hline\multirow{2}{*}{Planck (no lensing)} & $0.827^{+0.019}_{-0.017}$  & $0.327^{+0.008}_{-0.017}$  & $0.793^{+0.024}_{-0.009}$  & $0.051^{+0.001}_{-0.002}$  & $0.964^{+0.004}_{-0.005}$  & $0.665^{+0.013}_{-0.006}$  & <0.35  & \multirow{2}{*}{4217}  \\ [-0.2cm] 
 & $(0.830)$  & $(0.315)$  & $(0.810)$  & $(0.049)$  & $(0.968)$  & $(0.674)$  & (95\% CL)  \\ 
 \hline \vspace{0.1cm}
{\textbf{Combined Data}} &&&& \vspace{0.15cm} \\
\hline
\multirow{2}{*}{3$\times$2pt + Local $h_0$} & $0.780^{+0.016}_{-0.018}$  & $0.324^{+0.026}_{-0.023}$  & $0.752^{+0.028}_{-0.044}$  & --  & --  & $0.731^{+0.012}_{-0.013}$  & --  & \multirow{2}{*}{2720}  \\ [-0.2cm] 
 & $(0.775)$  & $(0.366)$  & $(0.702)$  & --  &--  & $(0.748)$  & -- \\ 
 \hline\multirow{2}{*}{3$\times$2pt + BAO + BBN} & $0.786^{+0.017}_{-0.016}$  & $0.314^{+0.011}_{-0.014}$  & $0.769^{+0.026}_{-0.027}$  & $0.048^{+0.001}_{-0.001}$  &-- & $0.676^{+0.009}_{-0.009}$  & -- & \multirow{2}{*}{5484}  \\ [-0.2cm] 
 & $(0.809)$  & $(0.296)$  & $(0.815)$  & $(0.048)$  & --  & $(0.673)$  & --  \\ 
 \hline\multirow{2}{*}{3$\times$2pt + Ext. Low-$z$} & $0.802^{+0.014}_{-0.013}$  & $0.302^{+0.007}_{-0.009}$  & $0.800^{+0.019}_{-0.019}$  & $0.050^{+0.012}_{-0.004}$  &-- & $0.702^{+0.101}_{-0.056}$  & --  & \multirow{2}{*}{8414}  \\ [-0.2cm] 
 & $(0.811)$  & $(0.295)$  & $(0.817)$  & $(0.034)$  & --  & $(0.591)$  & -- \\ 
 \hline\multirow{2}{*}{3$\times$2pt + Planck (no lensing)} & $0.804^{+0.013}_{-0.009}$  & $0.320^{+0.006}_{-0.019}$  & $0.779^{+0.030}_{-0.008}$  & $0.050^{+0.001}_{-0.002}$  & $0.967^{+0.004}_{-0.004}$  & $0.669^{+0.015}_{-0.005}$  & <0.43  & \multirow{2}{*}{6074}  \\ [-0.2cm] 
 & $(0.812\;\mathrm{b.f.})$  & $(0.318\;\mathrm{b.f.})$  & $(0.788\;\mathrm{b.f.})$  & $(0.050\;\mathrm{b.f.})$  & $(0.969\;\mathrm{b.f.})$  & $(0.670\;\mathrm{b.f.})$  & (95\% CL)   \\ 
  \hline\multirow{2}{*}{3$\times$2pt + All Ext.} & $0.812^{+0.008}_{-0.008}$  & $0.306^{+0.004}_{-0.005}$  & $0.804^{+0.008}_{-0.005}$  & $0.0487^{+0.0005}_{-0.0004}$  & $0.969^{+0.004}_{-0.003}$  & $0.680^{+0.004}_{-0.003}$  & <0.13  & \multirow{2}{*}{34041}  \\ [-0.2cm] 
 & $(0.815)$  & $(0.306)$  & $(0.807)$  & $(0.0486)$  & $(0.967)$  & $(0.681)$  & (95\% CL)  \\ 
 \hline
\hline
\end{tabular}
\end{table*}

\begin{table*}
\caption{\label{tab:postw} Summary of marginalized parameter constraints in \wcdm. The mean and 68\% CL are provided for each cosmological parameter, followed by the maximum posterior value in parentheses, except for neutrino mass, for which the 95\% upper bound is given. Parameters that are not significantly constrained are indicated by a dash. All data have been re-analyzed with model and prior choices matching the DES Y3 3$\times$2pt analysis.}
\resizebox{\textwidth}{!}{%
\setlength{\extrarowheight}{7pt}
\begin{tabular}{lcccccccccc}
\hline
\hline
\textbf{$w$CDM} & $S_8$ & $\Omega_{\mathrm{m}}$ & $\sigma_8$ & $\Omega_{\mathrm{b}}$ & $n_{\mathrm{s}}$ & $h$ & $w$ & $\sum m_{\nu}$ (eV) & FoM$_{\sigma_8,\Omega_{\mathrm{m}}}$ & FoM$_{w,\Omega_{\mathrm{m}}}$ \\ [0.2cm]
\hline
 \vspace{0.1cm}
{\textbf{DES Data}} &&&& \vspace{0.15cm}  \\
\hline
\multirow{2}{*}{3$\times$2pt} & $0.775^{+0.026}_{-0.024}$  & $0.352^{+0.035}_{-0.041}$  & $0.719^{+0.037}_{-0.044}$  & --  & --  & --  & $-0.98^{+0.32}_{-0.20}$  & --  & \multirow{2}{*}{1123}  & \multirow{2}{*}{115}  \\ [-0.2cm] 
 & $(0.780)$  & $(0.339)$  & $(0.733)$  & --  & --  & --  & $(-1.03)$  & --  \\ 
 \hline 3$\times$2pt  & $0.767^{+0.030}_{-0.023}$  & $0.312^{+0.036}_{-0.034}$  & $0.756^{+0.041}_{-0.053}$  & --  & --  & --  & $-1.24^{+0.34}_{-0.22}$  & --  & \multirow{2}{*}{1159}  & \multirow{2}{*}{117}  \\ [-0.2cm] 
 (NL Bias)& $(0.775)$  & $(0.312)$  & $(0.760)$  & --  & --  & --  & $(-1.23)$  & -- \\ 
 \hline 3$\times$2pt  & $0.794^{+0.020}_{-0.019}$  & $0.330^{+0.021}_{-0.022}$  & $0.759^{+0.035}_{-0.034}$  & --  & --  & -- & $-0.84^{+0.11}_{-0.10}$  & --  & \multirow{2}{*}{2426}  & \multirow{2}{*}{455}  \\ [-0.2cm] 
 + BAO + SNe& $(0.771)$  & $(0.345)$  & $(0.719)$  & --  & --  & --  & $(-0.86)$  & -- \\ 
 \hline \vspace{0.1cm}
{\textbf{External Data}} &&&& \vspace{0.15cm} \\
\hline
\multirow{2}{*}{Ext. Low-$z$} & $0.832^{+0.035}_{-0.038}$  & $0.287^{+0.012}_{-0.013}$  & $0.850^{+0.033}_{-0.037}$  & $0.057^{+0.013}_{-0.003}$  & --  & --  & $-0.93^{+0.05}_{-0.04}$  & --  & \multirow{2}{*}{2289}  & \multirow{2}{*}{1817}  \\ [-0.2cm] 
 & $(0.821)$  & $(0.282)$  & $(0.847)$  & $(0.059)$  & --  & --  & $(-0.92)$  & --  \\ 
 \hline Planck  & $0.794^{+0.025}_{-0.029}$  & $0.251^{+0.015}_{-0.056}$  & $0.876^{+0.071}_{-0.038}$  & $0.039^{+0.002}_{-0.009}$  & $0.964^{+0.005}_{-0.004}$  & $0.768^{+0.089}_{-0.035}$  & $-1.36^{+0.17}_{-0.26}$  & <0.46  & \multirow{2}{*}{957}  & \multirow{2}{*}{225}  \\ [-0.2cm] 
 \quad(no lensing)& $(0.779)$  & $(0.199)$  & $(0.956)$  & $(0.031)$  & $(0.962)$  & $(0.848)$  & $(-1.56)$  &(95\% CL)  \\ 
 \hline \vspace{0.1cm}
{\textbf{Combined Data}} &&&& \vspace{0.15cm} \\
\hline
3$\times$2pt  & $0.778^{+0.025}_{-0.028}$  & $0.328^{+0.032}_{-0.029}$  & $0.747^{+0.029}_{-0.037}$  & -- & --  & $0.731^{+0.013}_{-0.013}$  & $-1.00^{+0.29}_{-0.25}$  & --  & \multirow{2}{*}{1550}  & \multirow{2}{*}{167}  \\ [-0.2cm] 
+ Local $h_0$ & $(0.803)$  & $(0.341)$  & $(0.753)$  & -- &--  & $(0.724)$  & $(-0.85)$  & -- \\ 
 \hline 3$\times$2pt  & $0.790^{+0.019}_{-0.020}$  & $0.318^{+0.013}_{-0.015}$  & $0.768^{+0.027}_{-0.027}$  & $0.049^{+0.003}_{-0.003}$  & --  & $0.669^{+0.017}_{-0.020}$  & $-0.97^{+0.10}_{-0.08}$  & --  & \multirow{2}{*}{3733}  & \multirow{2}{*}{919}  \\ [-0.2cm] 
+ BAO + BBN & $(0.809)$  & $(0.314)$  & $(0.790)$  & $(0.051)$  & --  & $(0.657)$  & $(-0.92)$  & --  \\ 
 \hline 3$\times$2pt  & $0.803^{+0.016}_{-0.015}$  & $0.301^{+0.008}_{-0.008}$  & $0.801^{+0.021}_{-0.022}$  & $0.051^{+0.011}_{-0.004}$  & --  & $0.709^{+0.097}_{-0.049}$  & $-1.00^{+0.05}_{-0.04}$  & --  & \multirow{2}{*}{7941}  & \multirow{2}{*}{2662}  \\ [-0.2cm] 
 + Ext. Low-$z$ & $(0.824)$  & $(0.302)$  & $(0.821)$  & $(0.032)$  & --  & $(0.576)$  & $(-0.97)$  & --  \\ 
 \hline3$\times$2pt + Planck & $0.800^{+0.015}_{-0.011}$  & $0.303^{+0.020}_{-0.037}$  & $0.798^{+0.046}_{-0.026}$  & $0.047^{+0.004}_{-0.005}$  & $0.966^{+0.004}_{-0.004}$  & $0.691^{+0.036}_{-0.031}$  & $-1.090^{+0.128}_{-0.113}$  & <0.45 & \multirow{2}{*}{2634}  & \multirow{2}{*}{465}  \\ [-0.2cm] 
 \quad (no lensing)& $(0.797\;\mathrm{b.f.})$  & $(0.314\;\mathrm{b.f.})$  & $(0.779\;\mathrm{b.f.})$  & $(0.050\;\mathrm{b.f.})$  & $(0.969\;\mathrm{b.f.})$  & $(0.674\;\mathrm{b.f.})$  & $(-1.009\;\mathrm{b.f.})$  & (95\% CL)  \\  
 \hline 3$\times$2pt  & $0.812^{+0.008}_{-0.008}$  & $0.302^{+0.006}_{-0.006}$  & $0.810^{+0.010}_{-0.009}$  & $0.048^{+0.001}_{-0.001}$  & $0.968^{+0.003}_{-0.003}$  & $0.687^{+0.006}_{-0.007}$  & $-1.031^{+0.030}_{-0.027}$  & <0.17  & \multirow{2}{*}{21216}  & \multirow{2}{*}{7421}  \\ [-0.2cm] 
 + All Ext.& $(0.802)$  & $(0.298)$  & $(0.804)$  & $(0.048)$  & $(0.972)$  & $(0.686)$  & $(-1.001)$  & (95\% CL)  \\ 
 \hline
\hline
\end{tabular}%
}
\end{table*}

\subsection{\wcdm}

We also fit our data to the $w$CDM model, in order to test for evidence that
the dark energy equation of state departs from its cosmological-constant value
of $w=-1$. In $w$CDM, the dark energy density evolves with time with a
constant $w$, such that $\rho_{\mathrm{DE}}\propto (1+z)^{3(1+w)}$.
We show marginalized parameter posteriors for this model in Fig.~\ref{wcdm}
and parameter values in Table \ref{tab:postw} and Fig.~\ref{wcdmtabfig}.

We find similar levels of agreement between $\xi_{\pm}$ or
$\gamma_t + w(\theta)$ as in $\Lambda$CDM, and a similarly good fit to the data within the
$w$CDM model, but do not show constraints from these subsets of the data due
to increased prior influence and parameter volume effects. The DES Y3 3$\times$2pt
constraint on the matter density and dark energy equation of state parameter are
\begin{equation}
  \begin{aligned}
  \Omega_{\mathrm{m}}&=0.352^{+0.035}_{-0.041} \;\;(0.339),\\
  w &= -0.98^{+0.32}_{-0.20} \;\;(-1.03).
  \end{aligned}
\end{equation}

To determine if there is a preference for the $w$CDM model over the $\Lambda$CDM model, we compute the Bayes factor
 \be
R = \frac{P(\hat{\mathbf D}|\Lambda\mathrm{CDM})}{P(\hat{\mathbf D}|w\mathrm{CDM})}.
 \ee
A value of $R$ greater than unity implies that the $w$CDM model is not favored. We find $R=4.3$.  
This indicates that the late-universe large-scale structure probed by DES does not show evidence of needing the more complex dark energy density scenario of the $w$CDM model. 
We discuss further in Sec.~\ref{sec:ext} more stringent tests of the $\Lambda$CDM model that leverage data across the age of the Universe.

\subsection{Lens sample comparison}\label{lenscomp}

 \begin{figure}
\centering
\resizebox{\columnwidth}{!}{\includegraphics{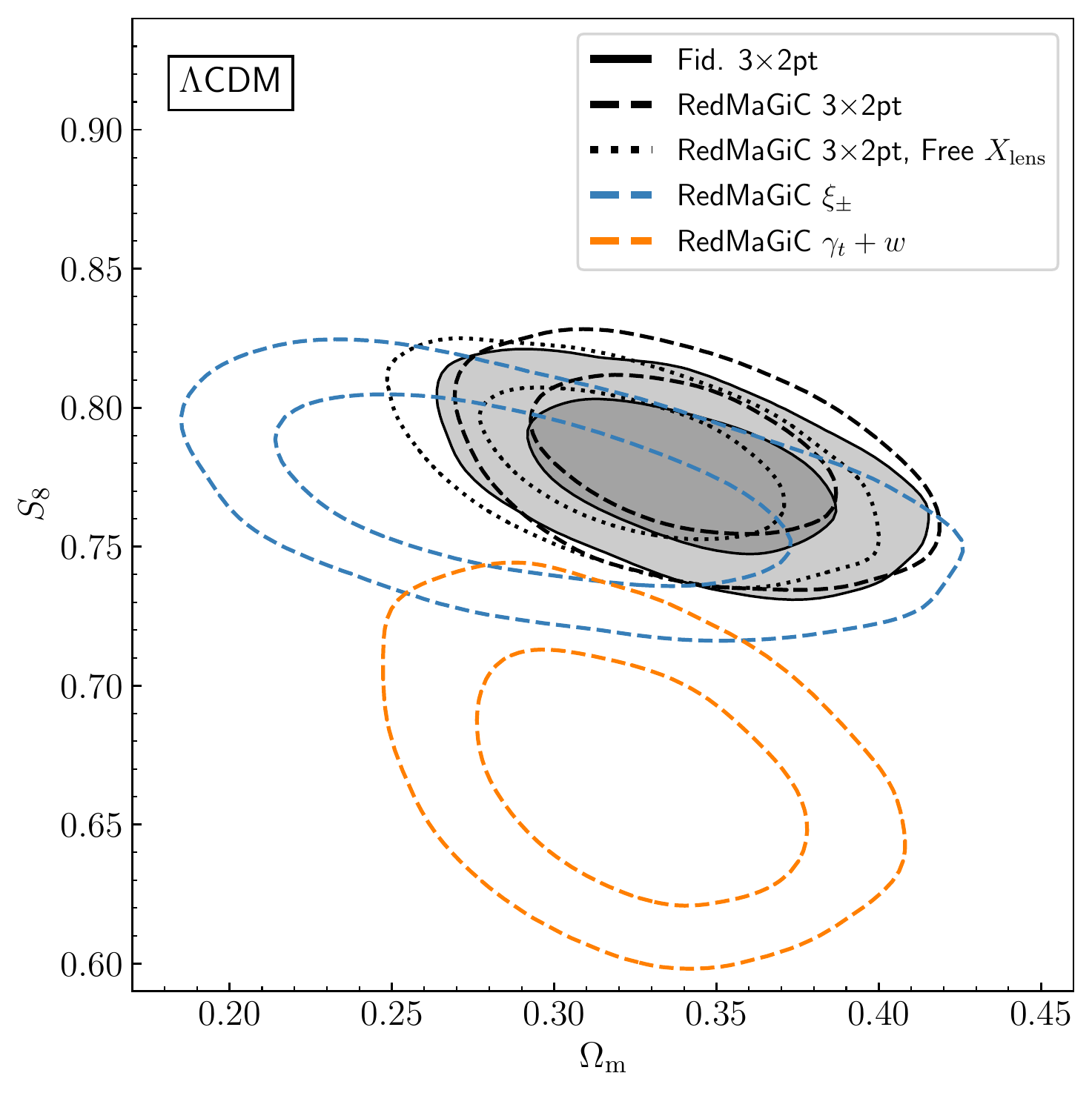}}
\caption{A comparison of the marginalized $\Lambda$CDM constraints of the two lens samples. Dashed contours show the cosmic shear (blue), galaxy--galaxy lensing and clustering (orange), and 3$\times$2pt (black) constraints based on the \redmagic\ lens sample. The 3$\times$2pt \redmagic\ constraints marginalizing over a free $X_{\rm lens}$ parameter are also shown (dotted black), and the 3$\times$2pt \maglim\ constraints (solid black). The inferred cosmological parameters from 3$\times$2pt are consistent in all three cases.  \label{lcdmcomp}}
\vspace{0cm}
\end{figure}
\begin{figure}
\centering
\resizebox{\columnwidth}{!}{\includegraphics{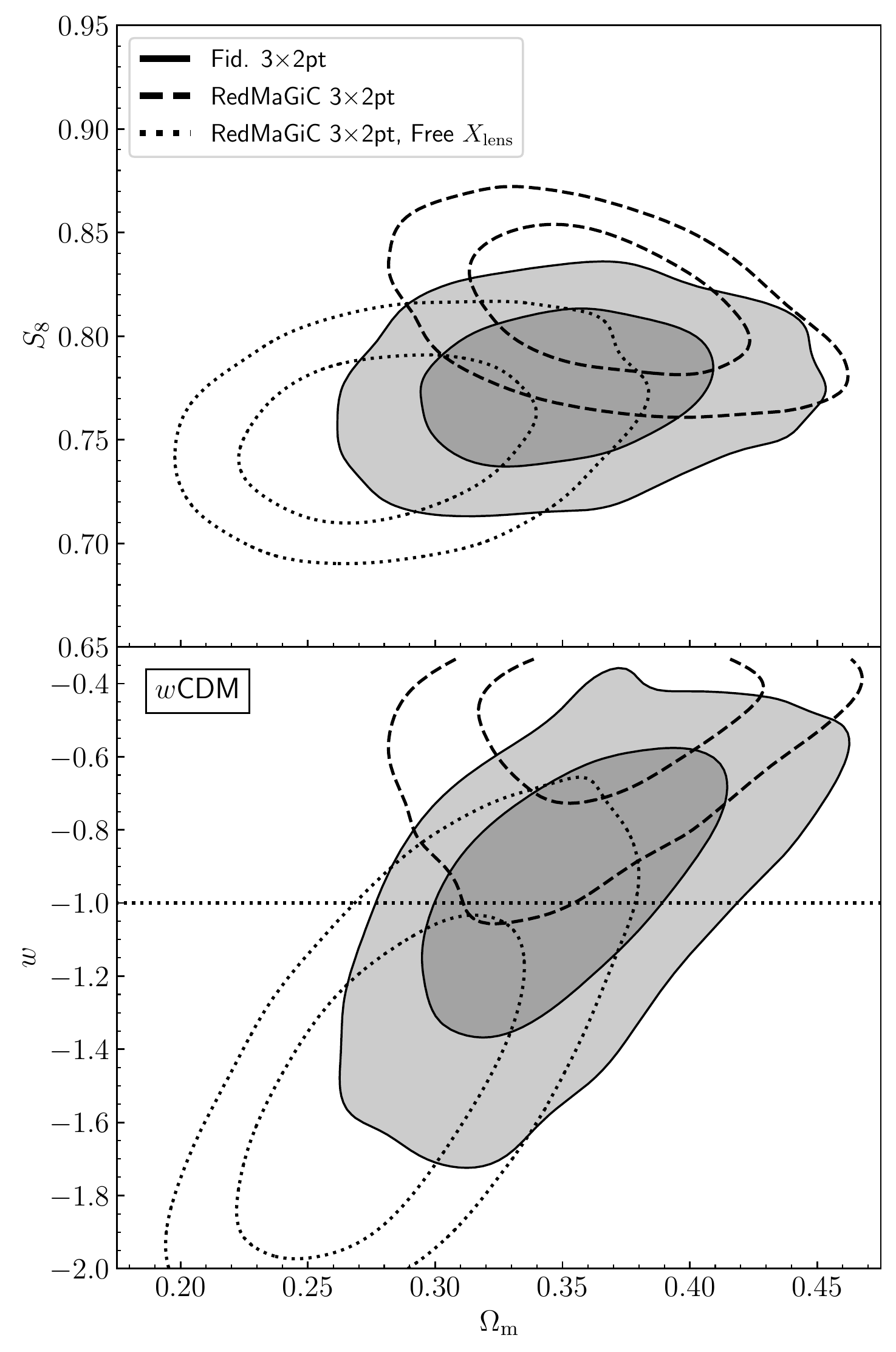}}
\caption{A comparison of the marginalized $w$CDM constraints of the two lens samples. Dashed black contours show the 3$\times$2pt constraints based on the \redmagic\ lens sample. The 3$\times$2pt \redmagic\ constraints marginalizing over a free $X_{\rm lens}$ parameter (dotted black) and the 3$\times$2pt \maglim\ constraints (solid black) are also shown. The inferred cosmological parameters from 3$\times$2pt are generally consistent, but in particular the \redmagic\ results are sensitive to the impact of $X_{\rm lens}$ in $w$CDM, showing substantial shifts in the inferred parameter values.\label{wcdmcomp}}
\vspace{-0.3cm}
\end{figure}

As optical surveys cover larger fractions of the sky and probe higher redshifts, photometric galaxy clustering becomes both more powerful and more difficult to  calibrate. 
Previous DES analyses used a luminous red sample of galaxies with constant comoving density, \redmagic. 
 To ensure robustness, we pursued two lens samples for DES Y3: a magnitude-limited lens sample, \maglim, and the \redmagic\ sample. The \redmagic\ sample was optimized for better understood and smaller photometric redshift errors.
The \maglim\ sample was optimized for $w$CDM constraints, balancing increased number density vs. less well-constrained photo-$z$s, while allowing selection to higher redshifts than possible with \redmagic. 
Comparing the inferred cosmological parameters of our models from these two very different samples, which have fewer than 20\% overlapping objects, allows us to infer potential uncorrected systematics from lens sample selection or photo-$z$ calibration of the lenses. 

Measurements based on the second redshift sample, \redmagic, also have an acceptable overall model fit to the $\Lambda$CDM and $w$CDM models.  The cosmic shear in this model is also consistent with the combination of galaxy--galaxy lensing and galaxy clustering. These measurements and model fits are shown in App. \ref{sec:redmagic}.
We find a PPD result for model goodness-of-fit $p(\xi_{\pm})=0.25$ and $p(w+\gamma_t)=0.04$, while the PPD result for consistency between the two model constraints is $p(\xi_{\pm}|w+\gamma_t)=0.02$. 
The joint 3$\times$2pt goodness-of-fit is $p(\xi_{\pm}+\gamma_t+w)=0.02$. 

The marginalized constraints of each individual probe and the 3$\times$2pt combination on $S_8$ and $\Omega_{\mathrm{m}}$ in $\Lambda$CDM are shown in Fig.~\ref{lcdmcomp}, while they are shown for $S_8$, $w$, and $\Omega_{\mathrm{m}}$ in $w$CDM in Fig.~\ref{wcdmcomp}. Both figures compare \redmagic\ results to the fiducial 3$\times$2pt using the \maglim\ sample. 
As described above, the cosmic shear or $\gamma_t+w(\theta)$ data alone are consistent with the 3$\times$2pt model fit, though the $\gamma_t+w(\theta)$ data on their own prefer a smaller $S_8$ value. This arises from the strong degeneracy between $\sigma_8$ and galaxy bias in $\gamma_t+w(\theta)$. Alone, it prefers a lower value of $\sigma_8$ and higher value of galaxy bias. Adding cosmic shear information effectively fixes the value of $S_8$ along that degeneracy, which brings the galaxy bias in 3$\times$2pt back down to a value more consistent with the DES Year 1 \redmagic\ galaxy bias constraints.

These results are consistent with the \maglim\ results and passed our
unblinding requirements, including having a sufficiently good model fit to
$\Lambda$CDM. However, after unblinding the results with \redmagic\ we
  found evidence of internal tension in the data.  Because $w(\theta)$ is not able to constrain
cosmology on its own, this has limited impact on the combination of galaxy
clustering and galaxy--galaxy lensing and no discernible impact on the
3$\times$2pt combination in $\Lambda$CDM.  Nevertheless, it is important
  to understand the source of this internal tension in \redmagic\ results and judge its impact on cosmological inference. To
  do so, we modeled this inconsistency
of the \redmagic\ clustering and galaxy--galaxy lensing amplitudes with a systematic
parameter $X_{\rm lens}$, which is related to the connection of the galaxy--galaxy lensing and galaxy clustering two-point functions to the matter two-point function:
\begin{equation}
\begin{aligned}
  w^{ii}(\theta) &= b_i^2 \xi^{ii}_{\mathrm{mm}}(\theta) \\
  \gamma^{ij}_t(\theta) &= X_{\rm lens} b_i \xi^{ij}_{\mathrm{mm}}(\theta) 
  \label{eq:X}
\end{aligned}
\end{equation}
where $b_i$ is the galaxy bias connecting the observable $\gamma_t$ or $w(\theta)$ to the matter correlation function ($\xi_{\mathrm{mm}}$) or spectrum and $X_{\rm lens}$ is the same for all redshift bins $i$. We expect $X_{\rm lens}=1$ in $\Lambda$CDM, if there are no systematic contributions to the signals. The fiducial model described in earlier sections is thus identical to the model including in Eq.~\ref{eq:X} with an additional constraint $X_{\rm lens}=1$.

We show the result of marginalizing over a free
$X_{\rm lens}$ in the \redmagic\ 3$\times$2pt analyses in Figs. \ref{lcdmcomp} and \ref{wcdmcomp}. We find a negligible impact on
the primary cosmological parameters in $\Lambda$CDM, particularly $S_8$.  We find 
 $X_{\rm lens} = 0.877^{+0.026}_{-0.019}$, strongly inconsistent with $X_{\rm lens}=1$ in $\Lambda$CDM. If we fix $X_{\rm lens}$ to this value in the \redmagic\ $\gamma_t+w(\theta)$ analysis, the contour in Fig.~\ref{lcdmcomp} shifts upward to agree with cosmic shear in $S_8$. The value of $X_{\rm lens}$ is correlated with the equation-of-state parameter $w$, so the \redmagic\ $w$CDM constraint is strongly affected by this potential systematic. Adding the single free parameter $X_{\rm lens}$ in $\Lambda$CDM leads to an improvement in $\chi^2$ of 25, while adding a free $w$ leads to an improvement in $\chi^2$ of 7. Thus, $X_{\rm lens}$ clearly leads to a better model fit.

After unblinding the results with \redmagic, but before unblinding those with
\maglim, we decided to use the \maglim\ sample for our fiducial
cosmological analysis if it showed no indication of this scale- and
redshift-independent effect that is present in \redmagic.  This potential
systematic was studied at length between the unblinding of the
\redmagic\ sample and the \maglim\ sample.  Studies of this effect are
discussed in much more detail in
\cite{y3-galaxyclustering,y3-2x2ptbiasmodelling,y3-2x2ptmagnification}.  We
have demonstrated that the effect (and its relative impact vs. the clustering
amplitude of the \maglim\ sample) is roughly independent of redshift, angular
scale, or position in the survey footprint.

After initial submission of this paper, we found that relaxing the goodness-of-fit requirement for the red galaxy model selection in \redmagic\ leads to a cosmological model fit consistent with $X_{\rm lens}=1$ and no significant change to the cosmological parameter results. This test suggests that a color-dependent photometric issue
is the source of $X_{\rm lens}!=1$, and is plausibly connected to
background subtraction. A specific fix for this systematic at the image level has not been identified, but these results pinning down the likely source of $X_{\rm lens}$ are described further in Ref.~\cite{y3-2x2ptbiasmodelling}. Further study of this effect and pipeline modifications will continue for the final DES Year 6 analyses.

\subsubsection{Summary of possible non-systematic causes of $X_{\rm lens}\neq 1$}

There are several classes of non-systematic explanations for $X_{\rm lens}$, all of which
we believe are implausible given our data. These possible explanations are:

\textit{Stochastic Bias}: While the effect of $X_{\rm lens}$ on clustering and galaxy--galaxy lensing looks very similar to stochasticity, a decorrelation between the galaxy and matter distributions, predictions from galaxy bias models make this interpretation unlikely. In configuration space, perturbative stochastic terms are expected to contribute only at small separations $r \sim R_{\star}$ (the Lagrangian size of halos), and to statistics that involve zero-lag correlators \cite{Desjacques:2016bnm}.

\textit{Lensing-is-low}: Ref.~\cite{Leauthaud:2016jdb} reported that the $\gamma_t$ signal around luminous red galaxies is lower than expected from a model conditioned on their autocorrelation, which resembles the pattern seen in our \redmagic\ sample. But with the possible exception of the large scale results in Ref.~\cite{2021MNRAS.502.2074L,2020MNRAS.491...51S}, the lensing-is-low result \cite{Leauthaud:2016jdb} applies to models that fit to scales sensitive to complexities of the small scale dark matter-galaxy connection. There is still debate within the lensing-is-low literature as to whether the effect can be accounted for by additional complexity in these small scale models \cite{2021MNRAS.502.3582Y}.

The DES Y1 results (which also used a \redmagic\ sample) do not support the lensing-is-low scenario, nor do the DES Y3 results for \maglim\ in the redshift range of the \redmagic\ sample. 
The DES Y3 results for the \redmagic\ sample show what could be interpreted as galaxy--galaxy lensing being 10-15\% percent lower than galaxy clustering at fixed cosmology (Planck 2015 in the case of Ref.~\cite{Leauthaud:2016jdb}; DES 3$\times$2pt cosmology in the DES Y3 results for $X_{\rm lens}$). 
However, the more plausible cause is that the clustering of the Y3 \redmagic\ sample is anomalously high, as indicated by internal consistency tests of the individual data vectors.
While we are still studying the $X_{\rm lens}<1$ anomaly, we currently do not believe that it supports a conclusion that galaxy--galaxy lensing is ``low.''

\textit{Fundamental physics}: Any dynamical modifications to either the Poisson equation or the shear equation generally changes the galaxy and matter distributions but their correlation is maintained, i.e. $X_{\rm lens}=1$ is maintained at linear scales. 
Beyond this possibility, any separation of the impact of relative ``bias''
between the two types of matter (apparent in lensing vs. clustering) at the
level of 15\% would require significant fluctuations in the dark matter field, which
would have substantial ramifications in other observables that we have not
seen. Therefore, we conclude that a fundamental-physics explanation for $X_{\rm lens}<1$
would probably have to be very fine-tuned.

\subsubsection{Potential systematics in $w(\theta)$ and $\gamma_t$ vs.\ $X_{\rm lens}\neq1$}

We now continue discussing the $X_{\rm lens}\neq 1$ anomaly by comparing \redmagic\ to
\maglim, and commenting on potential systematics in galaxy
clustering and galaxy--galaxy lensing as the cause of the anomaly.

We find the \redmagic\ sample shows $X_{\rm lens}<1$ at high significance at all scales
and redshifts. 
The highest two redshift bins of the \maglim\ sample, which have been removed from the analysis, also indicate $X_{\rm lens}<1$ at
high significance, which is clearly visible in the model fit in those two bins of Fig.~\ref{fig:wt}. 
In the redshift range overlapping the \redmagic\ sample, we find no evidence of a non-unity $X_{\rm lens}$ for \maglim. 
We discard the two high redshift bins for the \maglim\ sample as a conservative choice.
Based on our investigations so far and current understanding of theoretical extensions beyond $w$CDM, we do not believe these anomalies are indications of new physics. 
We have found plausible but unverified indications that the origin may lie in potential systematics, e.g., associated with the photometric uncertainty or background subtraction for large or faint objects, or in the de-reddening process. 

These issues are the subject of ongoing investigations, which will be crucial for understanding photometric clustering and its combination with galaxy--galaxy lensing in DES Y6 and beyond.
However, while these measurements are potentially impacted at a level we can measure by some as yet unidentified systematic, this does not have a significant impact on $\Lambda$CDM cosmology when the three two-point functions are combined within 3$\times$2pt.
For the \maglim\ sample our tests indicate that both the $\Lambda$CDM and $w$CDM constraints are robust.   
This self-calibration effect is one of the primary motivations for combining these different probes of the same underlying matter density field into the 3$\times$2pt observable. 

 \section{Comparison with other DES data}\label{sec:otherdes}
 
  \begin{figure}
\centering
\resizebox{\columnwidth}{!}{\includegraphics{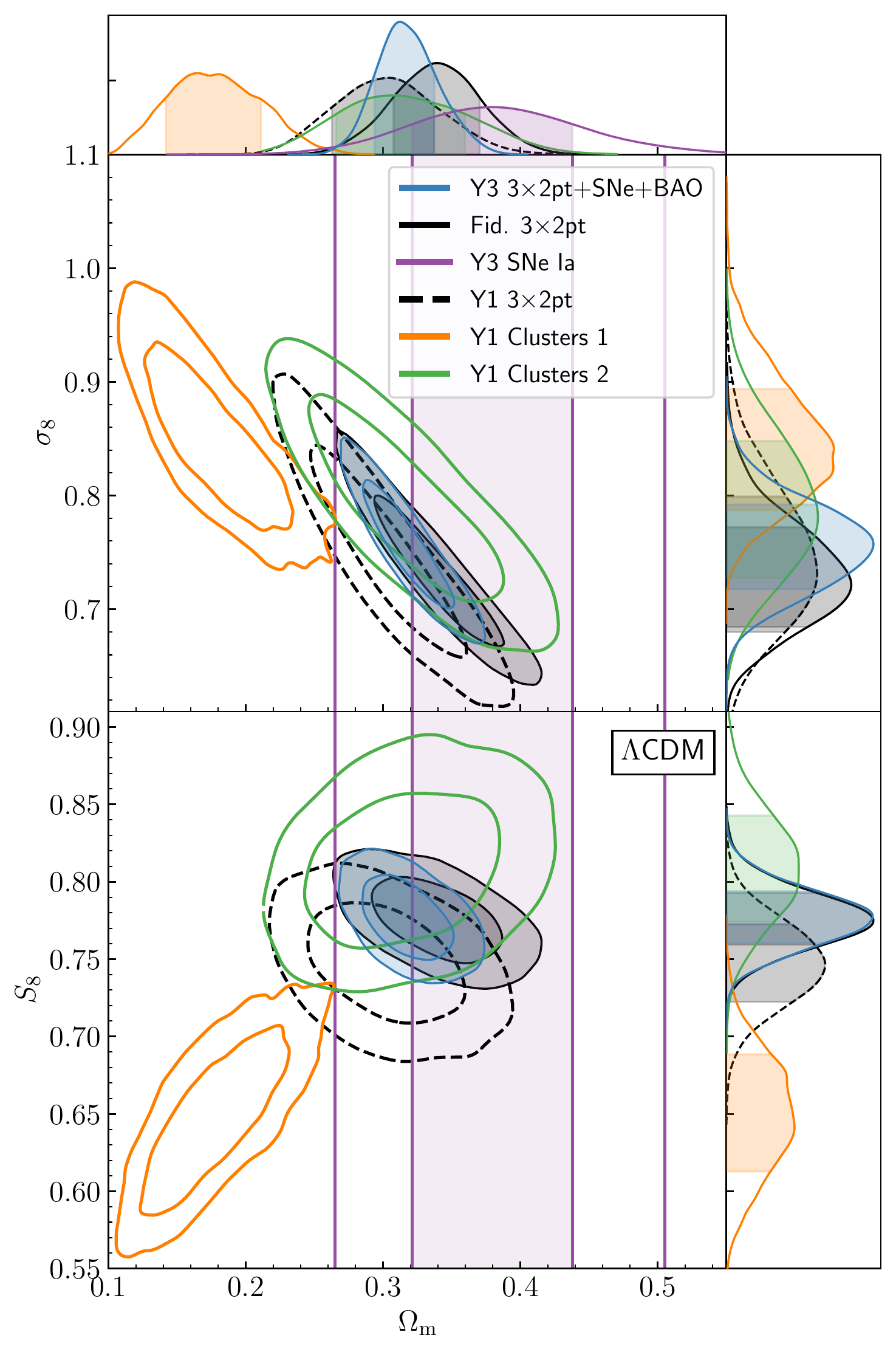}}
\caption{A comparison of the marginalized constraints on parameters in the
  $\Lambda$CDM model from a variety of DES probes: large-scale structure and
  weak lensing (3$\times$2pt; Y3 -- black solid, Y1 reanalyzed -- black dashed),
  type Ia supernovae (purple), galaxy cluster number counts and masses (orange
  and green), and BAO. The combination of DES
  Y3 3$\times$2pt, SNe Ia, and BAO is shown in blue. Going from Y1 to Y3, we
  find approximately a factor of two improvement in the 3$\times$2pt
  constraint in $\Omega_{\mathrm{m}}$-$S_8$ plane.  \label{descomp}}
\vspace{0cm}
\end{figure}

DES has produced competitive cosmological constraints using its four primary probes: galaxy clustering and weak gravitational lensing (3$\times$2pt), type Ia supernovae (SNe Ia), galaxy cluster counts and masses, and BAO. Together, these probes have been demonstrated to provide dark energy constraints that can be competitive with the best combined external constraints \cite{Abbott:2018wzc}. We describe each of them briefly below.

\textit{Type Ia supernovae}: The DES SNe Ia sample has 207 spectroscopically confirmed SNe 
in the redshift range $0.07<z<0.85$. The sample-building and analysis pipelines
are described in a series of papers that detail the SN search and
discovery \cite{Goldstein2015,Kessler2015,2018PASP..130g4501M}; simulations \cite{Kessler18}; photometry \cite{Brout18-SMP};
calibration \cite{Burke18,Lasker18}; spectroscopic
follow-up \cite{DAndrea18}; and selection bias \cite{2017ApJ...836...56K,SK16,2020MNRAS.494.4426S}. 
The methodology and systematic uncertainties are found in Ref.~\cite{Brout:2018jch}. These were used to constrain cosmology \cite{snepaper} and the Hubble constant \cite{Macaulay:2018fxi}. These analyses included additional external low-redshift SNe that we do not use in this analysis.
We compute the SNe likelihood using a module \cite{2018ApJ...859..101S} implemented in \textsc{CosmoSIS}, which reproduces the results in Ref.~\cite{snepaper}. The constraint from only DES SNe on $\Omega_{\mathrm{m}}$ is shown in Fig.~\ref{descomp} (purple).

\textit{Galaxy clusters}: The DES Y1 redMaPPer catalog consists of $\sim$6500 clusters with richness larger than $\lambda=20$ in the redshift range $z\in[0.2,0.65]$. The first cosmological analysis of DES clusters \cite{Abbott:2020knk} (`Clusters 1': orange contours in Fig.~\ref{descomp}), which combines cluster counts data and mass estimates from the stacked weak lensing analysis of \cite{desy1wl}, found a larger than 2$\sigma$ tension with the other DES probes. This is driven by low-richness systems, and has been interpreted as unmodeled systematics that affect the stacked weak lensing signal of the optically selected sample. 
 This interpretation is supported by the analysis of \cite{To:2020bhf} (`Clusters 2': green contours in Fig.~\ref{descomp}), which recovers results consistent with the other DES probes by combining cluster abundances with the large-scale auto-correlations of galaxy and cluster position and cross-correlations of cluster position with galaxy position and shear from DES Y1 data (4$\times$2pt+N). The conclusions of \cite{Abbott:2020knk} are further corroborated by the analysis of \cite{desxspt} which derive cosmological posteriors consistent with \cite{Abbott:2017wau} by analyzing the DES Y1 redMaPPer cluster abundances, but replacing the stacked weak lensing mass estimates of \cite{desy1wl} with multi-wavelength follow-up data from the SPT-SZ 2500 deg$^2$ survey \citep{SPTsz}. 

\textit{Baryon acoustic oscillations}: A sample of 7 million galaxies from the DES Y3 `Gold' catalog is selected in the redshift range $0.6 < z < 1.1$ \cite{y3-baosample} and used to measure the scale of the BAO feature in the distribution of galaxies at an effective redshift $z_{\rm eff}=0.835$ \cite{y3-BAOkp}. 
We use a likelihood from Ref.~\cite{y3-BAOkp} for the ratio of the angular 
diameter distance $D_A$ at $z_{\rm eff}$ and the sound horizon distance at the drag epoch, $r_d$, which is implemented in \textsc{CosmoSIS}.
The simulated galaxy catalogs used in the analysis to derive the uncertainty of the measurement are described in Ref.~\cite{y3-baomocks}.
While the BAO and 3$\times$2pt analyses probe common sky area and redshift range, and the measurements of this work include scales impacted by the BAO feature, the overlap in galaxy sample is small and the method for inferring the BAO distance ratio likelihood is insensitive to cosmology, so we neglect this non-zero correlation when combining the measurements. This will be further validated in future work that combines and studies all final DES Y3 probes.

\textit{DES Year 1 3$\times$2pt}: We reanalyze the DES Y1 3$\times$2pt data in the Y3 model and prior space, but do not update the scale cuts or marginalize over a free point mass for galaxy--galaxy lensing. We also make no changes in priors on systematic parameters (e.g., photo-$z$ or shear calibration parameters).

The comparison of these cosmological constraints in $\Lambda$CDM using the DES probes is shown in Fig.~\ref{descomp}. The combination of DES Y3 3$\times$2pt, SNe Ia, and BAO data is also shown in (blue). While the constraint in $S_8$ is driven primarily by 3$\times$2pt, there is substantial gain in other parameter dimensions due to the additional data. The marginalized parameter values are summarized in Tables \ref{tab:post} \& \ref{tab:postw} and Figs.~\ref{lcdmtabfig} \& \ref{wcdmtabfig}.

\section{Comparison with external data}\label{sec:ext}
 
It has been demonstrated that various combinations of low-redshift data and high-redshift data from the CMB can independently fit the $\Lambda$CDM model. However, the most stringent tests of the model will come from combining these data and testing whether the model can simultaneously fit the diverse set of cosmological probes available to us at all redshifts simultaneously. These data sets are sensitive to the growth of density perturbations, the expansion and geometry of the Universe, or both, and are sourced from a variety of very different physical processes. The combination of the independent external low-redshift probes with DES Y3 data further reduces the potential impact of any residual systematic effects in the low-redshift anchor of the test, while the combined DES probes have been carefully calibrated from the same data and consistently protected against confirmation bias. Both of these considerations give us further confidence, for complementary reasons, in testing the $\Lambda$CDM model across the age of the Universe.

\subsection{External data sets}

The likelihoods from data sets external to DES include: 

\textit{Type Ia supernovae}: The Pantheon sample \cite{pantheon} combines the distance measurements from 1048 SNe ranging from $0.01< z <2.3$, supplementing Pan-STARRS1 measurements with other available samples. 

\textit{Baryonic acoustic oscillations and redshift-space distortions}: We use the constraints from SDSS measurements of BAO and RSD in eBOSS DR16.\footnote{\url{https://svn.sdss.org/public/data/eboss/DR16cosmo/tags/v1_0_0/likelihoods/}} When using constraints on $f(z)\sigma_8(z)$, where $f(z)$ is the growth rate, from RSD, we use the released covariance matrices between the constraints from the BAO and RSD measurements. These measurements are expressed in terms of the Hubble distance $D_H$, sound horizon distance $r_d$, comoving distance $D_M$, and volume-average distance $D_V=(D_M^2 D_H z )^{1/3}$. The measurements include (from low to high redshift measurements):
\begin{itemize}
\itemsep0em 
\item The measurement of $D_V$ at an effective redshift of $z_{\mathrm{eff}} = 0.15$ using the Main Galaxy Sample (MGS) \cite{mgs_bao} and adding to $f\sigma_8$ measurement from Ref.~\cite{mgs_rsd}.    
\item A re-analysed version of BOSS DR12 measurements of $D_M$ and $D_H$ from BAO, and $f\sigma_8$ from RSD, at $z_{\mathrm{eff}} = 0.38$ and 0.51  \cite{bossdr12}. 
\item The eBOSS DR16 measurements of $D_M/r_d$, $D_H/r_d$ from BAO, and adding $f\sigma_8$ from the full-shape information, at $z_{\mathrm{eff}} = 0.698$ using Luminous Red Galaxies (LRG) \cite{ebossdr16_lrg1,ebossdr16_lrg2}.
\item The eBOSS DR16 measurements of $D_V/r_d$ when using BAO alone and $D_M/r_d$, $D_H/r_d$, $f\sigma_8$ when using BAO and the full-shape information, at $z_{\mathrm{eff}} = 0.845$ using Emission Line Galaxies (ELG) \cite{ebossdr16_elg},  
\item The eBOSS DR16 measurements of $D_M/r_d$, $D_H/r_d$ from BAO, and adding $f\sigma_8$ from the full-shape information, at $z_{\mathrm{eff}} = 1.48$ using the Quasar Sample (QSO) \cite{ebossdr16_qso1,ebossdr16_qso2}, 
\item The eBOSS DR16 measurements of $D_M$, $D_H$ at $z_{\mathrm{eff}} = 2.33$ using the Lyman-$\alpha$ forests \cite{ebossdr16_lya}. This data set only has information from BAO. 
\end{itemize}

\textit{CMB}: We use the likelihoods from the \textit{Planck} 2018 data release \cite{Aghanim:2019ame,Aghanim:2018eyx}. Our fiducial combination of \textit{Planck} likelihoods includes: 
\begin{itemize}
\itemsep0em 
\item The {\tt{Plik}} likelihood of the temperature power spectrum $C_{\ell}^{TT}$ in $30 \leq \ell \leq 2508$ and the $E$-mode power spectrum $C_{\ell}^{EE}$ and the cross power-spectrum between temperature and $E$-mode $C_{\ell}^{TE}$ in the range $30 \leq \ell \leq 1996$. 
\item The {\tt{Commander}} likelihood of the temperature power spectrum $C_{\ell}^{TT}$ in $2 \leq \ell \leq 29$. 
\item The {\tt{SimAll}} likelihood of the $E$-mode power spectrum $C_{\ell}^{EE}$ in $2 \leq \ell \leq 29$. 
\end{itemize}
We also use the likelihood of the lensing potential $\phi$ power spectrum $C_{\ell}^{\phi\phi}$ measured by \textit{Planck} in the range $8 \leq \ell \leq 400$, either in combination with our fiducial combination of \textit{Planck} likelihoods described above or alone. 
In the latter case we use the likelihood marginalized over the CMB power spectrum. \textit{Planck} CMB will refer to the primary CMB anisotropy data (without lensing) unless otherwise stated.

\textit{Big Bang nucleosynthesis}: We construct an $\Omega_{\mathrm{b}} h^2$ constraint based on observations of damped Lyman-$\alpha$ systems \cite{cooke18}. The primordial deuterium-to-hydrogen ratios measured from these systems can be translated to constraints on $\Omega_{\mathrm{b}} h^2$ via Big Bang nucleosynthesis (BBN) calculations, but different assumptions on the BBN physics, in particular on the rate of the $d(p,\gamma)^3\mathrm{He}$ nuclear reaction, yield different final constraints on $\Omega_{\mathrm{b}} h^2$. Our constraint conservatively incorporates the two major categories of such assumptions, namely the theoretical approach presented in Ref.~\cite{cooke18} and the experimental measurement-based approach from Ref.~\cite{pitrou21}. Specifically, we adopt the mean and the statistical uncertainty on $\Omega_{\mathrm{b}} h^2$ from Ref.~\cite{pitrou21}, and in addition introduce a systematic uncertainty defined by the difference between 1) the mean from Ref.~\cite{pitrou21} and 2) an inverse-variance weighted average of the two respective means \cite{cooke18,pitrou21}. This results in our adopted constraint of $100 \Omega_{\mathrm{b}} h^2 = 2.195 \pm 0.028$.

\textit{Local Hubble parameter}: We use a local $h$ prior from SH0ES \cite{Riess:2020fzl}, which constrains $h=0.732\pm0.013$  using a local distance ladder that depends on measurements of Cepheids and type Ia supernovae.

We use versions of these likelihoods implemented as modules in \textsc{CosmoSIS}, which are used to obtain the constraints presented in the following.
 
   \begin{figure}
\centering
\resizebox{\columnwidth}{!}{\includegraphics{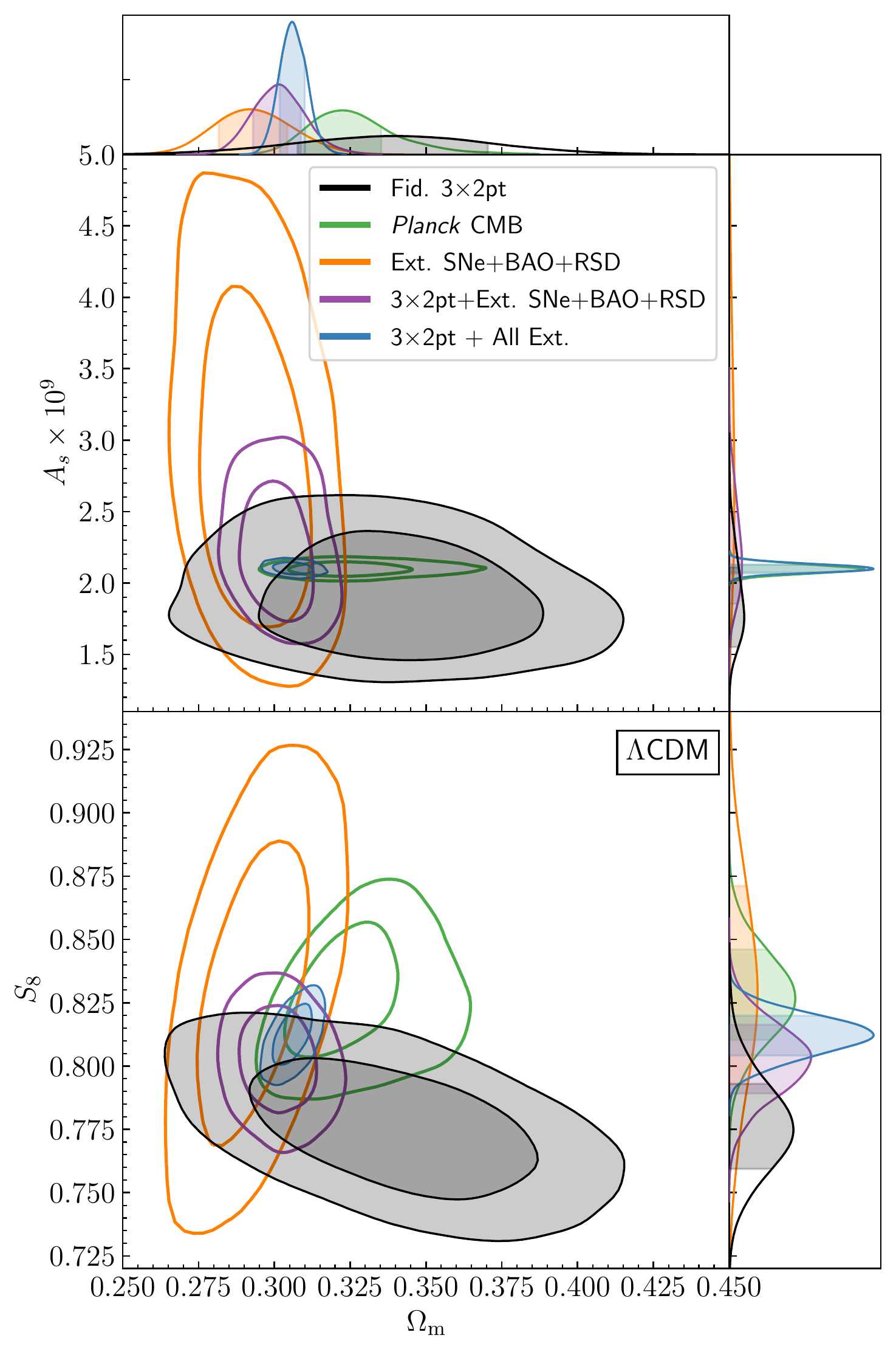}}
\caption{A comparison of marginalized constraints from three similarly constraining sets of cosmological probes in \lcdm. Combined external BAO, RSD, and SNe Ia data (Ext. Low-$z$) are shown in orange, the combination of DES galaxy clustering and weak lensing data (3$\times$2pt) is shown in black, and \textit{Planck} CMB (no lensing) data is shown in green. The three share a common parameter space in the $\Omega_{\mathrm{m}}$--$S_8$ plane at their 68\% CL bounds. The combination of Ext. Low-$z$ data with DES 3$\times$2pt is shown in purple and this combined additionally with \textit{Planck} CMB (w/ lensing) is shown in blue.   \label{extlcdm}}
\vspace{-0.5cm}
\end{figure}
  \begin{figure}
\centering
\resizebox{\columnwidth}{!}{\includegraphics{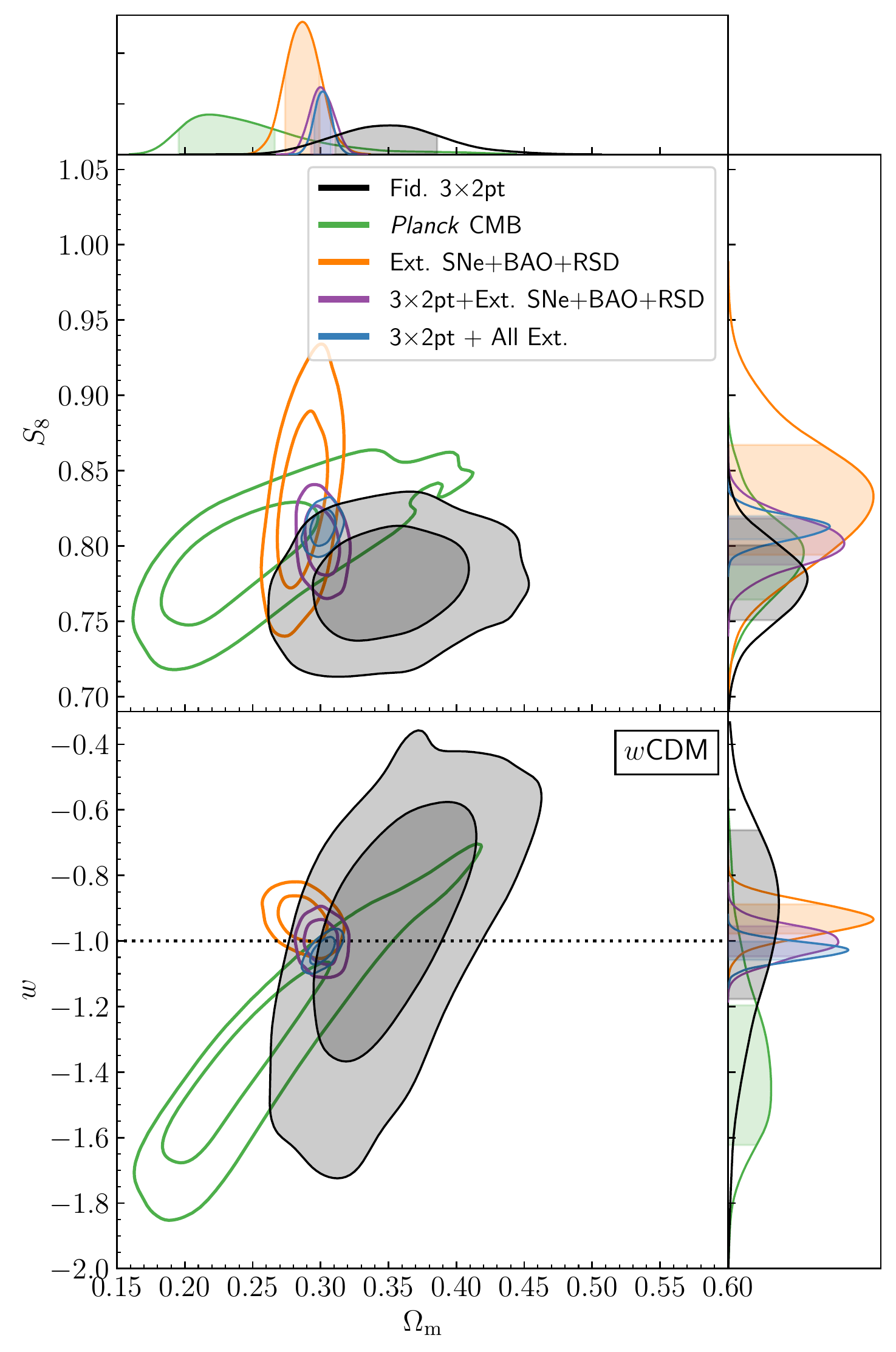}}
\caption{ A comparison of marginalized parameter constraints from three similarly constraining sets of cosmological probes in \wcdm. Combined external BAO, RSD, and SNe Ia data (Ext. Low-$z$) are shown in orange, the combination of DES galaxy clustering and weak lensing data (3$\times$2pt) is shown in black, and \textit{Planck} CMB (no lensing) data is shown in green. The combination of Ext. Low-$z$ data with DES 3$\times$2pt is shown in purple and this combined additionally with \textit{Planck} CMB (w/ lensing) is shown in blue.  \label{extwcdm}}
\vspace{-0.5cm}
\end{figure}
 
\subsection{High redshift vs. low redshift in \lcdm}\label{highlow}

  \begin{figure}
\centering
\resizebox{\columnwidth}{!}{\includegraphics{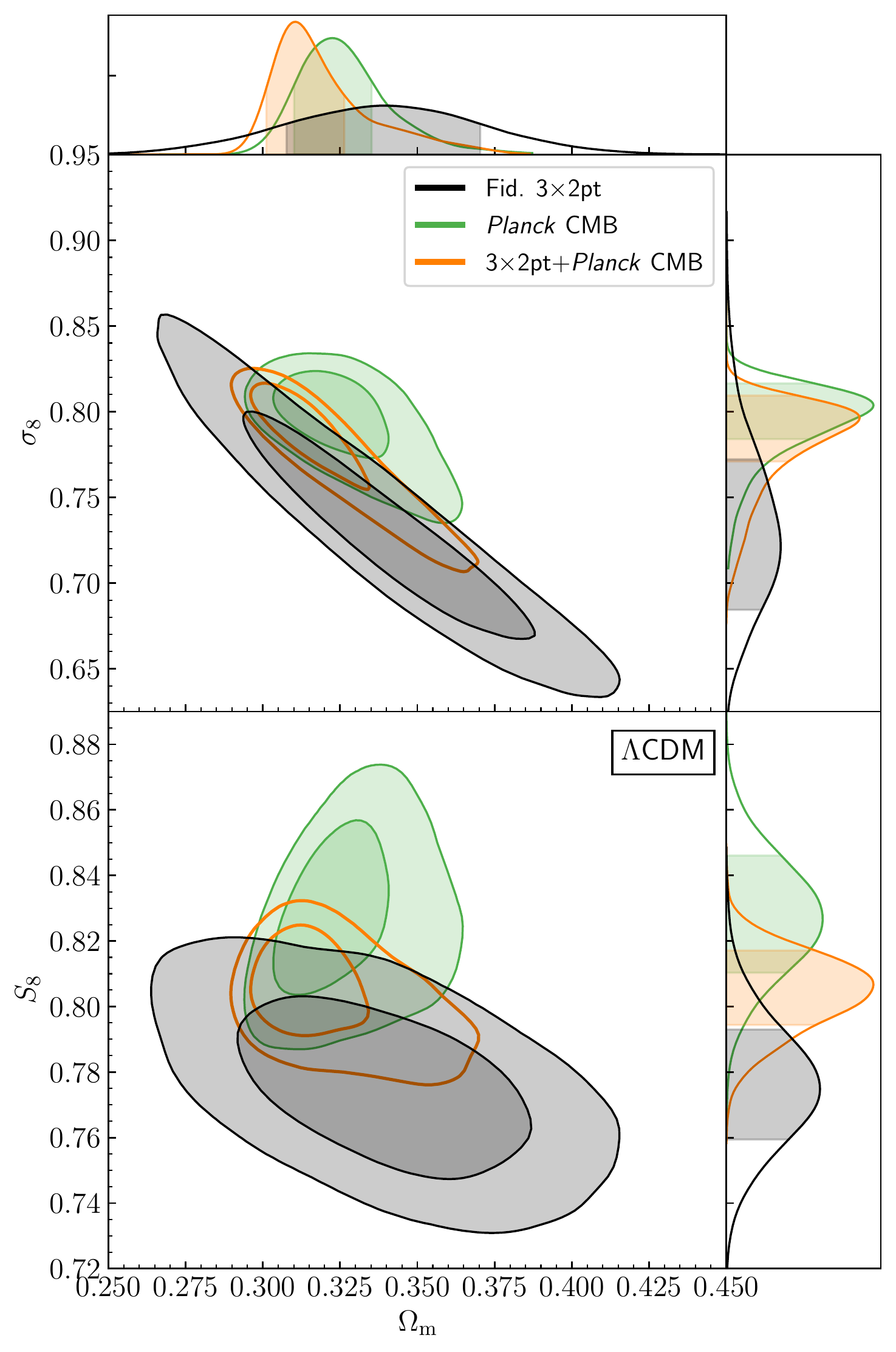}}
\caption{ A comparison of the marginalized parameter constraints in the \lcdm\ model from the Dark Energy Survey with predictions from \textit{Planck} CMB data (no lensing; green). We show the fiducial 3$\times$2pt (solid black) and the combined Y3 3$\times$2pt and \textit{Planck} (orange) results.
  \label{lcdmtest}}
\vspace{0cm}
\end{figure}
  \begin{figure}
\centering
\resizebox{\columnwidth}{!}{\includegraphics{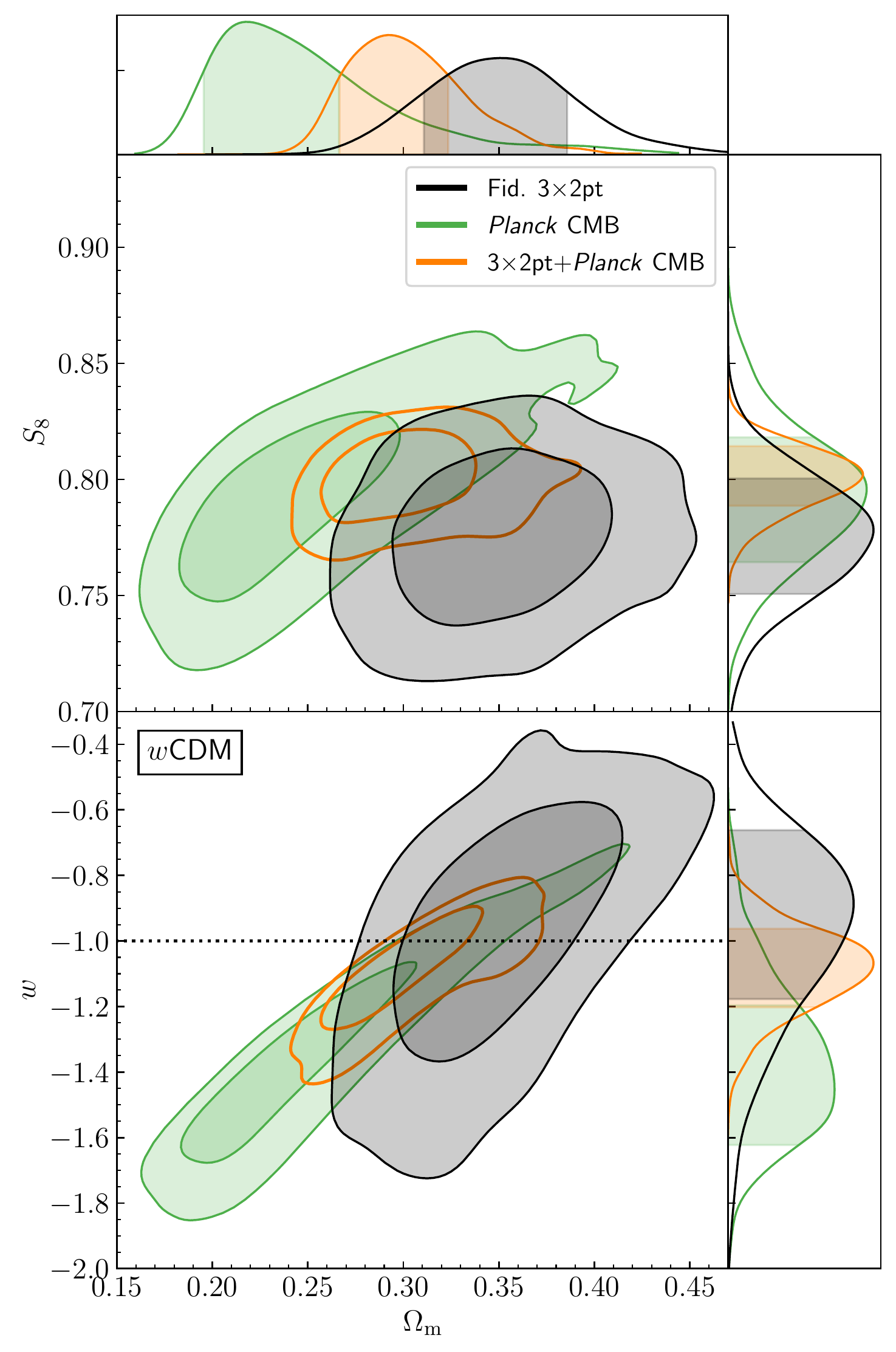}}
\caption{
A comparison of the marginalized parameter constraints in the \wcdm\ model from the Dark Energy Survey with predictions from \textit{Planck} CMB data (no lensing; green). We show the fiducial 3$\times$2pt (solid black) and the combined Y3 3$\times$2pt and \textit{Planck} (orange) results.
\label{wcdmtest}}
\vspace{0cm}
\end{figure}

One of the most stringent tests of $\Lambda$CDM is to compare the prediction of the state of the Universe and amplitude of perturbations from the epoch of recombination, which we can observe from the CMB, to the current day, which we observe with low-redshift surveys like DES. 
At the time of the CMB, the Universe was very hot and dense, and its physics was dominated by radiation. 
DES is most sensitive to a period in the Universe approximately eight billion years later, where perturbations have grown by several orders of magnitude and nonlinear growth is important. DES observes a volume of the Universe spanning nearly nine billion years of its evolution. 
The volumes probed by large low-redshift surveys provide significant additional information on potential changes to the evolution of perturbations or growth of the Universe over time, allowing them to strongly test the nature of dark energy. 

By taking precise measurements of the $\Lambda$CDM model from CMB observations and predicting what we should observe in terms of the amplitude of perturbations or matter density in the late Universe, we can test whether our observations from surveys like DES agree with those predictions. 
If they do not agree at high significance, we have demonstrated that $\Lambda$CDM cannot describe the full evolution of the Universe. 
There has been considerable debate about the tendency of late Universe measurements to prefer slightly lower matter density or amplitude of clustering relative to measurements from the CMB (e.g., \cite{Abbott:2017wau,Troxel:2018qll, Hikage:2018qbn,Asgari:2020wuj,Aghanim:2018eyx}).
As more powerful data becomes available, like the current DES Y3 analysis, we can determine whether these measurements converge towards or away from the \textit{Planck} CMB prediction.

We compare three similarly constraining and complementary subsets of available cosmological probes in $\Lambda$CDM and $w$CDM in Figs. \ref{extlcdm} \& \ref{extwcdm}. The external low-redshift SNe Ia, BAO, and RSD data primarily constrain $\Omega_{\mathrm{m}}$ and $w$, while the DES 3$\times$2pt data adds substantial information on $A_{\mathrm{s}}$ or $\sigma_8$, which helps to further constrain $\Omega_{\mathrm{m}}$ and $w$ through degeneracy-breaking of correlated parameters. These external low-redshift data sets complement the DES weak lensing and large-scale structure information by probing the growth and geometry of the cosmological model in fundamentally different ways. The CMB is able to tightly constrain both $\Omega_{\mathrm{m}}$ and $A_{\mathrm{s}}$ or $\sigma_8$ in $\Lambda$CDM, but is comparable in constraining power to DES 3$\times$2pt in $w$CDM, since it primarily has access to information limited to the surface of last scattering at $z\approx 1100$. The combination of DES 3$\times$2pt with the other low-redshift data provides substantial gain in  $A_{\mathrm{s}}$, $\sigma_8$, $\Omega_{\mathrm{m}}$, and $w$. We list marginalized parameter constraints for these probes in Tables \ref{tab:post} \& \ref{tab:postw}. 

\subsubsection{Consistency results}

We show the comparison of DES 3$\times$2pt and the \textit{Planck} CMB data for the $\Lambda$CDM and $w$CDM models in Figs. \ref{lcdmtest} \& \ref{wcdmtest}. 
Visually, we find better agreement in the overlap of the marginalized $\Omega_{\mathrm{m}}$--$S_8$ parameters with the DES Y3 3$\times$2pt data than found in the DES Y1 analysis \cite{Abbott:2017wau}, despite substantial improvements to the precision of both DES and \textit{Planck} predictions. 
This is qualitatively unchanged when using the more precise, optimized $\Lambda$CDM version of the analysis that uses more small scale information -- the DES contour shrinks, but asymmetrically in the direction of the CMB prediction. 

We evaluate the consistency of the DES and \textit{Planck} data in several ways, including shifts in parameter space and the Bayesian evidence. These are described further in Sec.~\ref{consistencya} and full results are provided in App.~\ref{consistency}. We find a parameter difference of 1.5$\sigma$ ($p=0.13$) in the cosmological model space and a Suspiciousness of $0.7\pm 0.1\sigma$, corresponding to $p=0.48\pm 0.08$. This generally leads to the conclusion that despite substantially increased precision from both experiments, we find no significant evidence against the $\Lambda$CDM model from comparing these data sets. Agreement between DES and \textit{Planck} in these metrics has improved relative to the comparison of DES Y1 3$\times$2pt and earlier \textit{Planck} results, which gave a parameter difference of $2.2\sigma$ and Suspiciousness of $2.4\pm 0.2\sigma$ \cite{y3-tensions}. The combined DES and \textit{Planck} CMB contour is shown in orange in Figs. \ref{lcdmtest} \& \ref{wcdmtest}.

We repeat this exercise for the full combined low-redshift data, including DES 3$\times$2pt, all BAO, and external SNe Ia and RSD data. This comparison is shown in Figs.~\ref{extlcdm} \& \ref{extwcdm}, and is highly complementary, as the external probes are sensitive to both growth and geometry in the model in ways the DES 3$\times$2pt data is not, and come from a variety of different experiments. We find better agreement between all of these low-redshift probes and \textit{Planck} CMB predictions than in the comparison with DES 3$\times$2pt data alone, with a parameter difference of 0.9$\sigma$ or $p=0.34$. These results indicate that we can combine all these available cosmic probes into a single joint result in the following subsection.

There are several reasonable motivations for caution in the interpretation of any strong evidence for or against cosmological consistency in tests like this.  
It is worth noting that while we have multiple redundant low-redshift sources of information for each main cosmological probe used, it would be useful to have a second, blinded large-scale CMB polarization experiment to increase confidence in the test at the high-$z$ limit. 
While polarization data is required to break degeneracies in the cosmological parameters with the optical depth $\tau$, we also repeat the caution from Ref.~\cite{Aghanim:2018eyx} against over-interpreting the \textit{Planck} polarization results and the sensitivity of the final parameter constraints to assumptions made in the construction of the likelihood, which can lead to a <1$\sigma$ shift toward the DES posterior relative to the fiducial \textit{Planck} likelihood. 
Similar shifts are seen based on certain analysis choices in the DES results as well, which are shown in App.~\ref{sec:robustness}. 
Neither the shift in \textit{Planck} posteriors or those from other analysis choices in DES contribute to a significant change in the final interpretation of the comparisons.
Finally, the DES Y3 analysis has uncovered potential systematics connected to photometry (e.g., Sec.~\ref{lenscomp}). While there is evidence that these do not impact the cosmological results, and thus would not impact this comparison of data sets, they have not been connected to a specific source.
However, these are examples of unresolved uncertainties that call for additional care in interpreting any statements about the consistency of early- and late-universe probes in $\Lambda$CDM, which should be addressed for future more precise analyses. 

\subsection{Joint cosmological constraints in \lcdm\ and \wcdm}

We find that external low-redshift (BAO+RSD+SNe Ia), \textit{Planck} CMB, and DES 3$\times$2pt data sets are able to provide three independent, highly complementary, and similarly powerful constraints on parameters related to dark matter and dark energy in the $\Lambda$CDM and $w$CDM models, as seen in Figs.~\ref{extlcdm} \& \ref{extwcdm}. Given the results of the above consistency tests, detailed in App.~\ref{consistency}, these data sets are each consistent with one another, and thus can be combined into a joint constraint on the models. We present these joint results in Figs.~\ref{extlcdm} \& \ref{extwcdm} and a summary in Figs. \ref{lcdmtabfig} \& \ref{wcdmtabfig} and Tables \ref{tab:post} \& \ref{tab:postw}. In the $\Lambda$CDM model, we find
\begin{equation}
  \begin{aligned}
    S_8              &= 0.812^{+0.008}_{-0.008} \;\;(0.815)\\[0.2cm]
    \Omega_{\mathrm{m}}&= 0.306^{+0.004}_{-0.005}\;\; (0.306)\\[0.2cm]
    \sigma_8         &= 0.804^{+0.008}_{-0.008} \;\;(0.807).
  \end{aligned}
\end{equation}
In the $w$CDM model,
\begin{equation}
  \begin{aligned}
    \sigma_8         &= 0.810^{+0.010}_{-0.009} \;\;(0.804), \\
  \Omega_{\mathrm{m}}&=0.302^{+0.006}_{-0.006} \;\;(0.298),\\
  w &= -1.031^{+0.030}_{-0.027} \;\;(-1.001).
  \end{aligned}
\end{equation}
We find $R=7.8$, indicating that there is also no preference for $w$CDM over $\Lambda$CDM in the full joint data analysis.

These data sets together are able to provide unprecedented precision on the cosmological parameters of the models. In $\Lambda$CDM, we are able to constrain $\sigma_8$, $S_8$, $h$, $\Omega_{\mathrm{b}}$, and $n_{\mathrm{s}}$ to less than 1\%; $\Omega_{\mathrm{m}}$ and $A_{\mathrm{s}}$ to about 1\%;  $\tau$ to about 10\%; and place an upper limit on the sum of neutrino masses of $\sum m_\nu < 0.13$ eV (95\% CL). In $w$CDM, we are able to constrain $n_{\mathrm{s}}$ to less than 1\%; $\Omega_{\mathrm{m}}$, $\Omega_{\mathrm{b}}$, $h$, and $A_{\mathrm{s}}$ to about 1-2\%;  $w$ to about 3\%; $\tau$ to about 10\%; and place an upper limit on the sum of neutrino masses of $\sum m_\nu < 0.17$ eV (95\% CL). Individually, the three subsets of data constrain $\sigma_8$ and $\Omega_{\mathrm{m}}$ in $\Lambda$CDM with FoM between 2000 and 4000, while combined, they reach a FoM of 34,000. This clearly demonstrates the highly complementary nature of these three independent data sets.

\subsection{Comparison of Lensing Probes}
 
\begin{figure}
\centering
\resizebox{\columnwidth}{!}{\includegraphics{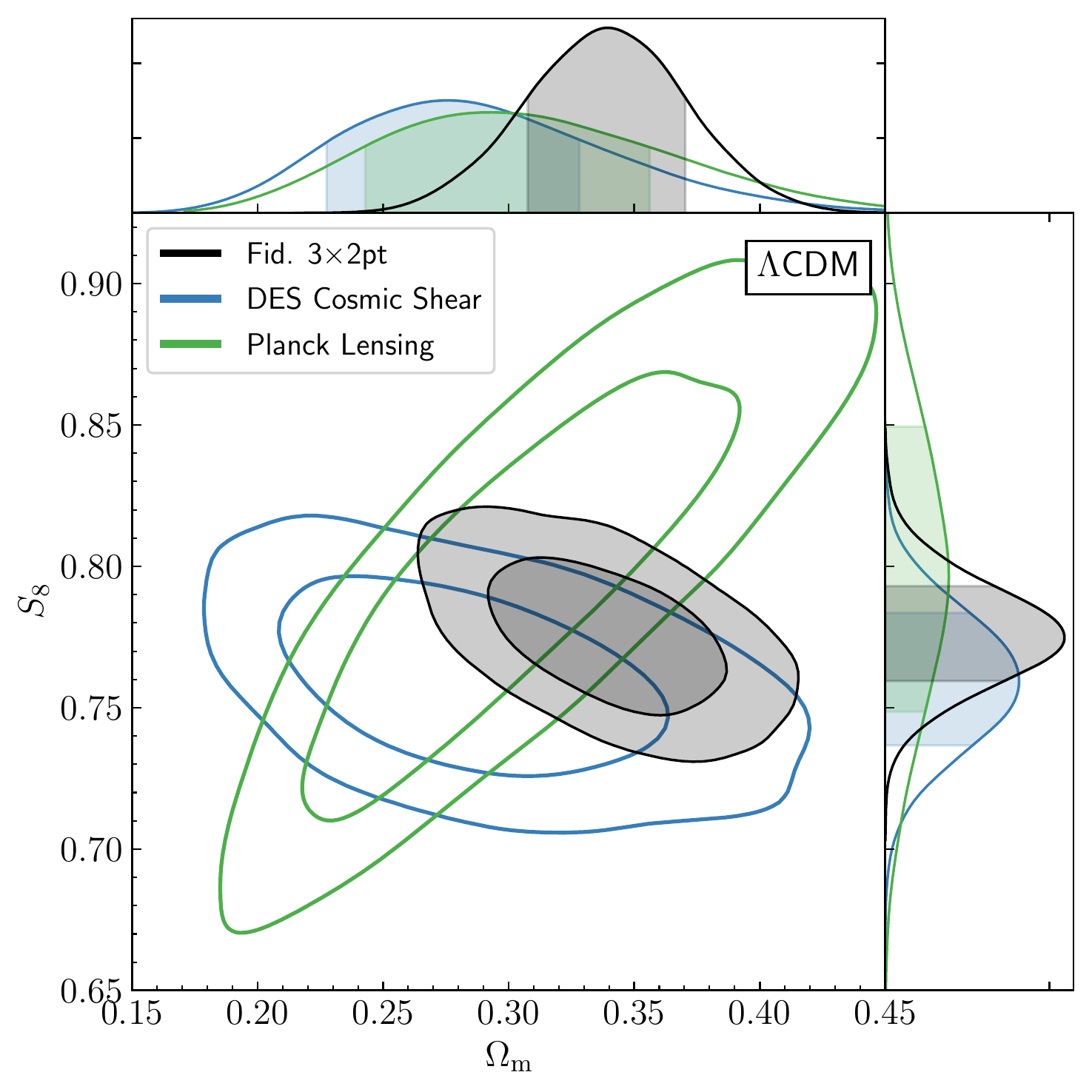}}
\caption{ A comparison of weak lensing constraints on the $\Lambda$CDM model. Weak lensing of the CMB is shown in green, weak lensing of galaxies in DES is shown in blue, and the combined DES 3$\times$2pt data is shown in black. 
  \label{lensing}}
\vspace{0cm}
\end{figure}

We are able to probe the distribution of large-scale structure via weak gravitational lensing (cosmic shear) in two very different ways, using either the shapes of galaxies or the CMB photons as tracers for the reconstruction of deflections in the path of the light.  
These probe the same physical phenomenon via independent sources and measurement methods, which are sensitive to different types of systematics. 
While the effective kernel of CMB lensing \cite{Aghanim:2018oex} is sensitive to higher redshift structure than galaxy lensing, their comparison provides a significant validation of the robustness of modern weak lensing results. 
We show this comparison in Fig.~\ref{lensing}, where we find very good agreement between the two cosmic shear measurements and the full 3$\times$2pt measurement from DES.

\begin{figure}
\centering
\resizebox{\columnwidth}{!}{\includegraphics{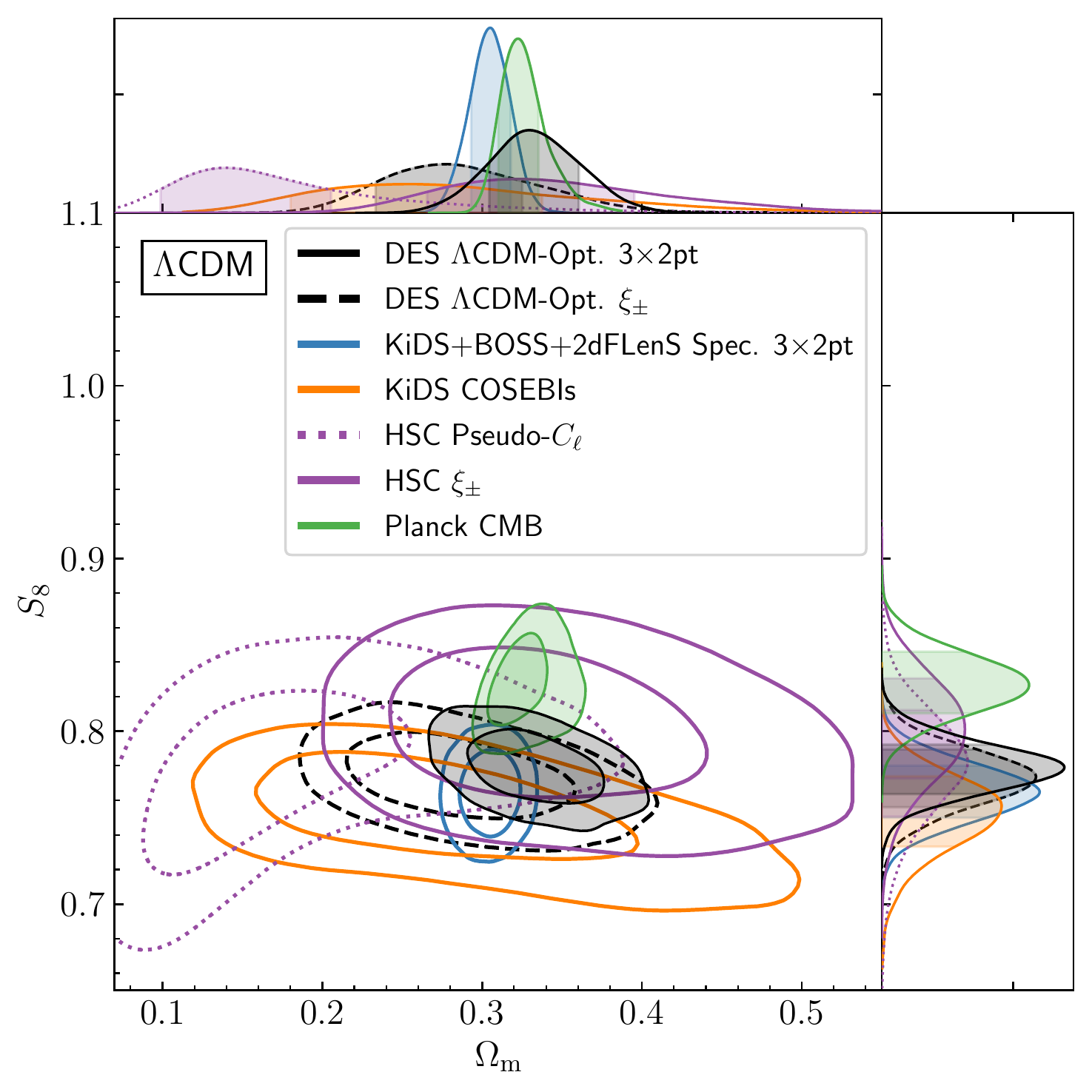}}
\caption{  The DES $\Lambda$CDM-optimized 3$\times$2pt and cosmic shear, HSC and KiDS cosmic shear, and KiDS lensing + BOSS+2dFLenS spectroscopic 3$\times$2pt data results are over-plotted for the $\Lambda$CDM model. Unlike other comparisons in this work, these external survey data have not been re-analyzed within a consistent model and prior space. Thus, no direct or rigorous comparison can be made about data consistency.
  \label{galaxylensing}}
\vspace{0cm}
\end{figure}

In addition to DES, other concurrent photometric surveys HSC \cite{Hikage:2018qbn,Hamana:2019etx} and KiDS \cite{Asgari:2020wuj,Heymans:2020gsg} are also pursuing precision weak lensing measurements using galaxy shapes. These three surveys span a range of depth and survey area tradeoffs, with HSC being deepest, DES widest, and KiDS using the widest wavelength coverage. We over-plot recent results from each of the surveys with our DES cosmic shear and 3$\times$2pt results in Fig.~\ref{galaxylensing}. Unlike other comparisons in this work, these external survey data have not been re-analyzed within a consistent model and prior space. Thus, no direct or rigorous comparison can be made about data consistency. We defer a detailed discussion of the consistency of concurrent photometric weak lensing surveys and their combination (e.g., Ref.~\cite{2019MNRAS.482.3696C}) to a future work. The apparent orthogonal direction of the KiDS+BOSS+2dFLenS 3$\times$2pt contours to the DES 3$\times$2pt contours is driven by the very strong constraint coming from spectroscopic clustering, similar to the orientation of the DES $\gamma_t$+$w(\theta)$ constraint.

\subsection{Constraints on the Hubble parameter}

\begin{figure}
\centering
\resizebox{\columnwidth}{!}{\includegraphics{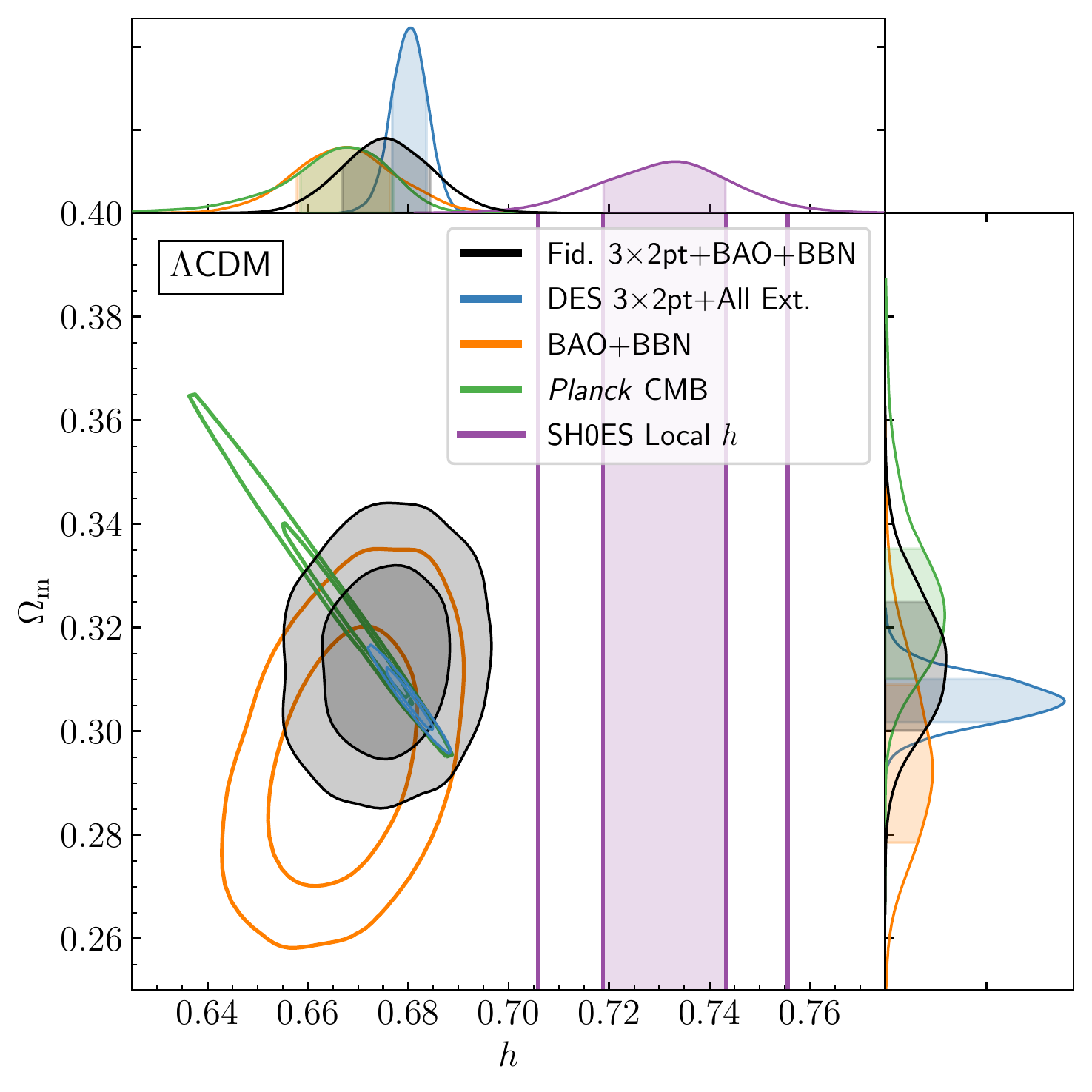}}
\caption{Marginalized constraints on $h$ and $\Omega_{\mathrm{m}}$ in the $\Lambda$CDM model are compared to the SH0ES local determination of $h$. \textit{Planck} CMB data and the  combination of  BAO and BBN data provide comparable uncertainties on $h$ compared to the local constraint. Adding DES 3$\times$2pt to BAO and BBN improves the constraint on $h$ slightly due to 3$\times$2pt providing additional information on $\Omega_{\mathrm{m}}$, while the combination of DES 3$\times$2pt and all non-local external data provide a constraint on $h$ that is a factor of 3-4 more powerful than the local determination. }
  \label{hfig}
\vspace{0cm}
\end{figure}

There is an interesting disagreement in local measurement of the Hubble parameter $h$ and marginalized constraints on $h$ from cosmological experiments.
Multiple local measurements prefer a higher value of the expansion velocity, such as, most prominently, the astronomical distance ladder (e.g., $h= 0.732\pm 0.013$ \citep{Riess:2020fzl} with Cepheid variable stars; $h= 0.733\pm 0.040$ \citep{Huang:2020} with Mira variable stars), or masers (e.g., $h= 0.739\pm 0.030\,{\rm
  km/s/Mpc}$ \citep{Pesce:2020}).
These local measurements stand in contrast to constraints from the CMB by \textit{Planck}, which prefer $h=0.665^{+0.013}_{-0.006}$ (when the neutrino mass density is varied) \citep{Aghanim:2018eyx}.
However, there are also local measurements with lower values reported ($h= 0.696\pm 0.019$ \citep{Freedman:2020} with tip of the red giant branch distance ladder;
  $h= 0.674^{+0.041}_{-0.032}$ with strong lensing when combining the TDCOSMO+SLACS data set \citep{Birrer:2020tdcosmoIV}).
The Hubble tension may indicate new physics and it is crucial to improve measurements, revisit assumptions \citep[e.g.,][]{Efstathiou:2020, Birrer:2020tdcosmoIV}, check for consistencies among different measurements, and invest in novel, independent methods and probes \citep{Chen:2017rfc, Feeney:2018mkj}.

We can also constrain the value of $h$ independently of CMB data using a combination of BAO, BBN constraints on $\Omega_{\mathrm{b}} h^2$, and DES 3$\times$2pt measurements. Constraints on $h$ and $\Omega_{\mathrm{m}}$ in $\Lambda$CDM are summarized in Fig.~\ref{hfig}.
The determination of $h$ using BAO and BBN is of similar constraining power to that of the CMB and agrees very well with the CMB constraint on $h$. Adding DES 3$\times$2pt data slightly improves the constraint on $h$ and shifts it to higher values by about 1$\sigma$. 
Combining DES 3$\times$2pt data with BAO, RSD, SNe Ia, and \textit{Planck} CMB (w/ lensing) leads to a marginalized constraint on $h$ 
\begin{equation}
  \begin{aligned}
    h         &= 0.680^{+0.004}_{-0.003} \;\;(0.681)
  \end{aligned}
\end{equation}
that is 3-4 times more powerful than any current local measurement of $h$. Constraints on other cosmological parameters are summarized in Tables  \ref{tab:post} \& \ref{tab:postw}.
We find no significant impact on the other cosmological parameters by adopting this high-redshift anchor for the expansion rate vs a local prior on the expansion rate from Ref.~\cite{Riess:2020fzl}. The final joint constraint on $h$ is consistent with the \textit{Planck}- or BAO+BBN-only constraints and slightly less than 4$\sigma$ offset relative to the local $h$ by SH0ES.

\subsection{Neutrino Mass}
\label{sec:neutrinos}

\begin{figure}
\centering
\resizebox{\columnwidth}{!}{\includegraphics{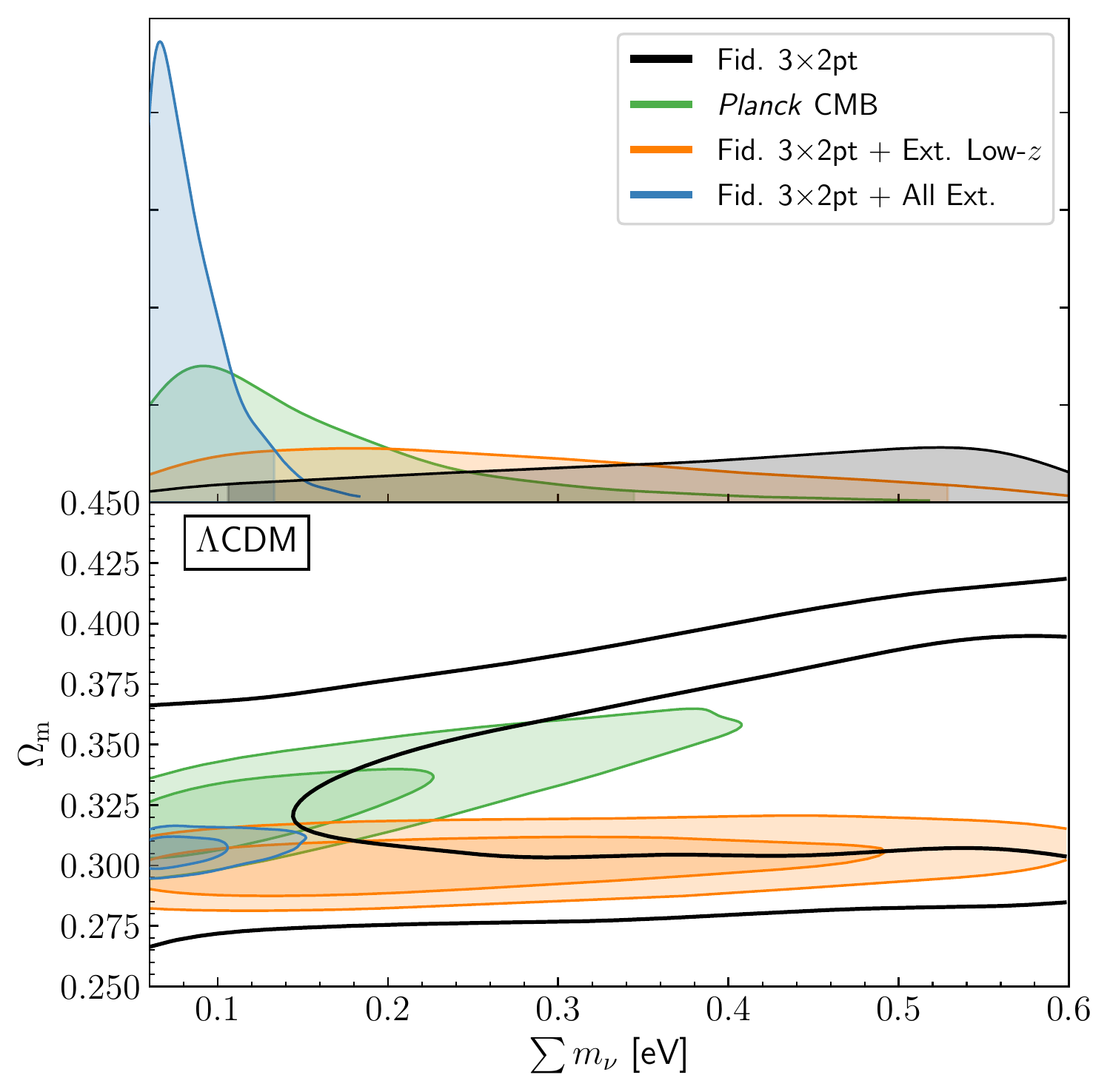}}
\caption{Marginalized constraints on the sum of neutrino masses in the $\Lambda$CDM model. We show  the DES fiducial 3$\times$2pt constraints (black), DES 3$\times$2pt combined with external BAO, RSD, and SNe Ia (orange), \textit{Planck} CMB constraints (green), and DES 3$\times$2pt combined with all of these external data sets. The upper panel shows the one-dimensional marginalized posteriors for $\sum m_{\nu}$, with shaded 95\% confidence regions. The lower panel shows 68 and 95\% CL for $\Omega_{\mathrm{m}}$ and $\sum m_{\nu}$. }
  \label{mnu}
\vspace{0cm}
\end{figure}

Figure \ref{mnu} shows marginalized constraints on the sum of neutrino masses, where neutrino mass and density $\Omega_{\nu}$ are related via $\sum m_{\nu} = 93.14 \Omega_{\nu}h^2$ eV. We model massive neutrinos as three degenerate species of equal mass. As expected, DES does not constrain neutrino mass: whether alone or in combination with external BAO, RSD, and supernova data, the marginalized posterior of the sum of neutrino masses is bounded by its prior. Given this, DES data is expected to add very little direct information on neutrinos.

Beyond constraints on neutrino mass themselves, a motivation for looking at the neutrino mass constraint is to highlight a feature of the relationship between \textit{Planck} and low-$z$ constraints on $\Omega_{\mathrm{m}}$, as shown in e.g. Fig.~\ref{extlcdm}. As has been previously discussed \cite{Aghanim:2019ame}, a geometric degeneracy means that CMB-only constraints are unable to distinguish between $\Omega_{\mathrm{m}}$ and $\sum m_{\nu}$, but combining CMB data with low redshift expansion history constraints from BAO and/or SNe can break that degeneracy. This also illustrates that the parts of the \textit{Planck} CMB posterior at higher $\Omega_{\mathrm{m}}$ also have relatively high neutrino mass. This is further supported by the behavior of constraints when neutrino mass is fixed, as shown in App.~\ref{sec:robustness}.  

Lowering the clustering amplitude has a similar impact on $\Omega_{\mathrm{m}}$, which we see as an increase in the upper limit on $\sum m_{\nu}$ of 23\% when combining DES 3$\times$2pt with \textit{Planck}. Combining the DES 3$\times$2pt data, other low-redshift data, and \textit{Planck} CMB, we find an upper limit 
\begin{equation}
  \begin{aligned}
   \sum m_{\nu}<0.13 \;\mathrm{eV\; (95\% \;CL)},
  \end{aligned}
\end{equation}
nearly a factor of three reduction from the CMB-only constraint.

\section{Conclusions}\label{sec:con}

We have described the 3$\times$2pt measurements, calibration, modeling, and analysis from the first three years of DES data. 
The substantial improvement in statistical power of the DES Y3 data, which cover an area of sky about three times that of DES Y1, has required substantial improvements in almost every part of the data processing, analysis and inference. The specific improvements relative to the DES Y1 analysis are detailed in App.~\ref{changes}, but we also briefly summarize here. At the catalog level, some of the most important updates for this analysis include improved PSF modeling and two complementary lens sample selections. We have substantially revised our shear and redshift inference and calibration processes. This includes more realistic image simulations to derive corrections on shear and redshift bias due to blending and detection, and a redshift inference process that combines spectroscopic and deep, multi-band photometric redshifts from DES deep-field data, cross-clustering between source and high-quality photo-$z$ and spectroscopic samples, and small-scale galaxy--galaxy lensing shear ratio information. Finally, we have updated several components of the analysis, including blinding, modeling of non-Limber and RSD contributions to the clustering signal, mitigation of nonlocal effects in $\gamma_t$, improved validation of the covariance matrix, and improved, calibrated metrics for evaluating the internal and external consistency of data sets.

The statistical power of the DES Year 3 data set has posed unique challenges for precision cosmological inference, some of which were unforeseen. 
We have identified some puzzling results from our photometric lens samples and galaxy clustering, which have not been identified with a clear source. 
This can be seen at very large significance as an apparent disagreement in the clustering and lensing amplitudes at all redshifts and angular scales for \redmagic, and in the highest redshifts of \maglim\ not used in the fiducial analysis. 
After unblinding our analysis, it was necessary to make two important revisions to the fiducial analysis plan: 1) we made the \maglim\ galaxy sample our fiducial lens sample owing to the decorrelation of the lensing and clustering amplitudes inferred from the \redmagic\ sample, and 2) we dropped the two highest redshift bins of the \maglim\ sample as they contributed to a very poor fit to all models considered in this paper. 
Sections~\ref{lcdmsec} \& \ref{lenscomp} provide the detailed rationale. These decisions were made after extensive, careful investigations of possible systematics, which did not reveal problems that we could address. 
Further investigations are already underway and may reveal the source of these issues, but our robustness tests so far indicate that any potential changes to the results will lie well within our quoted uncertainties.

We have achieved a factor of two improvement in statistical power relative to
the DES Y1 3$\times$2pt analysis in the $\sigma_8$--$\Omega_{\mathrm{m}}$
marginalized parameter plane, providing competitive cosmological constraints
relative to both the combination of all other external non-lensing
low-redshift data and the predictions from the \textit{Planck} CMB data. We find
consistent cosmological constraints from cosmic shear and the combination of
galaxy clustering and galaxy--galaxy lensing, as well as consistent
cosmological results from 3$\times$2pt utilizing two lens samples,
\maglim\ and \redmagic. For the fiducial 3$\times$2pt analysis in
$\Lambda$CDM, we find constraints on the clustering amplitude
$S_8=0.776^{+0.017}_{-0.017}$~(0.776) and matter density $\Omega_{\mathrm{m}} =
0.339^{+0.032}_{-0.031}$~(0.372). In the $w$CDM model, we find
$\Omega_{\mathrm{m}} =
0.352^{+0.035}_{-0.041} $~(0.339), and dark energy equation of state parameter
$w=-0.98^{+0.32}_{-0.20} $~($-1.03$).
      
The low-redshift measurements of the matter clustering amplitude by some galaxy surveys have tended to find lower variance relative to the prediction from \textit{Planck} CMB anisotropies, which may indicate some inconsistency between the low- and high-redshift Universe within the $\Lambda$CDM model, with claims of up to 2--3$\sigma$ significance. 
The DES Y3 3$\times$2pt analysis is an ideal experiment to test whether this is a real problem with the $\Lambda$CDM model. 
There have been substantial improvements in constraining power for both DES and \textit{Planck} since the DES Y1 3$\times$2pt analysis, yet we continue to find that DES 3$\times$2pt data and the combination of 3$\times$2pt with BAO and external SNe Ia and RSD data that the low-redshift Universe are consistent with predictions in the $\Lambda$CDM model from measurements at the time of the CMB from \textit{Planck}. We find all three independent data set combinations (DES 3$\times$2pt; BAO, RSD, and SNe Ia; and \textit{Planck} CMB) to be mutually consistent within $\Lambda$CDM.
Despite caveats on the precision with which we can make this statement discussed in Sec.~\ref{highlow}, this is the most powerful test of the standard cosmological model to date, comparing predictions from measurements of acoustic peaks in the early plasma of the Universe when it was 380,000 years old to measurements of large-scale structure from low-redshift surveys like DES spanning nearly nine billion years of cosmic evolution to the current day. 

Combining DES 3$\times$2pt, CMB, BAO, RSD, and SNe Ia data allows us to place the most precise constraints on the $\Lambda$CDM and $w$CDM models to date. We find $S_8 = 0.812^{+0.008}_{-0.008}$~(0.815) and $\Omega_{\mathrm{m}} = 0.306^{+0.004}_{-0.005} $~(0.306) in $\Lambda$CDM; $\sigma_8 = 0.812^{+0.008}_{-0.008} $~(0.804), $\Omega_{\mathrm{m}} = 0.302^{+0.006}_{-0.006} $~(0.298), and $w=-1.031^{+0.030}_{-0.027} $~($-1.00$) in $w$CDM. Additionally, we find an independent constraint on the Hubble parameter combining DES 3$\times$2pt, BAO, and BBN data of $h=0.676^{+0.009}_{-0.009} \;(0.673)$, which is consistent with the \textit{Planck} prediction for $h$. From the combination of DES 3$\times$2pt, CMB, BAO, RSD, and SNe Ia data, we find $h=0.680^{+0.004}_{-0.003} \;(0.681)$. This is slightly closer to the local $h$ measurement by SH0ES than \textit{Planck}, but a factor of three to four more constraining than either the local or \textit{Planck} measurement of $h$. We are also able to constrain the sum of neutrino masses to be $\sum m_{\nu}<0.13$ eV (95\% CL) in $\Lambda$CDM.

While we have shown that the inferred cosmological constraints from the fiducial analysis are robust, there remains significant work to fully characterize the underlying causes of these potential systematics and examine other potential theoretical causes in extended model spaces beyond the dark energy models considered in this work. 
Further understanding the potential systematic issues related to differences between photometric clustering and galaxy--galaxy lensing; improvements in how we deal with shear calibration and redshift inference in the presence of blending; and finding ways to improve systematic floors in our redshift inference are all important next steps. 
This continued followup work will be critical to the final DES Year 6 analyses and future `Stage IV' photometric surveys like the Euclid Space Telescope,\footnote{\url{https://www.euclid-ec.org}} the Nancy G. Roman Space Telescope,\footnote{\url{https://roman.gsfc.nasa.gov}} and the Vera C. Rubin Observatory Legacy Survey of Space and Time (LSST).\footnote{\url{https://www.lsst.org}} 

The novel advances required for the DES Y3 analyses, summarized in the accompanying 29 papers \cite{y3-gold,y3-deepfields,y3-shapecatalog,y3-piff,y3-simvalidation,y3-imagesims,y3-balrog,y3-lenswz,y3-sompz,y3-sompzbuzzard,y3-sourcewz,y3-2x2ptaltlenssompz,y3-hyperrank,y3-shearratio,y3-generalmethods,y3-covariances,y3-blinding,y3-inttensions,y3-tensions,y3-samplers,y3-2x2maglimforecast,y3-galaxyclustering,y3-gglensing,y3-massmapping,y3-2x2ptmagnification,y3-2x2ptbiasmodelling,y3-2x2ptaltlensresults,y3-cosmicshear1,y3-cosmicshear2}, set the stage for these future precision low-redshift large-scale structure and weak lensing studies. DES has utilized only half its final data set in the DES Y3 3$\times$2pt and BAO analyses, and future SNe Ia and galaxy cluster analyses promise even larger improvements in statistical power. The legacy analyses of the full DES data will be a focus of the next several years leading up to the start of Stage IV dark energy experiments.

\section*{Acknowledgments}
  
Funding for the DES Projects has been provided by the U.S. Department of Energy, the U.S. National Science Foundation, the Ministry of Science and Education of Spain, 
the Science and Technology Facilities Council of the United Kingdom, the Higher Education Funding Council for England, the National Center for Supercomputing 
Applications at the University of Illinois at Urbana-Champaign, the Kavli Institute of Cosmological Physics at the University of Chicago, 
the Center for Cosmology and Astro-Particle Physics at the Ohio State University,
the Mitchell Institute for Fundamental Physics and Astronomy at Texas A\&M University, Financiadora de Estudos e Projetos, 
Funda{\c c}{\~a}o Carlos Chagas Filho de Amparo {\`a} Pesquisa do Estado do Rio de Janeiro, Conselho Nacional de Desenvolvimento Cient{\'i}fico e Tecnol{\'o}gico and 
the Minist{\'e}rio da Ci{\^e}ncia, Tecnologia e Inova{\c c}{\~a}o, the Deutsche Forschungsgemeinschaft and the Collaborating Institutions in the Dark Energy Survey. 

The Collaborating Institutions are Argonne National Laboratory, the University of California at Santa Cruz, the University of Cambridge, Centro de Investigaciones Energ{\'e}ticas, 
Medioambientales y Tecnol{\'o}gicas-Madrid, the University of Chicago, University College London, the DES-Brazil Consortium, the University of Edinburgh, 
the Eidgen{\"o}ssische Technische Hochschule (ETH) Z{\"u}rich, 
Fermi National Accelerator Laboratory, the University of Illinois at Urbana-Champaign, the Institut de Ci{\`e}ncies de l'Espai (IEEC/CSIC), 
the Institut de F{\'i}sica d'Altes Energies, Lawrence Berkeley National Laboratory, the Ludwig-Maximilians Universit{\"a}t M{\"u}nchen and the associated Excellence Cluster Universe, 
the University of Michigan, NFS's NOIRLab, the University of Nottingham, The Ohio State University, the University of Pennsylvania, the University of Portsmouth, 
SLAC National Accelerator Laboratory, Stanford University, the University of Sussex, Texas A\&M University, and the OzDES Membership Consortium.

Based in part on observations at Cerro Tololo Inter-American Observatory at NSF's NOIRLab (NOIRLab Prop. ID 2012B-0001; PI: J. Frieman), which is managed by the Association of Universities for Research in Astronomy (AURA) under a cooperative agreement with the National Science Foundation.

The DES data management system is supported by the National Science Foundation under Grant Numbers AST-1138766 and AST-1536171.
The DES participants from Spanish institutions are partially supported by MICINN under grants ESP2017-89838, PGC2018-094773, PGC2018-102021, SEV-2016-0588, SEV-2016-0597, and MDM-2015-0509, some of which include ERDF funds from the European Union. IFAE is partially funded by the CERCA program of the Generalitat de Catalunya.
Research leading to these results has received funding from the European Research
Council under the European Union's Seventh Framework Program (FP7/2007-2013) including ERC grant agreements 240672, 291329, and 306478.
We  acknowledge support from the Brazilian Instituto Nacional de Ci\^encia
e Tecnologia (INCT) do e-Universo (CNPq grant 465376/2014-2).

This manuscript has been authored by Fermi Research Alliance, LLC under Contract No. DE-AC02-07CH11359 with the U.S. Department of Energy, Office of Science, Office of High Energy Physics. 

This research used resources of the National Energy Research Scientific Computing
Center, a DOE Office of Science User Facility supported by the Office of
Science of the U.S. Department of Energy under Contract No. DE-AC02-05CH11231. This work also used resources on Duke Compute Cluster (DCC), the CCAPP condo of the Ruby Cluster at the Ohio Supercomputing Center \cite{OhioSupercomputerCenter1987}, and computing resources at SLAC National Accelerator Laboratory. We also thank the staff of the Fermilab Computing Sector for their support. Plots in this manuscript were produced partly with \textsc{Matplotlib} \cite{Hunter:2007}, and it has been prepared using NASA's Astrophysics Data System Bibliographic Services.

\bibliography{refs,../../../des_bibtex/y3kp}

 \appendix
 
\section{Summary of Associated Papers}\label{sec:papers}

This paper is built on the results presented in 29 accompanying papers. A useful way to navigate these is to divide them up into five categories: 

{\it Catalog Papers}: The link between the raw images and the two-point functions from which cosmology is extracted is a set of catalogs. The Gold catalog (\citet*{y3-gold}) uses coadd images to identify galaxies and their properties. This is a first step to almost all ensuing work. Estimating the redshifts of those galaxies hinges in large part on the much deeper catalog of galaxies from the DES deep fields with overlapping near-infrared photometry presented in~\citet*{y3-deepfields}. Finally, the shear catalog is presented in \citet*{y3-shapecatalog}, which uses a new PSF measurement described in \citet*{y3-piff}.

{\it Simulations}: To test our models and calibration of the data, we rely on large suites of cosmological and image-level simulations. The Buzzard simulations (\citet*{y3-simvalidation}) are generated from cosmological N-Body simulations, populated with realistic galaxy samples, and then used in end-to-end analyses to stress test our modeling and methods. To calibrate blending and detection biases, we generated multiple image simulations, described in \citet*{y3-imagesims}, where shear and redshift biases are evaluated. Balrog (\citet*{y3-balrog}) is a tool developed to inject realistic images of galaxies into real DES images to evaluate the survey selection function. This is critical for multiple purposes, including our photometric redshift inference. The full DES detection and measurement pipelines are run on the simulations described in \citet*{y3-imagesims} and \citet*{y3-balrog}.

{\it Photometric Redshifts}: The redshift distributions for the \redmagic\ and \maglim\ lens galaxy samples are validated using cross-correlations with spectroscopic galaxies in \citet*{y3-lenswz}, which informs our priors on the uncertainty of the redshift distribution. The redshift inference process for the source galaxies is a much more involved process. The overview of this work is presented in \citet*{y3-sompz}, which builds on the self-organizing map formalism developed in \citet*{y3-sompzbuzzard}, and incorporates constraints from the cross-correlation with both \redmagic\ and spectroscopic galaxies described in \citet*{y3-sourcewz}. A followup analysis applying this methodology to the \maglim\ lens redshift distributions is described in \citet*{y3-2x2ptaltlenssompz}. The sampling of the resulting realizations of the redshift distribution are described in \citet*{y3-hyperrank}. Finally, we also add additional information from the posterior from the {\it shear ratio} likelihood, which is most sensitive to photo-$z$ parameters and is described in \citet*{y3-shearratio}.

{\it Analysis}: The modeling outlined in the text is described and tested in detail in \citet*{y3-generalmethods}. One of the most important pieces of the analysis is the generation of the covariance matrix of the two-point functions. The way in which this is generated and tested is described in \citet*{y3-covariances}. As mentioned in the text, we remain blinded to the data throughout the analysis to avoid unconscious bias; this blinding has multiple levels, but one of the key tiers is described in \citet*{y3-blinding}. Methods to evaluate the internal consistency of our data and to evaluate consistency with external data sets were calibrated and described in \citet*{y3-inttensions} \& \citet*{y3-tensions}. Efficient sampling of the likelihood can save months of analysis time, and ways to optimally, and more importantly robustly, sample our likelihood are summarized in \citet*{y3-samplers}. Finally, the optimization of the \maglim\ lens sample is described in \citet*{y3-2x2maglimforecast}.

{\it Results}: The galaxy clustering weights and measurement are presented in \citet*{y3-galaxyclustering}; galaxy--galaxy lensing in \citet*{y3-gglensing} and in \citet*{y3-shearratio} for the smaller-scale {\it shear ratio} measurements; and the weak lensing convergence mass map in \citet*{y3-massmapping}. The results of combining clustering with galaxy--galaxy lensing are presented in three papers, one of which focuses on magnification (\citet*{y3-2x2ptmagnification}), another on \redmagic\ and the bias model (\citet*{y3-2x2ptbiasmodelling}), and the last on the \maglim\ sample (\citet*{y3-2x2ptaltlensresults}). Finally, the cosmic shear results are presented in two papers, one of which focuses on observational systematics (\citet*{y3-cosmicshear1}) and the other on biases in theoretical modeling (\citet*{y3-cosmicshear2}).

\section{Differences relative to DES Year 1 analysis}\label{changes}

In this Appendix, we summarize the major differences in the DES Y3 analysis relative to DES Y1. 

\textit{Data Processing}: DES Y3 contains significantly more data than DES Y1 \citep{2018ApJS..239...18A}, which are processed with an improved version of the DESDM system \citep{2018PASP..130g4501M}. Photometric calibration is performed with the forward global calibration method \citep{Burke18}, which significantly improves the relative photometric calibration compared to the calibration techniques applied in DES Y1 \citep{y1-gold}. The Y3 catalogs also introduce more robust morphological classification based on multi-epoch fitting and improved flagging to enhance the quality of the galaxy sample. Further details on improvements to the DES data set and object catalogs can be found in \citet{y3-gold}.

\textit{Catalog-level}: We have produced a calibrated deep-field data reduction for use in the analysis. We have constructed only one shape catalog (\mcal; \cite{y3-shapecatalog}), but produced two very different lens samples to compare cosmological results \cite{y3-galaxyclustering,y3-2x2ptbiasmodelling,y3-2x2ptaltlensresults}. We use a new PSF model (PIFF) \cite{y3-piff}.

\textit{Simulation \& Calibration}: We have produced the Balrog simulation to characterize the wide-field survey selection function of objects derived from the deep-field coadd images \cite{y3-balrog}. We have produced a new suite of image simulations for shear calibration that are more realistic and better match the data \cite{y3-imagesims}. We have produced a new methodology for shear calibration that explicitly accounts for the impact of blending and detection biases as a function of redshift, including modifications to the effective $n(z)$ \cite{y3-imagesims}. We have developed a new redshift inference and calibration framework, mixing spectroscopic and photometric redshift information, to produce Bayesian posterior $n(z)$ samples \cite{y3-sompzbuzzard,y3-sompz,y3-lenswz,y3-sourcewz,y3-2x2ptaltlenssompz,y3-shearratio,y3-hyperrank}.

\begin{figure*}
\centering
\resizebox{\textwidth}{!}{\includegraphics{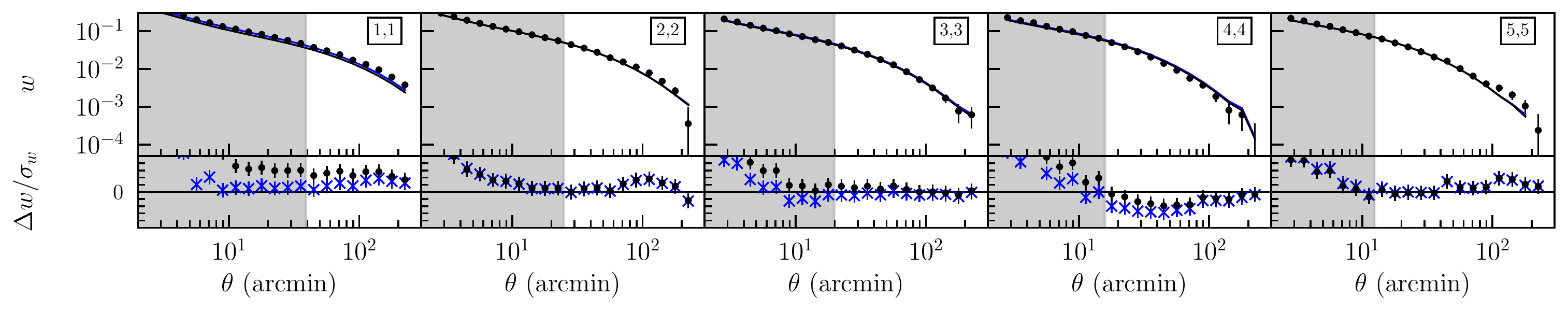}}
\caption{The measured $w(\theta)$ correlation functions for each tomographic
  bin combination used in the \redmagic\ analysis, which is indicated by the
  $i,j$ label in each set of panels. The best-fit $\Lambda$CDM model from the
  analysis using \redmagic\ is plotted as the solid line in the top part of each
  panel, while the bottom parts of each panel shows the fractional difference
  between the measurements and the model prediction,
  $(w^{\textrm{obs.}}-w^{\textrm{th.}})/\sigma_w$ (with $y$-axis range $\pm
  5\sigma$). The best-fit model with fixed $X_{\rm lens}$ is shown in black, while the
  best-fit model marginalizing over $X_{\rm lens}$ is shown in blue. Both the top and
  bottom part of each panel includes 1$\sigma$ error bars. Small angular
  scales where the linear galaxy bias assumption breaking down are not used in
  the cosmological analysis; these scales are indicated by grey shading.
}
\label{fig:wtrm}
\end{figure*}

\textit{Modeling}: We have improved the rigor of constraining potential bias from model approximations, with explicit accuracy goals at the $\chi^2$ and parameter levels \cite{y3-generalmethods}. We have included non-Limber and redshift-space distortion (RSD) contributions to the clustering theory \cite{y3-generalmethods}. We have included the impact of lens magnification \cite{y3-generalmethods,y3-2x2ptmagnification}. We have made updates to the covariance modeling and improved the rigor of validation tests in both $\chi^2$ and parameter space \cite{y3-covariances}. We utilize a new intrinsic alignment model that allows for `red', `blue', and mixed alignment modes to second order in perturbation theory \cite{2019PhRvD.100j3506B}. We have developed a nonlinear galaxy bias model for the analysis \cite{y3-2x2ptbiasmodelling}. We have updated our methodology for determining scale cuts using explicit accuracy goals in both $\chi^2$ and parameter space \cite{y3-generalmethods}. 

\textit{Analysis}: We have utilized a new summary-statistic-level blinding scheme \cite{y3-blinding}. We have introduced a new shear-ratio likelihood \cite{y3-shearratio}. We have introduced a new cross-clustering redshift likelihood \cite{y3-sompz,y3-sourcewz}. We have implemented a method for sampling over the full-shape $n(z)$ samples \cite{y3-hyperrank}. We analytically marginalize over uncertainty in the lens sample clustering weights \cite{y3-galaxyclustering}. We analytically marginalize over a point-mass contribution to $\gamma_t$ to mitigate non-local effects \cite{y3-generalmethods}. We marginalize over the widths of the lens $n(z)$ \cite{y3-galaxyclustering,y3-lenswz}. We have updated and calibrated a new set of internal and external consistency tests \cite{y3-inttensions,y3-tensions}. We have updated requirements on sampling and evidence precision and now use the \texttt{PolyChord} sampler \cite{y3-samplers}.

\begin{table}
\caption{The parameter differences for the \redmagic\ analysis relative to Table \ref{tab:params}.}
\begin{center}
\begin{tabular*}{\columnwidth}{ l  @{\extracolsep{\fill}} c  c}
\hline
\hline
\multicolumn{2}{l}{{\bf Lens Galaxy Bias} (\redmagic\ )} \\
$b_{i} (i\in[1,5])$   & Flat  & (0.8, 3.0) \\
\hline
\multicolumn{2}{l}{{\bf Lens magnification} (\redmagic\ )} \\
$C_{\rm l}^1 $ & Fixed &  $0.63$ \\
$C_{\rm l}^2 $ & Fixed &  $-3.04$ \\
$C_{\rm l}^3 $ & Fixed &  $-1.33$ \\
$C_{\rm l}^4 $ & Fixed &  $2.50$ \\
$C_{\rm l}^5 $ & Fixed &  $1.93$ \\
\hline
\multicolumn{2}{l}{{\bf Lens \photoz\ } (\redmagic\ )} \\
$\Delta z^1_{\rm l}$  & Gaussian  & ($0.006, 0.004$) \\
$\Delta z^2_{\rm l}$  & Gaussian  & ($0.001, 0.003$) \\
$\Delta z^3_{\rm l}$  & Gaussian  & ($0.004, 0.003$) \\
$\Delta z^4_{\rm l}$  & Gaussian  & ($-0.002, 0.005$) \\
$\Delta z^5_{\rm l}$  & Gaussian  & ($-0.007, 0.010$) \\
$\sigma^i_{z,\rm l} (i\in[1,4])$  & Fixed  & $1.0$ \\
$\sigma^5_{z,\rm l}$  & Gaussian  & ($1.23, 0.054$) \\
\hline
\hline
\end{tabular*}
\end{center}
\label{tab:params2}
\end{table}

\begin{figure*}
\centering
\resizebox{\textwidth}{!}{\includegraphics{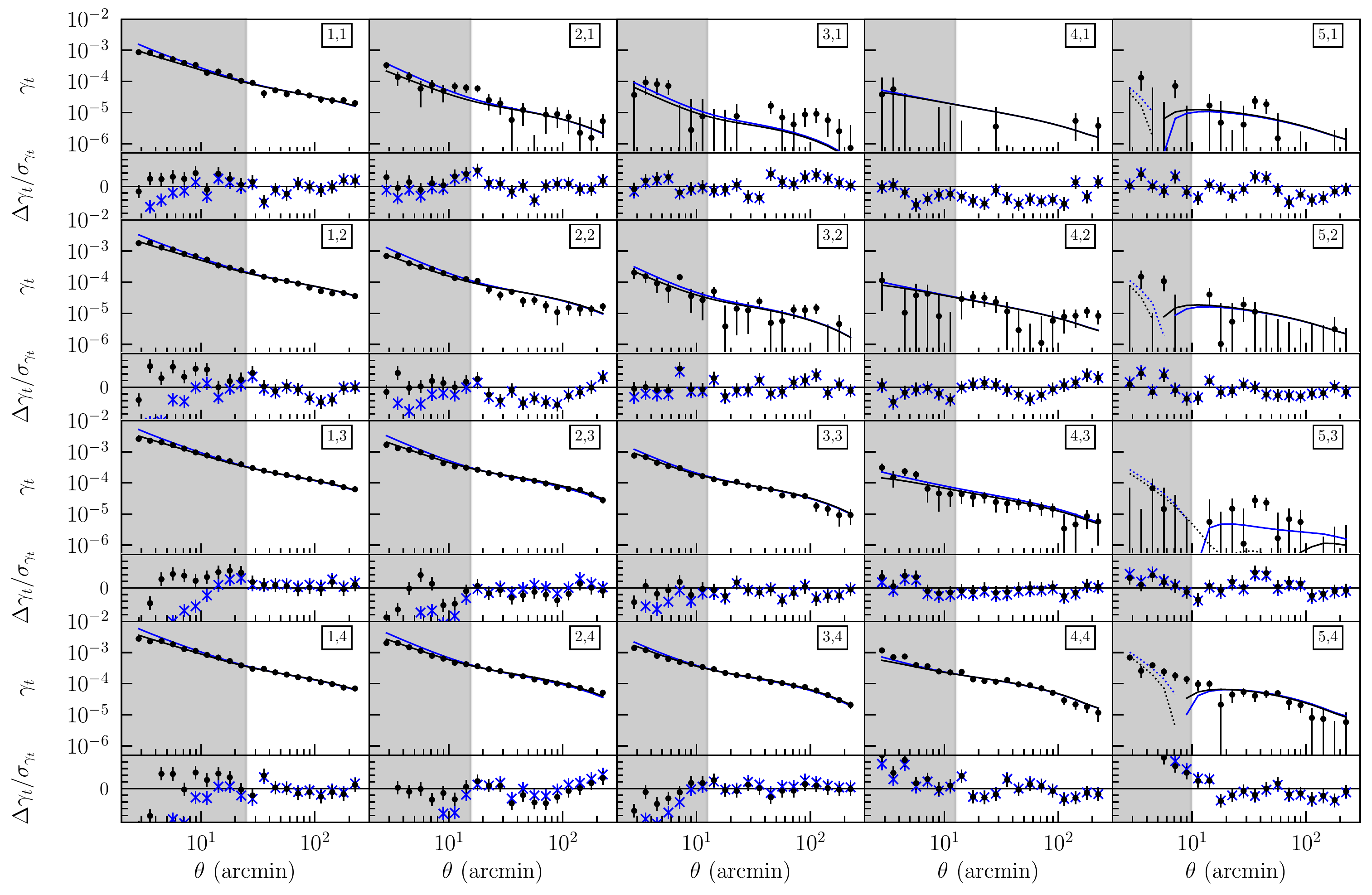}}
\caption{The measured $\gamma_{t}(\theta)$ correlation functions for each
  tomographic bin combination using the \redmagic\ sample, with labels as
  described in Fig.~\ref{fig:gt}. The best-fit $\Lambda$CDM model from the
  analysis with fixed $X_{\rm lens}$ is plotted as the solid line in the top
  part of each panel, with dotted curves indicating a negative model fit. The
  best-fit model marginalizing over $X_{\rm lens}$ is shown in blue. The bottom part of
  each panel shows the fractional differences between the measurements and the
  model prediction,
  $(\gamma_t^{\textrm{obs.}}-\gamma_t^{\textrm{th.}})/\sigma_{\gamma_t}$ (with
  $y$-axis range $\pm 5\sigma$). In both panels, 1$\sigma$ error bars are
  included. Angular scales not used in the cosmological analysis are indicated
  by grey shading, which are excluded on small scales where the linear galaxy
  bias assumption breaks down.}
\label{fig:gtrm}
\end{figure*}

\section{Differences in \redmagic\ analysis}\label{sec:redmagic}

The \redmagic\ lens analysis differs in several ways due to the different redshift range, binning, and sample selection relative to the \maglim\ lens analysis. The parameterization and prior changes relative to the fiducial analysis are listed in Table \ref{tab:params2}. In particular, we do not vary the width of the redshift distributions in the first four \redmagic\ lens bins, since constraints from clustering on the $n(z)$ agree well with the predictions from the \redmagic\ photo-$z$ algorithm \cite{y3-lenswz}. The magnification parameters are also fixed to different values based on measurements from Balrog \cite{y3-2x2ptmagnification}. We also show the \redmagic\ galaxy clustering and galaxy--galaxy lensing data vectors and best-fit cosmological model in Figs. \ref{fig:wtrm} \& \ref{fig:gtrm}.

  \begin{figure}
\centering
\resizebox{\columnwidth}{!}{\includegraphics{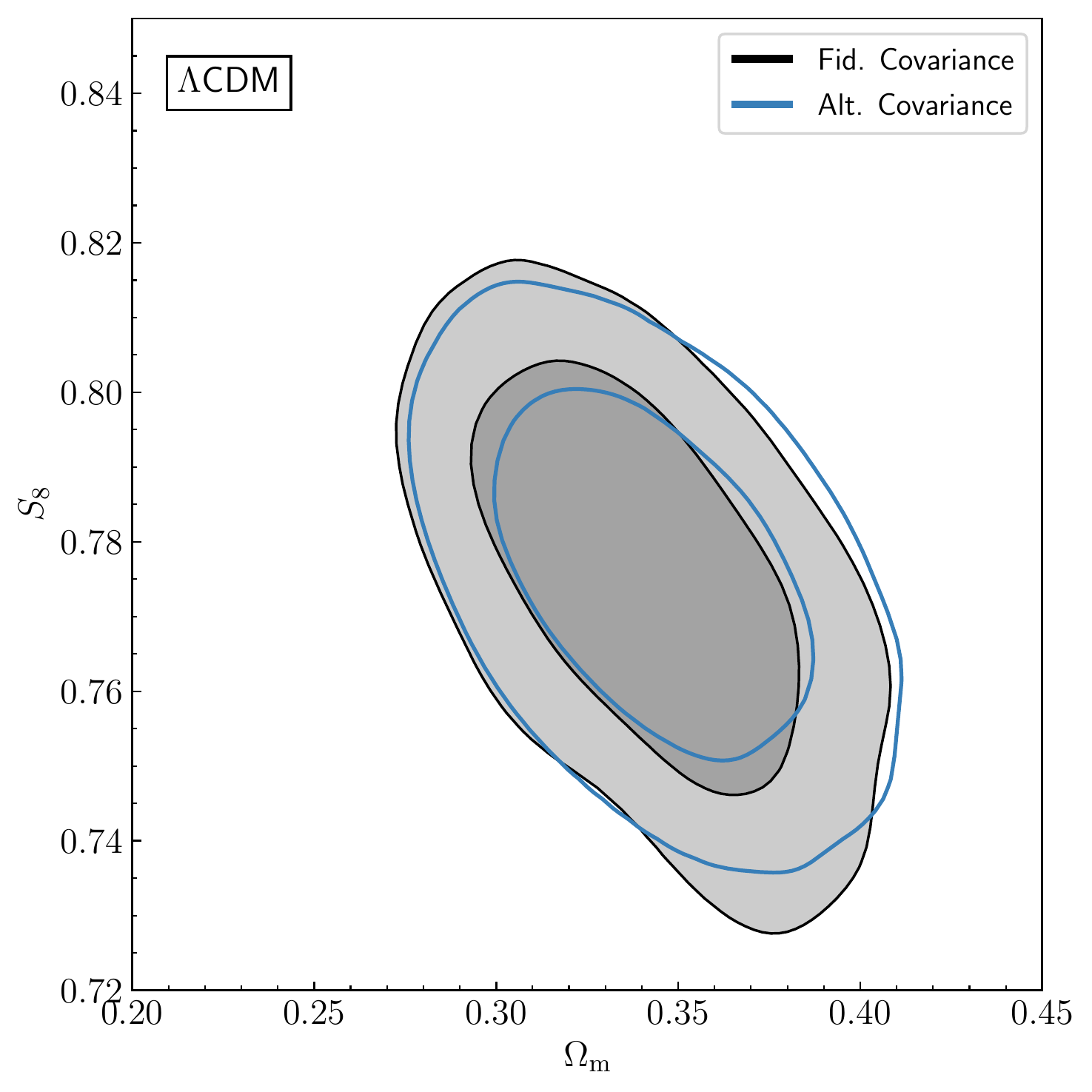}}
\caption{ A test of the convergence of the theoretical 3$\times$2pt covariance, showing marginalized parameter constraints using two iterations of the covariance that use the best-fit cosmological parameters from the previous analysis. 
  \label{cov}}
\end{figure}

\section{Observer Bias \& Validation Process}\label{sec:blind}

The role of the observer in determining how to proceed in an analysis or measurement and when to accept a result as final has been demonstrated to contribute to uncharacterized bias in results (e.g., \cite{doi:10.1146/annurev.nucl.55.090704.151521, 2011arXiv1112.3108C}). 
To protect against this, all cosmologically relevant measurements, calibrations, model validation, and fiducial analysis plans were performed blinded. This was done in three stages, to provide redundant protection: 
\begin{enumerate}
\item The ellipticity in the shape catalog was blinded by a random factor in $b\in [0.9,1.1]$, transforming $\eta_i \rightarrow b\eta_i$, where $\eta_i = 2 e_i \arctanh(e)/e$ and $e^2=e_1^2+e_2^2$. The factor $b$ is generated in a random way by hashing a short string phrase, and this phrase was known only to the people making the catalog.
\item The two-point correlation functions were coherently shifted by an unknown vector in cosmological parameter space ($\Omega_{\mathrm{m}}$, $w$), as described in Ref.~\cite{y3-blinding}. This produces a new data vector corresponding to an unknown $w$CDM cosmology, and has the benefit of leaving all three two-point correlation functions internally consistent.
\item All parameter values and the axes of posterior plots inferred from chains were randomly shifted.
\end{enumerate}
These protections were removed one by one as the analysis matured. 

First, the catalog-level blinding was removed when the validation and planned calibration of the catalog were finalized, to allow null tests and non-cosmological measurements to be remade for the papers while the data vectors were still blinded. In practice, this also only occurred after model validation tests had been completed and other analysis and calibration plans were also finalized. 

Once the data vector, shear and redshift calibration, all model validation and choices, and plans for testing internal and external consistency were finalized, we proceeded with a pre-determined set of tests on the blinded results. The enumerated tests below were those that would have led us to reconsider whether to proceed with unblinding and pursue potential systematic causes. In addition to this, other validation checks were also performed at this point on the blinded data vector and posteriors, but were not required for unblinding.
\begin{enumerate}
\item Verify the final independent redshift posteriors are consistent (i.e., from SOMPZ, the shear ratio, and clustering cross-correlations).  
\item Verify no posteriors of systematic parameters concentrate at the edge of their priors in ways that are not understood.
\item Verify the goodness-of-fit for each subset of the data (i.e., cosmic shear and galaxy clustering+galaxy--galaxy lensing). We used the posterior predictive distribution (PPD) with a quantitative requirement $p>0.01$.
\item Verify that cosmic shear and galaxy clustering+galaxy--galaxy lensing are consistent with each other, with the same quantitative requirement.
\item If any of the previous tests failed in $\Lambda$CDM, a passing condition in $w$CDM would also be sufficient. 
\end{enumerate}

The theoretical covariance matrix associated with the 3$\times$2pt data is calculated with an assumption about the true cosmological and other parameters in our model that may be different from what we find after unblinding our data. Using a covariance matrix that assumes the wrong cosmology can bias the cosmological inference process. To mitigate this, we recalculate the covariance matrix at the best-fit cosmology of the initial 3$\times$2pt analysis and run all final chains with this covariance. We confirm that this process has converged by comparing the result of this analysis with an analysis based on yet a third covariance calculated at the second best-fit cosmology. This is shown in Fig.~\ref{cov}.

We unblinded the \redmagic\ sample before the \maglim\ sample, finding it to pass all unblinding checks. After unblinding it became apparent from additional internal consistency (PPD) tests that we had statistically significant evidence for a potential systematic error in the clustering part of the data vector. This was explored at length in parallel to the final validation of the \maglim\ sample leading up to its unblinding. We verified, though did not require, that there was no evidence that \maglim\ suffered from the same effect before unblinding it. The \maglim\ sample, however, did fail the $\chi^2$ criterion above for both models after updating the parameter values for the covariance matrix after the initial unblinding chains, with an excess $\chi^2 \approx 100$. This was not seen before unblinding due partly to a poor choice of galaxy bias values for the initial covariance, leading to the updated covariance matrix being 40-50\% smaller in the $w(\theta)$ block.

We had agreed that it was implausible that our data should be able to distinguish a non-$w$CDM model at such a large $\chi^2$ before unblinding, and so we had planned to explore potential causes for the poor model fit before proceeding in such a case, correcting any problems that were then discovered before unblinding. Unfortunately, this effect was hidden due to the initial covariance before parameter updates being weaker in the $w(\theta)$ blocks, and so this was a post-unblinding correction. However, the solution (imposing an upper redshift cut for the \maglim\ sample) is something we would almost surely have pursued before unblinding had this problem been apparent then. These lens sample issues and related tests are summarized further in Sec.~\ref{lenscomp} and in Refs. \cite{y3-galaxyclustering,y3-2x2ptbiasmodelling,y3-2x2ptaltlensresults}.

The list of changes made to the analyses post-unblinding are as follows. For \redmagic, we investigated how the lens weights were calculated due to indications that an alternative method that used a principal component (PC) basis of the observing condition maps produced a non-trivial shift in the clustering signal in the direction to correct the observed excess clustering parameterized by $X_{\rm lens}$. However, it was determined that this basis was contaminated by true large-scale structure modes. A simplified basis limited to the first 50 PCs to remove any contaminated modes was used in the final analysis that gives consistent results with the original weights used at unblinding. An additive component was added to the $w(\theta)$ covariance block of both lens samples that accounts for potential over-correction and differences between the two weights methods. The \maglim\ sample used this final weighting and covariance when unblinded. As discussed previously, the highest two redshift bins of the \maglim\ sample were also removed to resolve a very poor $\chi^2$ fit to any dark energy model considered and a strong indication of inconsistency between $w(\theta)$ and $\gamma_t$ approaching $X_{\rm lens}=0.55$ in the highest redshift bin. There was no indication of the impact of $X_{\rm lens}$ for \maglim\ generally within the redshift range overlapping \redmagic. This change in redshift limit resulted in a shift of $0.78\sigma$ in the $\Omega_{\mathrm{m}}$--$S_8$ plane, almost fully in the direction of increasing $\Omega_{\mathrm{m}}$, relative to the parameter values found at unblinding. This is consistent with shifts observed when removing parts of the redshift range of the data, so not clearly evidence of systematic impact on the parameter values.

\section{Alternative analysis choices and robustness}
\label{sec:robustness}

  \begin{figure}
\centering
\resizebox{\columnwidth}{!}{\includegraphics{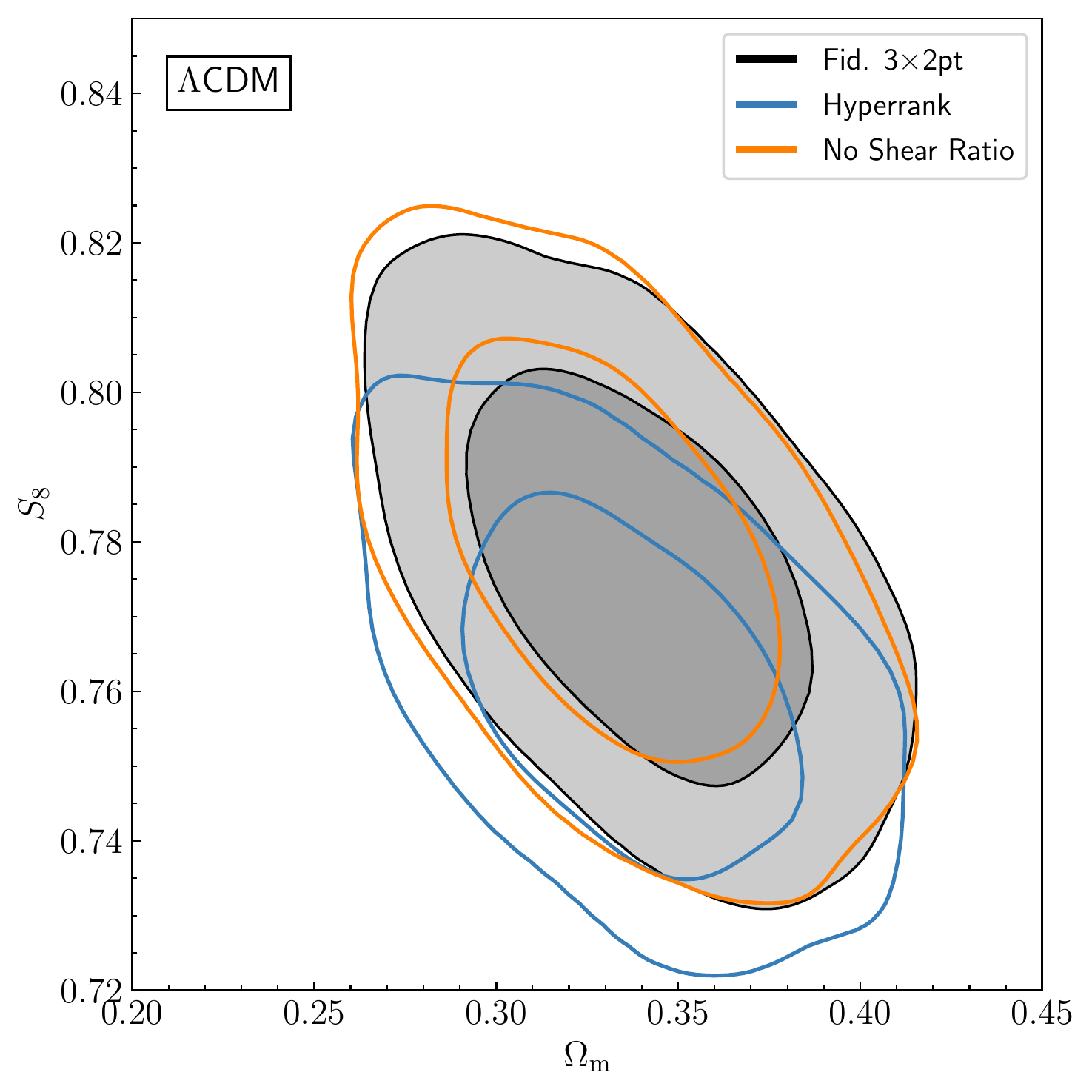}}
\caption{ A test of the impact of alternative redshift analysis choices on the inferred cosmology from 3$\times$2pt. We compare the fiducial 3$\times$2pt analysis (black) to an analysis where we marginalize over the ensemble of $n(z)$ realizations directly via Hyperrank instead of their effective mean redshifts (blue) and to an analysis where we remove the shear-ratio data (orange). 
  \label{lcdmalt}}
\end{figure}
  \begin{figure}
\centering
\resizebox{\columnwidth}{!}{\includegraphics{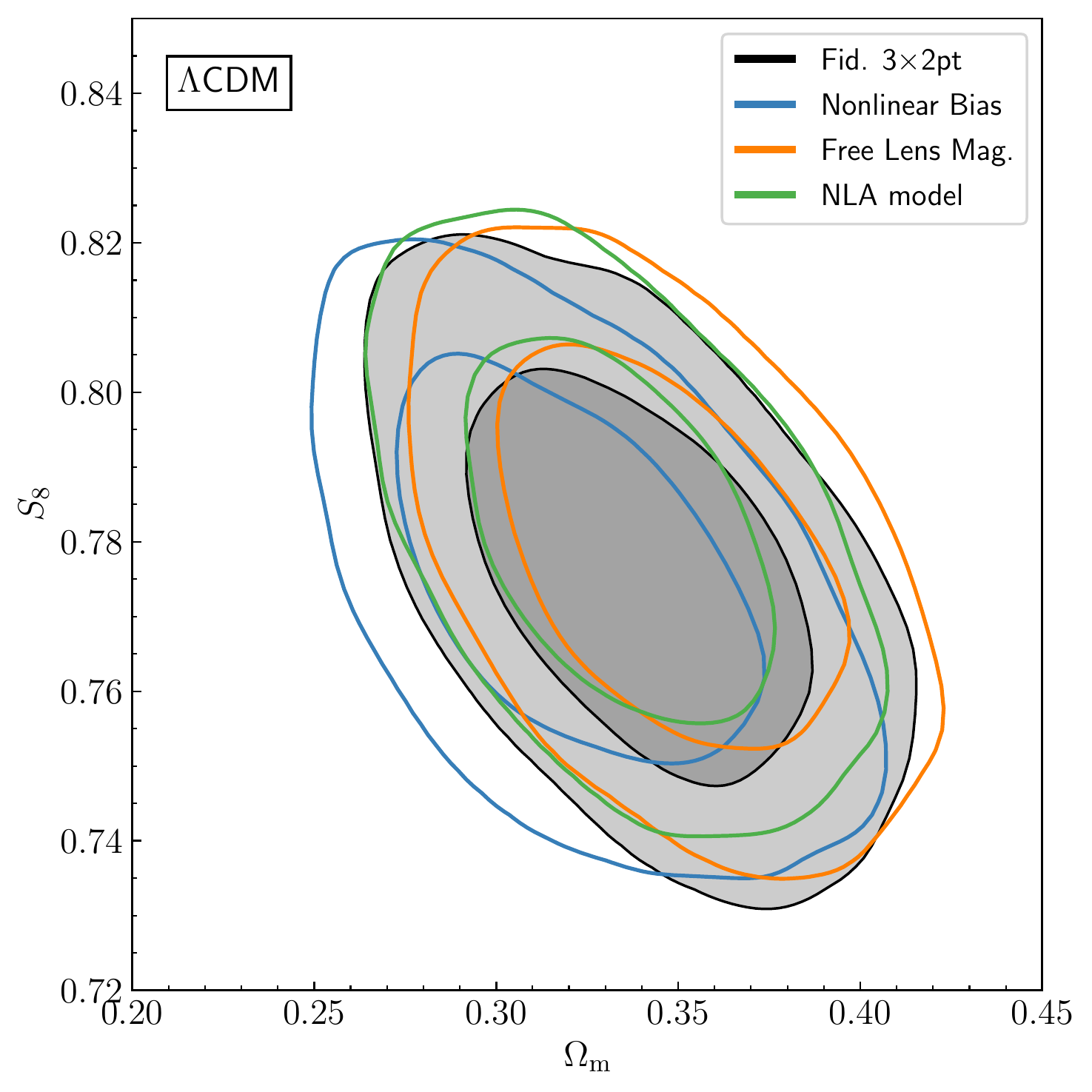}}
\caption{ A test of the impact of alternative analysis choices on the inferred cosmology from 3$\times$2pt. We compare the fiducial 3$\times$2pt analysis (black) to an analysis where we marginalize over a nonlinear bias model using smaller scales in $\gamma_t$ and $w(\theta)$ in blue, an analysis that marginalizes over free lens magnification bias parameters in orange, and an analysis that uses the NLA IA model in green.
  \label{lcdmalt2}}
\end{figure}
  \begin{figure}
\centering
\resizebox{\columnwidth}{!}{\includegraphics{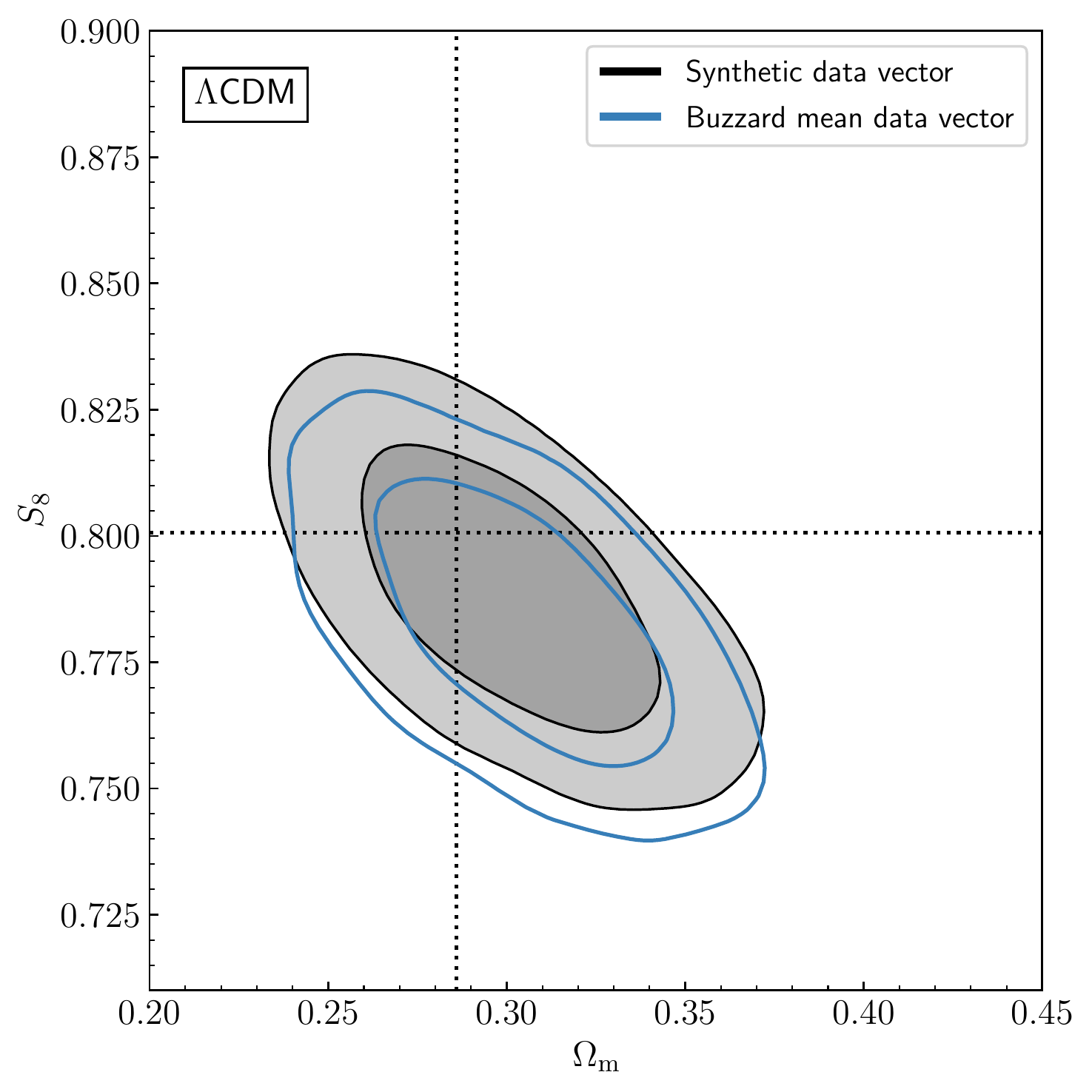}}
\caption{ The validation of the model and inference pipeline on the Buzzard simulation suite, where the true cosmology is indicated by the cross. An analysis of a synthetic data vector (black) at the true Buzzard cosmology based on the true redshift distributions with fixed shear and photo-$z$ parameters is compared to a full analysis of the mean data vector of 18 simulation realizations (blue) including all nuisance parameters and $n(z)$ distributions inferred in the same way we do using the real survey data.
  \label{simfig}}
\end{figure}
  \begin{figure}
\centering
\resizebox{\columnwidth}{!}{\includegraphics{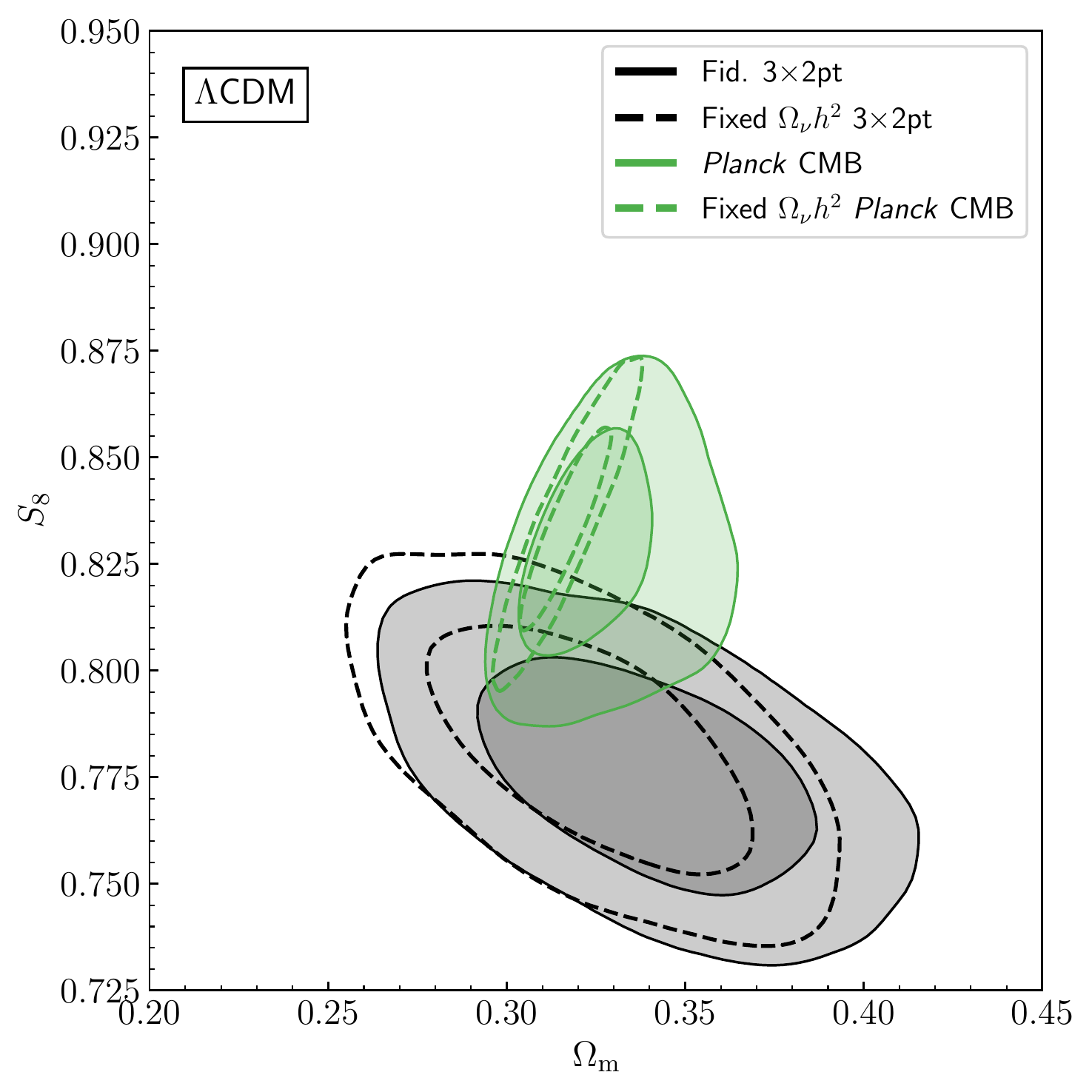}}
\caption{  A comparison of the fiducial 3$\times$2pt analysis (black solid) with one
  that fixes the neutrino mass density to fixed is minimum value (black dashed). We
  make a similar comparison for the \textit{Planck} CMB data (green solid and dashed).
  \label{fixednu}}
\end{figure}

\begin{table*}
\caption{The \maglim\ galaxy bias constraints (mean with 68\% CL) from the fiducial linear bias analysis and the nonlinear bias analysis. Due to a nontrivial second peak in the posterior of the $b^2_4$, we also show in parentheses the 1D marginalized peak value.}
\begin{center}
\begin{tabular}{ l c c c c c c c c }
\hline
\hline
 \quad & $b_1$ & $b_2$ & $b_3$ & $b_4$ & $b^2_1$ & $b^2_2$ & $b^2_3$ & $b^2_4$   \\
\hline
Fid. Linear Bias & $1.49_{-0.10}^{+0.10}$ & $1.69_{-0.11}^{+0.11}$ & $1.91_{-0.12}^{+0.12}$ & $1.79_{-0.12}^{+0.11}$   &-- &-- &-- &--  \\
Nonlinear Bias & $1.44_{-0.08}^{+0.08}$ & $1.60_{-0.10}^{+0.10}$ & $1.85_{-0.11}^{+0.10}$ & $1.74_{-0.11}^{+0.11}$ & $0.12_{-0.15}^{+0.14}$ & $0.06_{-0.26}^{+0.45}$ & $-0.06_{-0.27}^{+0.34}$ & $0.40_{+0.02}^{+0.62}$ ($0.71_{-0.30}^{+0.31}$) \\
\hline
\hline
\end{tabular}
\end{center}
\label{tab:bias}
\end{table*}

In addition to the validation described in the main text of this work and the associated papers described in App. \ref{sec:papers}, we discuss several other analysis modifications in this Appendix that test the robustness of our result.

\subsection{Photometric redshifts} 

In Fig.~\ref{lcdmalt} we show the fiducial 3$\times$2pt analysis compared to two variations in how we use information related to the photo-$z$s of our source sample. First, we simply remove the shear-ratio part of the data vector, which helps to constrain the photo-$z$ parameters, but does not contribute directly to constraining cosmological parameters. Second, we marginalize over the full ensemble of $n(z)$ realizations. Before unblinding, we decided to simplify the way we use the redshift information for the sources and marginalize over the more typical set of parameters that encode shifts in the mean redshift of the average $n(z)$ from the ensembles. This choice was driven by the additional computational expense of marginalizing over the $n(z)$ realizations directly, but was demonstrated in simulated analyses to produce consistent cosmological results at DES Y3 precision. To test this, we run a single chain to show the potential impact of this choice on the real data. While all three results are consistent, with a shift of 0.53$\sigma$ between Hyperrank and marginalizing over $\Delta z$, it is likely that future analyses will need to marginalize over the full ensemble of $n(z)$ realizations directly using something like the Hyperrank process. Finally, we utilize the SOMPZ framework to rederive the redshift distributions of the magnitude limited sample \cite{y3-2x2ptaltlenssompz}, instead of relying on the fiducial DNF redshifts. The 3 × 2pt cosmology inferred from this alternate redshift distribution is fully consistent with the fiducial result.

\begin{table*}
    \centering
    \caption{Summary of internal goodness-of-fit tests with the posterior predictive distribution for analyses with the \redmagic\ and \maglim\ samples. The first and second columns indicate the subset of data $d$ considered. Realizations of $d$ are generated for model parameters drawn from its own posterior and compared to actual observations of $d$, following the method developed in Ref.~\cite{y3-inttensions}. The third column shows the (calibrated) probability-to-exceed for the full data test used for comparison. The other columns indicate the probability-to-exceed for different subsets used for comparison.}
    \begin{tabular}{l c c c c c c c c}
        \hline
        \hline
        Goodness-of-fit test & $d$ & $p(d | d)$  & $p(\xi_\pm | d)$ & $p( w +\gamma_{t} | d)$ & $p(\xi_{+} | d)$ & $p(\xi_{-} | d)$ & $p(\gamma_{t} | d)$ & $p(w | d)$ \\
        \hline
        \textbf{\maglim} \\
        3$\times$2pt  & ${\xi_\pm,\gamma_{t},w}$ & 0.023 & 0.209 & 0.209 & 0.243 & 0.353 & 0.009 & 0.296 \\
        2$\times$2pt & ${\gamma_{t},w}$ & 0.019 & -- & 0.019 & -- & -- & 0.013 & 0.356 \\
        Cosmic shear & ${\xi_\pm}$ & 0.218 & 0.218 & -- & 0.220 & 0.359 & -- & -- \\
        \hline
        \textbf{\redmagic} \\
        3$\times$2pt  & ${\xi_\pm,\gamma_{t},w}$ & 0.005 & 0.242 & 0.001 & 0.260 & 0.410 & 0.029 & 0.041 \\
        2$\times$2pt & ${\gamma_{t},w}$ & 0.038 & -- & 0.038 & -- & -- & 0.187 & 0.072 \\
        Cosmic shear & ${\xi_\pm}$ & 0.255 & 0.255 & -- & 0.253 & 0.397 & -- & -- \\
        \hline
        \hline
    \end{tabular}
    \label{tab:ppd1}
\end{table*}

\begin{table}
    \centering
    \caption{Summary of internal consistency tests with the posterior predictive distribution for analyses with the \redmagic\ and \maglim\ samples. We consider the consistency between different two-point functions, as listed in the first column. The probability-to-exceed $p(a|b)$ is obtained by generating realizations of $a$ for model parameters drawn from the posterior of $b$, and comparing these realizations to the actual observations of $a$. The third and fourth columns indicate the probability-to-exceed for the \redmagic\ and \maglim\ samples respectively.}
    \begin{tabular}{p{4cm} l c c}
        \hline
        \hline
        Consistency test & $p$ & \redmagic & \maglim \\
        \hline
        Cosmic shear \textit{vs} galaxy--galaxy lensing and clustering & ${p(\xi_\pm|\gamma_{t},w)}$ & 0.019 & 0.297 \\
        galaxy--galaxy lensing \textit{vs} cosmic shear and clustering & ${p(\gamma_t|\xi_{\pm},w)}$ & 0.016 & 0.004  \\
        Clustering \textit{vs} cosmic shear and galaxy--galaxy lensing & ${p(w|\xi_{\pm},\gamma_{t})}$ & 0.008 & 0.526 \\
        \hline
        \hline
    \end{tabular}
    \label{tab:ppd2}
\end{table}

\subsection{Nonlinear bias modelling}\label{nlbias}

The fiducial analysis of galaxy clustering and galaxy--galaxy lensing assumes a linear galaxy bias model, which requires removing significant small-scale information in our data vector that exhibits substantial nonlinear behavior. 
This keeps the analysis simpler with fewer free parameters, but potentially wastes this additional information to further constrain cosmology on small scales. 
In our fiducial model, we only utilize large scales above 8 $h^{-1}$Mpc for $w(\theta)$ and 6 $h^{-1}$Mpc for $\gamma_t$ measurements. 
To remedy this, we have developed a nonlinear galaxy bias model and analysis that can utilize this smaller scale information from the lens sample at the expense of marginalizing over additional bias parameters. This is similar to going to smaller scales in cosmic shear and having to marginalize over additional baryonic effect freedom. In both cases, the improvements are limited by needing to simultaneously constrain these additional parameters.

Our model is a hybrid 1-Loop effective field theory model, having five free parameters, as detailed in Refs. \cite{y3-2x2ptbiasmodelling,y3-simvalidation, y3-2x2ptaltlensresults,y3-generalmethods}.  
In Ref.~\cite{2020PhRvD.102l3522P} we validated this model using 3D correlation function measurements in \redmagic\ and \maglim\ mock catalogs and find that this nonlinear bias model agrees with significantly higher signal-to-noise measurements at better than 2\% above scales of 4 $h^{-1}$Mpc. 
For cosmological inferences we fix three of these parameters based on theoretical considerations and validation on these mock catalogs in order to reduce parameter degeneracies and potential parameter volume effects.
We have validated the analysis using multiple buzzard simulation realizations\cite{y3-2x2ptbiasmodelling,y3-simvalidation, y3-2x2ptaltlensresults} (see Sec.~\ref{sec:sim} for details), finding that this model gives a cosmological bias of less than 0.3$\sigma$ in the $\Omega_{\mathrm{m}}$--$S_8$ plane for both the $w(\theta)+\gamma_t$ and 3$\times$2pt probes.

In Fig.~\ref{lcdmalt2} we show the cosmological constraints when applying this model to our data. The galaxy bias constraints are summarized in Table \ref{tab:bias}. We find cosmological constraints in $w$CDM to be consistent with the fiducial analysis. We find a similar improvement in constraining power for $\Lambda$CDM with this nonlinear bias analysis as with the $\Lambda$CDM-optimized analysis that instead incorporates additional small scale information in cosmic shear. With DES Year 6 data, it is possible we could better optimize these choices of how to utilize small-scale information in $\Lambda$CDM analyses to substantially improve constraining power.

\begin{table*}
    \centering
    \caption{A comparison of metrics testing the consistency of independent data sets within the $\Lambda$CDM model. We show results in units of $\sigma$, with a derived probability-to-exceed $p$ shown in parentheses. Overall, we find no significant (defined as $p<0.01$) evidence of disagreement between the DES 3$\times$2pt, external low-redshift, or \textit{Planck} CMB data sets. The details of the metrics and derivation of $p$ are described fully in Ref.~\cite{y3-tensions}.}
    \begin{tabular}{l l  c c  c c c}
    \hline
    \hline
    Data set 1 & Data set 2 & Parameter difference & Suspiciousness & Eigentension & $Q_\mathrm{UDM}$ & $\ln$(Evidence ratio)\\
    \hline
    DES 3$\times$2pt & \textit{Planck} CMB & $1.5 \sigma~ (0.13)$ & $(0.7 \pm 0.1) \sigma \ (0.48 \pm 0.08)$ & $1.2\sigma ~ (0.23)$ & $1.6\sigma ~ (0.10)$  & $5.9\pm0.5$ \\
    DES 3$\times$2pt & Ext. Low-$z$ & $1.3 \sigma~ (0.11)$ & $(1.6 \pm 0.3) \sigma \ (0.13 \pm 0.07)$  & $0.9\sigma ~ (0.40)$ & $1.4 \sigma ~(0.11)$  & $1.6\pm0.4$ \\
    Ext. Low-$z$ & \textit{Planck} CMB & $1.1 \sigma ~(0.27)$ & -- & $1.5\sigma ~ (0.13)$ & $0.6 \sigma ~(0.54)$& -- \\
    DES 3$\times$2pt + Ext. Low-$z$ & \textit{Planck} CMB & $0.9 \sigma~ (0.34)$ & -- & $1.2\sigma ~ (0.25)$ & $0.8 \sigma~ (0.42)$ & -- \\
    DES $\Lambda$CDM-Opt. 3$\times$2pt & \textit{Planck} CMB & $1.9 \sigma~ (0.06)$ & -- & $1.3\sigma ~ (0.21)$ & -- & -- \\
    DES Hyperrank 3$\times$2pt & \textit{Planck} CMB & $2.2 \sigma ~(0.02)$ & -- & $2.0\sigma ~ (0.04)$ & -- & -- \\
    DES redMaGiC 3$\times$2pt & \textit{Planck} CMB & $1.5 \sigma  ~(0.13)$ & -- & $1.0 \sigma ~ (0.30)$ & -- & -- \\ 
    \hline
    \hline
\end{tabular}
\label{tab:consistency}
\end{table*}

\subsection{Lens magnification bias parameters}

In the fiducial analysis, we fix the lens magnification coefficients $C_{l}^{i}$ to values derived in Ref.~\cite{y3-2x2ptmagnification}. In Fig.~\ref{lcdmalt2} we show the impact of freeing these coefficients with a wide flat prior between $-6$ and 10 in each lens bin. We find the cosmological constraints in $\Lambda$CDM to be consistent with the fiducial analysis. Further investigations into the lens magnification coefficients and their impact on cosmological constraints can be found in Ref.~\cite{y3-2x2ptmagnification}. 

\subsection{Intrinsic alignment models}

In the fiducial analysis, we use the full TATT intrinsic alignment model. In Fig.~\ref{lcdmalt2} we show the impact of limiting the intrinsic alignment model to the more commonly used NLA model with free redshift power-law evolution. This is the same as fixing the parameters $A_2=b_{\mathrm{TA}}=0$ in the TATT model. We find consistent cosmological parameters, with a gain of only 13\% relative to TATT in the $\sigma_8$--$\Omega_m$ parameter plane. The use of NLA was not motivated a priori as the fiducial IA model, but due to the IA amplitude parameters being constrained to be small in TATT, $A_2=b_{\mathrm{TA}}=0$ is not a poor approximation.

\subsection{Simulation validation tests}

We reproduce the fiducial analysis on a suite of 18 Buzzard simulations described in Sec.~\ref{sec:sim}. This is shown in Fig.~\ref{simfig}, where the true cosmology is indicated by the cross. We compare two simulated analysis. 
The first analysis uses a synthetic, noiseless data vector based on the true $n(z)$ from Buzzard and without marginalizing over shear or photo-$z$ parameters.
The second reproduces the full analysis on the mean data vector of all 18 simulation realizations, which marginalizes over all nuisance parameters and uses an photo-$z$ $n(z)$ that is inferred from the same process we apply to the real survey data.
We find that the two simulated analyses agree very well with each other and with the true cosmology for each simulated data vector.

\subsection{Neutrino mass} 

Finally, we also show a version of the analysis that fixes the neutrino mass density at the minimum allowed mass in Fig.~\ref{fixednu}, comparing to a similar fixed neutrino mass density of the \textit{Planck} CMB data.

\section{Details of Internal and External Consistency}\label{consistency}

Results of the final internal consistency tests are reported in Tables~\ref{tab:ppd1} and \ref{tab:ppd2}. Table~\ref{tab:ppd1} reports results of the goodness-of-fit tests where realizations of a subset of the data vector are generated for model parameters drawn from its own posterior. We show results for analyses using the \maglim\ lens sample and for \redmagic. Table~\ref{tab:ppd2} reports extra consistency tests between the two-point functions that were not included as unblinding criteria. In these consistency tests, the data vector is split in two, and one part is used to predict the other. For instance, the first row of Table~\ref{tab:ppd2} shows the result of a test where realizations of cosmic shear are generated for model parameters drawn from the $2\times2$pt posterior, and compared to observations of cosmic shear. All pre-defined unblinding requirements were met for these internal data combinations, and both 3$\times$2pt and the cosmologically-constraining subsets of the data for both lens samples show consistency with each other and with $\Lambda$CDM. However, there persists evidence of potential issues with the consistency of some parts of the combination of galaxy clustering and galaxy--galaxy lensing relative to the best-fit model of the full 3$\times$2pt or cosmic shear data. In particular, for \redmagic\ this seems slightly more likely to be sourced primarily from the clustering data.

We report detailed consistency metrics between external data set pairs in Table~\ref{tab:consistency}. The details of the metrics and derivation of probability-to-exceed $p$ are described fully in Ref.~\cite{y3-tensions}. Overall, we find no significant (defined as $p<0.01$) evidence of disagreement between the DES 3$\times$2pt, external low-redshift, or \textit{Planck} CMB data sets shown in Figs.~\ref{extlcdm} \& \ref{extwcdm}. We show both a method measuring parameter differences and a method based on the evidence (`Suspiciousness') for the two primary comparisons of DES 3$\times$2pt with external data sets. The evidence ratio is also shown, and all three metrics give qualitatively consistent results. 

All data set combinations are largely in agreement, according to all tension metrics. 
Some differences between methods are likely due to non-Gaussianity of both posteriors involved, as well as the different approaches that the methods employ in treating the impact of priors on tension quantifications, but in all cases are below a fraction of a sigma.
We also note that the Eigentension result for the case of DES 3$\times$2pt vs complementary external low-redshift probes may not be capturing the full tension; the two data sets have comparable constraining powers, for which case Eigentension is not optimized.

\end{document}